\documentclass[preprintnumbers, floatfix, letterpaper, superscriptaddress,nofootinbib,aps]{revtex4-2}
\usepackage{graphicx}
\usepackage{microtype}
\usepackage{amsmath}
\usepackage{amssymb}
\usepackage{subfigure}
\usepackage{hyperref}
\usepackage{url}
\usepackage{xcolor}
\usepackage{color}
\usepackage{mathrsfs}
\usepackage{amsfonts}
\usepackage{latexsym}
\usepackage{ragged2e}
\usepackage{textcomp}
\usepackage{phaistos}
\usepackage[utf8]{inputenc}
\usepackage[normalem]{ulem}
\usepackage{float}
\usepackage{natbib}

\makeatletter
\renewcommand\@makefnmark{\hbox{\@textsuperscript{\normalfont\color{purple}\@thefnmark}}}
\renewcommand\@makefntext[1]{%
  \parindent 1em\noindent
            \hb@xt@1.8em{%
                \hss\@textsuperscript{\normalfont\@thefnmark}}#1}
\makeatother

\usepackage{caption}
\DeclareCaptionJustification{justified}{\leftskip=0pt \rightskip=0pt \parfillskip=0pt plus 1fil}
\definecolor{vividviolet}{rgb}{0.62, 0.0, 1.0}
\definecolor{amaranth}{rgb}{0.9, 0.17, 0.31}
\definecolor{palatinateblue}{rgb}{0.15, 0.23, 0.89}
\definecolor{brightpink}{rgb}{1.0, 0.0, 0.5}
\definecolor{cornflowerblue}{rgb}{0.39, 0.58, 0.93}
\definecolor{deepcarminepink}{rgb}{0.94, 0.19, 0.22}
\definecolor{radicalred}{rgb}{1.0, 0.21, 0.37}

\hypersetup{ linktoc=all,
    colorlinks, linkcolor={palatinateblue},
    citecolor={brightpink}, urlcolor={amaranth}
}

\graphicspath{{Images/}}

\def\sideremark#1{\ifvmode\leavevmode\fi\vadjust{\vbox to0pt{\vss
 \hbox to 0pt{\hskip\hsize\hskip1em
 \vbox{\hsize1.5cm\tiny\raggedright\pretolerance10000
 \noindent #1\hfill}\hss}\vbox to8pt{\vfil}\vss}}}%
\begin{document}

\title{Purely Electric, Magnetic, and Dyonic Black Holes in Einstein-Euler-Heisenberg Theory}

\author{Haoyuan \surname{Luo}}
\email{2770521341@qq.com}
\affiliation{School of Science, Jiangsu University of Science and Technology, 212100, Zhenjiang, China.}

\author{Nan \surname{Cao}}
\email{1913562551@qq.com}
\affiliation{School of Science, Jiangsu University of Science and Technology, 212100, Zhenjiang, China.}

\author{Xiao Yan \surname{Chew}}
\email{xiao.yan.chew@just.edu.cn}
\affiliation{School of Science, Jiangsu University of Science and Technology, 212100, Zhenjiang, China.}

\author{Kok-Geng \surname{Lim}}
\email{K.G.Lim@soton.ac.uk}
\affiliation{University of Southampton Malaysia, 79100 Iskandar Puteri, Malaysia.}

\author{Chao \surname{Chen}}
\email{cchao012@just.edu.cn}
\affiliation{School of Science, Jiangsu University of Science and Technology, 212100, Zhenjiang, China.}

\author{Dong-han \surname{Yeom}}
\email{innocent.yeom@gmail.com}
\affiliation{Department of Physics Education, Pusan National University, Busan 46241, Republic of Korea}
\affiliation{Research Center for Dielectric and Advanced Matter Physics, Pusan National University, Busan 46241, Republic of Korea}
\affiliation{Leung Center for Cosmology and Particle Astrophysics, National Taiwan University, Taipei 10617, Taiwan }

\begin{abstract}
We investigate static, spherically symmetric charged black holes in Einstein gravity coupled to Euler--Heisenberg (EH) nonlinear electrodynamics, including purely electric, purely magnetic, and dyonic configurations. Rather than adopting the Hamiltonian formulation based on the auxiliary electromagnetic invariant $\mathcal{P}$, we work directly with the physical electromagnetic invariant $\mathcal{F}$ in the Einstein-Euler--Heisenberg Lagrangian, thereby describing all charged configurations without introducing auxiliary variables. Within this approach, we derive an exact analytical solution for the purely electric case, recover the purely magnetic solution directly from the field equations, and construct the dyonic solutions numerically. We systematically study the horizon structure, causal properties, and thermodynamics of these solutions. While the purely electric branch exhibits the familiar Reissner--Nordstr\"om horizon structure, the purely magnetic branch naturally admits a novel three-horizon configuration consisting of one event horizon and two inner horizons. The dyonic solutions continuously interpolate between the electric and magnetic limits and exhibit either one- or three-horizon configurations, depending on the magnetic-to-electric charge ratio and the EH coupling. We further show that the EH nonlinear interaction significantly modifies the horizon structure and thermodynamic properties of charged black holes while leaving the central curvature singularity unresolved. These results demonstrate that EH nonlinear electrodynamics gives rise to qualitatively new causal structures beyond Einstein--Maxwell theory.
\end{abstract}

\date{\today}

\maketitle



\section{Introduction}

The direct detection of gravitational waves from the binary black-hole merger GW150914 by the LIGO and Virgo Collaborations marked the beginning of gravitational-wave astronomy and provided the first direct evidence for coalescing black-hole binaries \cite{LIGOScientific:2016lio}. Since then, successive observing runs of the LIGO--Virgo--KAGRA Collaboration have revealed a rich population of compact-binary mergers \cite{LIGOScientific:2016aoc,LIGOScientific:2020iuh,KAGRA:2021vkt}. Complementing these discoveries, the Event Horizon Telescope Collaboration obtained the first horizon-scale image of the supermassive black hole M87 in 2019 and subsequently imaged Sagittarius A*, the supermassive black hole at the center of our Galaxy in 2022 \cite{EventHorizonTelescope:2019dse,EventHorizonTelescope:2019ths,EventHorizonTelescope:2022wkp,EventHorizonTelescope:2022wok}. These observations have established black holes as unique laboratories for testing gravity in the strong-field regime and have stimulated extensive studies of black-hole solutions beyond the Einstein--Maxwell (EM) theory.

Within EM theory, the Reissner--Nordstr\"om (RN) spacetime is the unique asymptotically flat, static electrovacuum black-hole solution under suitable regularity assumptions \cite{Israel:1967za,osti_4225613,POMazur_1982,Heusler_1996}. In contrast to the Schwarzschild solution, the RN geometry possesses both an outer event horizon and an inner Cauchy horizon. While the event horizon defines the observable boundary of the black hole, the Cauchy horizon gives rise to several well-known pathologies. In particular, the infinite blueshift of infalling perturbations triggers the mass inflation instability, driving the inner horizon toward a curvature singularity \cite{Simpson:1973ua,Poisson:1990eh,Ori:1991zz,Bonanno:1994ma}. These results strongly suggest that the Cauchy horizon of the stationary RN solution is dynamically unstable and motivate the search for charged black holes beyond linear electrodynamics.

One promising avenue is provided by nonlinear electrodynamics (NED), which naturally modifies the electromagnetic interaction in the strong-field regime. An early example is the Bardeen black hole, demonstrating that nonlinear electromagnetic effects can regularize the spacetime geometry. Subsequently, E. Ay\'on-Beato and A. Garc\'ia constructed the first regular black-hole solution derived directly from a nonlinear electromagnetic Lagrangian \cite{Ayon-Beato:1998hmi}, whereas the Hayward black hole was obtained by first postulating a regular metric and subsequently reconstructing the corresponding NED theory \cite{Hayward:2005gi}. Despite their success in resolving spacetime singularities, these models are generally regarded as effective descriptions because the nonlinear electromagnetic Lagrangian is either reconstructed from a prescribed geometry or introduced phenomenologically, rather than derived from a fundamental theory of electromagnetic interactions \cite{Bronnikov:2000vy,Zhou:2022yio,Bokulic:2023afx,BUENO2025139260}. Comprehensive reviews of black holes in NED can be found in Refs.~\cite{Fan:2016hvf,Sorokin:2021tge,Bronnikov:2022ofk,Lan:2023cvz}.

A more physically motivated NED theory is Born--Infeld (BI) electrodynamics, originally proposed to remove the divergent self-energy of point charges \cite{Born:1934gh}. Owing to its elegant mathematical structure and its intimate connection with string theory, BI electrodynamics has attracted sustained interest in gravitational physics \cite{Tseytlin:1999dj,Gibbons:2001gy}. Nevertheless, despite its theoretical appeal, BI electrodynamics has not been experimentally established as the fundamental theory describing electromagnetic interactions \cite{Carley:2006zz,Franklin:2011zz,Breton:2012yt}. Instead, quantum electrodynamics (QED) provides the experimentally verified description of electromagnetic interactions, whose leading nonlinear corrections are encoded in the Euler--Heisenberg (EH) effective Lagrangian \cite{Heisenberg:1936nmg}.

The EH theory predicts several genuinely quantum phenomena absent in classical electrodynamics, including photon--photon scattering and vacuum birefringence \cite{Euler:1935qgl}. Recent polarization measurements of neutron stars and magnetars provide increasing observational support for these nonlinear vacuum effects \cite{Mignani:2016fwz,Stewart:2025cmt,Taverna:2026syy}, thereby strengthening the physical motivation for investigating Einstein gravity coupled to EH electrodynamics.

NED coupled to gravity gives rise to a considerably richer class of black-hole solutions than the EM theory, including regular black holes, modified causal structures, and spacetimes with multiple horizons. Besides the familiar one- and two-horizon configurations, suitably engineered NED can even support black holes possessing several distinct horizons \cite{Nojiri:2017kex,Gao:2021kvr}. However, such solutions are typically obtained by reconstructing or designing the nonlinear electromagnetic Lagrangian to realize a prescribed spacetime geometry. It therefore remains an open question whether similarly rich horizon structures can arise naturally within a physically motivated effective theory such as EH electrodynamics.

Although Einstein--Euler--Heisenberg (EEH) black holes have attracted increasing attention in recent years \cite{Amaro:2020xro,Magos:2020ykt,Breton:2021mju,Fu:2021akc,Magos:2023nnb,Abbas:2023nra,Sekhmani:2024vsu,Hamil:2024njs,Jafarzade:2025byr,MONDAL2025116859,Ditta:2025ezx,Ditta:2026yvp}, most existing studies formulate the theory in terms of the auxiliary electromagnetic invariant $\mathcal{P}$ within the Hamiltonian framework, which considerably simplifies the field equations. While computationally convenient, this approach obscures the direct relation between the spacetime geometry and the original EEH Lagrangian expressed in terms of the physical electromagnetic invariant $\mathcal{F}$. This motivates us to revisit the EEH theory directly at the Lagrangian level, without introducing auxiliary variables.

Hence, we work with the EEH field equations directly in terms of the physical electromagnetic invariant $\mathcal{F}$ in this paper. The electromagnetic field equations reduce to an exactly solvable cubic equation, allowing us to derive an exact analytical solution for the purely electric black hole in terms of Gaussian hypergeometric functions and to recover the purely magnetic solution directly from the original field equations. For the dyonic configuration, the coupled ordinary differential equations (ODEs) do not appear to admit analytical solutions, and thus the corresponding black-hole solutions are constructed numerically.

Building on this formulation, we perform a systematic analysis of the horizon structure, spacetime geometry, causal structure, and thermodynamic properties of purely electric, purely magnetic, and dyonic EEH black holes. We show that the purely electric solutions possess a horizon structure qualitatively similar to that of the RN black hole, whereas the purely magnetic solutions naturally admit a novel three-horizon phase consisting of one event horizon and two inner horizons. The dyonic solutions continuously interpolate between these limiting cases and exhibit one-, two-, and three-horizon configurations depending on the magnetic-to-electric charge ratio and the EH coupling. Unlike previously constructed multi-horizon solutions obtained by engineering nonlinear electrodynamics, the rich horizon structure found here emerges naturally from the physically motivated EH effective theory. These results demonstrate that quantum nonlinear electromagnetic corrections significantly enrich the causal structure and thermodynamic properties of charged black holes beyond Einstein--Maxwell theory.

The remainder of this paper is organized as follows. In Sec.~\ref{sec:th}, we introduce the EEH theory and some basic properties of static and spherically symmetric black holes, derive the corresponding ODEs from equations of motion. In Sec.~\ref{sec:Re}, we investigate the purely electric, purely magnetic, and dyonic black-hole solutions, presenting both analytical and numerical results. Finally, in Sec.~\ref{sec:Conc}, we summarize our main results and discuss possible directions for future research.

\section{Theoretical Framework}
\label{sec:th}

\subsection{Einstein--Euler--Heisenberg theory}

Euler--Heisenberg (EH) electrodynamics provides the leading nonlinear correction to classical Maxwell theory arising from quantum-electrodynamic vacuum polarization in the weak-field, low-frequency regime. When coupled to Einstein gravity, it offers a simple effective framework for investigating how quantum nonlinear electromagnetic effects modify the geometry and thermodynamics of charged black holes.

The Einstein--Euler--Heisenberg (EEH) theory is described by the following action
\begin{equation}
S=\frac{1}{16\pi}\int d^4x\,\sqrt{-g}
\left[R-\mathcal{L}_{\rm EH}(\mathcal{F},\mathcal{G})\right]\,,
\end{equation}
where the one-loop EH effective Lagrangian $\mathcal{L}_{\rm EH}(\mathcal{F},\mathcal{G})$ is given by \cite{Heisenberg:1936nmg}
\begin{equation}
\mathcal{L}_{\rm EH}(\mathcal{F},\mathcal{G})
=
\mathcal{F}
-\epsilon\left(
\mathcal{F}^2+\frac{7}{4}\mathcal{G}^2
\right)\,.
\end{equation}
Here, $\epsilon$ is the EH nonlinear coupling parameter, and
\begin{equation}
\mathcal{F}\equiv F_{\mu\nu}F^{\mu\nu}\,,
\qquad
\mathcal{G}\equiv \tilde{F}^{\mu\nu}F_{\mu\nu} \,,
\end{equation}
are the two electromagnetic invariants. The electromagnetic field tensor and
its Hodge dual are defined by
\begin{equation}
F_{\mu\nu}=\partial_\mu A_\nu-\partial_\nu A_\mu\,,
\qquad
\tilde{F}^{\mu\nu}
=
\frac{1}{2}\eta^{\mu\nu\rho\sigma}F_{\rho\sigma}\,,
\end{equation}
where $\eta^{\mu\nu\rho\sigma}$ is the completely antisymmetric Levi-Civita
tensor. For purely electric or purely magnetic configurations, $\mathbf{E}\cdot\mathbf{B}=0$, so that $\mathcal{G}=0$, whereas the $\mathcal{G}^2$ correction becomes indispensable for dyonic configurations. In the limit $\epsilon=0$, the theory reduces to the standard
EM theory. 

Varying the action with respect to the metric and gauge potential yields
\begin{align}
R_{\mu\nu}-\frac{1}{2}g_{\mu\nu}R
&=
2\left(
\mathcal{L}_{\mathcal{F}}
F_{\mu\lambda}F_{\nu}{}^{\lambda}
+
\mathcal{L}_{\mathcal{G}}
F_{\mu\lambda}\tilde{F}_{\nu}{}^{\lambda}
\right)
-\frac{1}{2}g_{\mu\nu}\mathcal{L}_{\rm EH} \,,
\\
\nabla_\mu\left(
\mathcal{L}_{\mathcal{F}}F^{\mu\nu}
+
\mathcal{L}_{\mathcal{G}}\tilde{F}^{\mu\nu}
\right)
&=0\,,
\label{eq:EOM_gauge}
\end{align}
where
\begin{equation}
\mathcal{L}_{\mathcal{F}}
\equiv
\frac{\partial\mathcal{L}_{\rm EH}}{\partial\mathcal{F}}\,,
\qquad
\mathcal{L}_{\mathcal{G}}
\equiv
\frac{\partial\mathcal{L}_{\rm EH}}{\partial\mathcal{G}}\,.
\end{equation}

To construct static and spherically symmetric black-hole solutions, we employ
the metric ansatz
\begin{equation}
\label{eq:metric}
ds^2
=
-e^{-2\sigma(r)}N(r)dt^2
+\frac{dr^2}{N(r)}
+r^2\left(d\theta^2+\sin^2\theta\,d\varphi^2\right)\,,
\end{equation}
where
\begin{equation}
N(r)=1-\frac{2m(r)}{r}\,. \label{funcN}
\end{equation}
The function $m(r)$ is the Misner-Sharp mass function, and the ADM mass is
obtained from
\begin{equation}
M=\lim_{r\rightarrow\infty}m(r)\,.
\end{equation}

The most general static and spherically symmetric gauge potential compatible
with electric and magnetic charges is
\begin{equation}
A_\mu dx^\mu
=
V(r)\,dt+P\cos\theta\,d\varphi\,,
\end{equation}
where $P$ denotes the magnetic charge.

Substituting the metric and gauge-field ans\"atze into the Einstein equations
gives
\begin{align}
\frac{1-N}{r^2}
-\frac{N'}{r}
-P^2\left(
\frac{1}{r^4}-\frac{2\epsilon P^2}{r^8}
\right)
&=
e^{2\sigma}V'^2
\left(
1+6\epsilon e^{2\sigma}V'^2
+\frac{10\epsilon P^2}{r^4}
\right)\,,
\label{eq:gtt}
\\
\frac{1-N}{r^2}
+\frac{2N\sigma'}{r}
-\frac{N'}{r}
-P^2\left(
\frac{1}{r^4}-\frac{2\epsilon P^2}{r^8}
\right)
&=
e^{2\sigma}V'^2
\left(
1+6\epsilon e^{2\sigma}V'^2
+\frac{10\epsilon P^2}{r^4}
\right)\,,
\label{eq:grr}
\\
\frac{N''}{2}
+\left(\frac{1}{r}-\frac{3\sigma'}{2}\right)N'
+\left(
\sigma'^2-\frac{\sigma'}{r}-\sigma''
\right)N
-P^2\left(
\frac{1}{r^4}-\frac{6\epsilon P^2}{r^8}
\right)
&=
e^{2\sigma}V'^2
\left(
1+2\epsilon e^{2\sigma}V'^2
-\frac{10\epsilon P^2}{r^4}
\right)\,.
\label{eq:gthth}
\end{align}
Here and below, a prime denotes differentiation with respect to $r$.

Subtracting Eq.~\eqref{eq:gtt} from Eq.~\eqref{eq:grr} gives
\begin{equation}
N(r)\sigma'(r)=0\,.
\end{equation}
Outside the horizons, where $N(r)\neq0$, this implies that $\sigma$ is a
constant. Asymptotic flatness and a rescaling of the time coordinate allow us
to set
\begin{equation}
\sigma(r)=0\,.
\label{eq:sigma}
\end{equation}

The gauge field Eq.~\eqref{eq:EOM_gauge} reduces to
\begin{equation}
\left[
r^2V'
\left(
1+4\epsilon V'^2+\frac{10\epsilon P^2}{r^4}
\right)
\right]'
=0\,.
\end{equation}
After one integration, we obtain
\begin{equation}
V'^3
+\frac{1}{4\epsilon}
\left(
1+\frac{10\epsilon P^2}{r^4}
\right)V'
=
-\frac{Q}{4\epsilon r^2}\,,
\label{eq:dyn_cubic}
\end{equation}
where $Q$ is the electric charge

Using Eq.~\eqref{funcN}, Eq.~\eqref{eq:gtt} can be written as
\begin{equation}
\label{eq:dyn_mprime}
m'(r)
=
\frac{P^2}{2r^2}
\left(
1-\frac{2\epsilon P^2}{r^4}
\right)
+
\frac{r^2V'^2}{2}
\left(
1+6\epsilon V'^2
+\frac{10\epsilon P^2}{r^4}
\right)\,.
\end{equation}
Thus, Eqs.~\eqref{eq:dyn_cubic} and~\eqref{eq:dyn_mprime} constitute a minimal set of ordinary differential equations for constructing the black-hole solutions. Eq.~\eqref{eq:gthth} is not independent and is automatically satisfied as a consequence of the preceding equations and the contracted Bianchi identity.

Defining the radial electric field $E(r)$ by
\begin{equation}
E(r)\equiv -V'(r)\,,
\end{equation}
then Eq.~\eqref{eq:dyn_cubic} becomes the cubic equation,
\begin{equation}
E(r)^3+p_1(r)E(r)+p_2(r)=0\,,
\label{eq:E_cubic}
\end{equation}
where
\begin{equation}
p_1(r)
=
\frac{1}{4\epsilon}
\left(
1+\frac{10\epsilon P^2}{r^4}
\right)\,,
\qquad
p_2(r)
=
-\frac{Q}{4\epsilon r^2}\,.
\end{equation}

The discriminant $\Delta(r)$ of Eq.~\eqref{eq:E_cubic} is defined as
\begin{equation}
\Delta(r)
=
\left(\frac{p_2}{2}\right)^2
+
\left(\frac{p_1}{3}\right)^3\,.
\end{equation}
The Cardano quantities are defined as
\begin{equation}
E_1^\pm(r)
=
\sqrt[3]{
-\frac{p_2}{2}
\pm
\sqrt{
\left(\frac{p_2}{2}\right)^2
+
\left(\frac{p_1}{3}\right)^3
}
}\,,
\label{eq:discri}
\end{equation}
and the three formal roots are given by
\begin{align}
E_1(r)
&=
E_1^+(r)+E_1^-(r)\,,
\label{eq:root1}
\\
E_2(r)
&=
-\frac{1}{2}E_1(r)
+\frac{\sqrt{3}}{2}i
\left[E_1^+(r)-E_1^-(r)\right]\,,
\label{eq:root2}
\\
E_3(r)
&=
-\frac{1}{2}E_1(r)
-\frac{\sqrt{3}}{2}i
\left[E_1^+(r)-E_1^-(r)\right]\,.
\label{eq:root3}
\end{align}

For real $p_1$ and $p_2$, the sign of $\Delta$ determines the nature of the
roots:
\begin{equation}
\begin{cases}
\Delta>0: & \text{one real root and one complex-conjugate pair}\,,\\
\Delta=0: & \text{three real roots, at least two of which coincide}\,,\\
\Delta<0: & \text{three distinct real roots}\,.
\end{cases}
\end{equation}
For the physical EH coupling, i.e., $\epsilon>0$, one has $p_1>0$ and
therefore
\begin{equation}
\Delta
=
\left(\frac{p_2}{2}\right)^2
+
\left(\frac{p_1}{3}\right)^3
>0\,.
\end{equation}
Consequently, the cubic equation possesses a unique real electric-field branch, namely $E_1(r)$, while $E_2(r)$ and $E_3(r)$ form a complex-conjugate pair.

Although Cardano-type expressions have appeared in previous studies of EEH black holes
\cite{Costa:2013tha,Costa:2013xba,Kruglov:2017ymn,Guerrero:2020uhn}, the present formulation will be used to construct and compare the purely
electric $(P=0)$, purely magnetic $(Q=0)$, and dyonic $(P\neq0,Q\neq0)$
solutions directly in terms of the original electromagnetic invariant
$\mathcal{F}$.

\subsection{Einstein--Maxwell limit: Reissner--Nordstr\"om black hole}

In the limit $\epsilon=0$, the EEH theory reduces to the Einstein-Maxwell (EM) theory. The field equations become
\begin{align}
m'(r)
&=
\frac{P^2}{2r^2}
+\frac{r^2V'^2}{2}\,,
\\
V'(r)
&=-\frac{Q}{r^2} \,.
\label{RN_ode2}
\end{align}
Choosing the gauge $V(\infty)=0$, the electric potential $V(r)$ is
\begin{equation}
V(r)=\frac{Q}{r}\,.
\end{equation}

Substituting the electromagnetic field into the mass equation gives
\begin{equation}
m'(r)
=
\frac{Q^2+P^2}{2r^2}\,,
\end{equation}
and therefore
\begin{equation}
m(r)
=
M-\frac{Q^2+P^2}{2r}\,.
\end{equation}
The metric function $N(r)$ for Reissner-Nordstr\"om (RN) black hole is
\begin{equation}
N(r)
=
1-\frac{2M}{r}
+\frac{Q^2+P^2}{r^2}\,.
\end{equation}
Introducing the dimensionless total charge-to-mass ratio $q$,
\begin{equation}
q=\frac{\sqrt{Q^2+P^2}}{M}\,, \label{q-ratio}
\end{equation}
the horizon radii $r_\pm$ are obtained by solving $N(r)=0$,
\begin{equation}
\frac{r_\pm}{M}
=
1\pm\sqrt{1-q^2}\,.
\end{equation}
Here, $r_+$ is the event horizon, whereas $r_-$ is the inner, or Cauchy, horizon. For $q=1$, the two horizons coincide and the black hole becomes
extremal.

\subsection{Comparison with the auxiliary invariant formulation}

Most previous studies of EEH black holes
\cite{Amaro:2020xro,Magos:2020ykt,Breton:2021mju,Fu:2021akc,Magos:2023nnb,Abbas:2023nra,Sekhmani:2024vsu,Hamil:2024njs,Jafarzade:2025byr,MONDAL2025116859,Ditta:2025ezx,Ditta:2026yvp}
adopt the Hamiltonian formulation of NED, in which the dynamics is expressed in terms of the auxiliary electromagnetic tensor
$P_{\mu\nu}$ and its associated invariant $\mathcal{P}$,
\begin{equation}
\mathcal{H}(\mathcal{P},\mathcal{O})
=
-\frac12P^{\mu\nu}F_{\mu\nu}
-\mathcal{L}_{\rm EH}\,,
\qquad
P_{\mu\nu}
=
\frac{\partial\mathcal{L}_{\rm EH}}
{\partial F^{\mu\nu}}\,,
\qquad
\mathcal{O}
=
-P_{\mu\nu}\tilde P^{\mu\nu}\,.
\end{equation}
For a static, spherically symmetric, purely electric configuration,
$\mathcal{O}=0$, the generalized Maxwell equation becomes
\begin{equation}
\nabla_{\mu}P^{\mu\nu}=0
\quad\Longrightarrow\quad
\left(r^2P^{rt}\right)'=0\,,
\end{equation}
which immediately gives
\begin{equation}
P^{rt}=\frac{Q}{r^2},
\end{equation}
where $Q$ is the electric charge. Hence,
\begin{equation}
\mathcal{P}
=
-P_{\mu\nu}P^{\mu\nu}
=
-2P_{rt}P^{rt}
=
\frac{2Q^2}{r^4}\,,
\end{equation}
where the factor of two follows from the antisymmetry of
$P_{\mu\nu}$.

The physical electromagnetic invariant is then obtained from the constitutive relation
\begin{equation}
\mathcal{P}
=
-\mathcal{L}_{\mathcal F}^{\,2}\mathcal{F}
=
-\left(1-2\epsilon\mathcal{F}\right)^2\mathcal{F}\,,
\end{equation}
which cannot be inverted analytically. Therefore, previous studies resort to a perturbative expansion in the EH coupling,
\begin{equation}
\mathcal{F}
=
-\mathcal{P}
+
4\epsilon\mathcal{P}^2
+
O(\epsilon^2)
\approx
-\frac{2Q^2}{r^4}
\left(
1-\frac{8\epsilon Q^2}{r^4}
\right)\,,
\end{equation}
leading to the approximate electric field
\begin{equation}\label{eq:elec_field_approx}
E(r)
=
\frac{Q}{r^2}
\left(
1-\frac{4\epsilon Q^2}{r^4}
\right)
+
O(\epsilon^2)\,,
\end{equation}
which coincides with the result reported in Ref.~\cite{Magos:2020ykt}. In contrast, the formulation adopted in the present work is based directly on the physical electromagnetic invariant $\mathcal{F}$, allowing the electric field to be obtained exactly without invoking a perturbative expansion.

\subsection{Properties of the black holes}

\subsubsection{Scaling symmetry}

The EEH field equations are invariant under the scaling transformation
\begin{equation}
    r\rightarrow\lambda r,\qquad
    (M,Q,P)\rightarrow\lambda(M,Q,P),\qquad
    \epsilon\rightarrow\lambda^2\epsilon \,,
\end{equation}
where $\lambda>0$ is a constant. By contrast, in the EM limit, the corresponding transformation takes the simpler form
\begin{equation}
    r\rightarrow\lambda r,\qquad
    (M,Q,P)\rightarrow\lambda(M,Q,P)\,,
\end{equation}
without requiring the transformation of an additional coupling parameter.
Thus, the dimensionful EH coupling introduces a new physical scale and breaks the fixed-coupling scaling symmetry of the RN solution.

It is therefore convenient to introduce the dimensionless nonlinear
coupling
\begin{equation}
    \tilde{\epsilon}\equiv\frac{\epsilon}{M^2}\,,
\end{equation}
which remains invariant under the EEH scaling transformation. Similarly, we introduce several dimensionless parameters:
\begin{equation}
    q_e=\frac{Q}{M}\,,\qquad
    q_m=\frac{P}{M}\,,\qquad
    \beta=\frac{P}{Q}\,.
\end{equation}
Together with Eq.~\eqref{q-ratio}, the parameters $(q,q_e,q_m,\beta)$ provide an equivalent parametrization of the electric and magnetic charges, where the first three quantities measure the corresponding charge-to-mass ratio and $\beta$ characterizes the relative strength of the magnetic and electric charges.

\subsubsection{Curvature invariants}

To characterize the geometry of the EEH black holes, we evaluate the Ricci scalar $R$ and the Kretschmann scalar $K$. These quantities provide
coordinate-independent measures of the spacetime curvature and are particularly useful for identifying curvature singularities.

For the metric ansatz~\eqref{eq:metric}, the Ricci scalar is
\begin{align}
R={}&
\frac{2(1-N)}{r^2}
-\frac{4N'}{r}
-N''
+\frac{6N\sigma'}{r}
+3N'\sigma'
+2N\sigma''
-2N\sigma'^2 \,,
\end{align}
whereas the Kretschmann scalar is
\begin{equation}
K={}
\frac{4(1-N)^2}{r^4}
+\frac{2N'^2}{r^2}
+\frac{2\left(N'-2N\sigma'\right)^2}{r^2}
+
\left[
N''-3N'\sigma'
+2N\left(\sigma'^2-\sigma''\right)
\right]^2\,.
\end{equation}

Since $\sigma(r)=0$, these expressions reduce to
\begin{align}
R
&=
-N''
-\frac{4N'}{r}
+\frac{2(1-N)}{r^2} \,,
\label{eq:Ricci_geometric}
\\
K
&=
N''^2
+\frac{4N'^2}{r^2}
+\frac{4(1-N)^2}{r^4}\,.
\label{eq:scalar_Kretsh}
\end{align}
Using the field equations, the Ricci scalar can equivalently be written as
\begin{equation}
R
=
8\epsilon
\left(
\frac{P^4}{r^8}
+\frac{5P^2V'^2}{r^4}
+V'^4
\right)\,.
\label{eq:scalar_Ricci}
\end{equation}
This expression explicitly shows that the Ricci scalar vanishes in the
EM limit $(\epsilon=0)$, as expected from the tracelessness of the energy--momentum tensor in the EM theory.

\subsubsection{Horizon area and Hawking temperature}

The thermodynamic properties of a static and spherically symmetric black
hole can be characterized by its horizon area $A_H$ and Hawking temperature
$T_H$,
\begin{equation}
    A_H=4\pi r_H^2,\qquad
    T_H=\frac{1}{4\pi} e^{-\sigma_H}N'(r_H) \,,
\end{equation}
where $r_H$ denotes the radius of the event horizon and
$\sigma_H=\sigma(r_H)$. Since $\sigma(r)=0$ for the present solutions,
\begin{equation}
    T_H=\frac{1}{4\pi} N'(r_H)\,.
\end{equation}

To quantify deviations from the Schwarzschild and RN solutions, we introduce
the reduced horizon area and reduced Hawking temperature,
\begin{equation}
    a_H=\frac{A_H}{16\pi M^2}\,,
    \qquad
    t_H=8\pi M T_H\,.
\end{equation}
Both quantities are equal to unity for the Schwarzschild black hole and
therefore provide convenient dimensionless measures of the effects of
electric and magnetic charges and nonlinear electrodynamics.

For comparison, the corresponding RN expressions are
\begin{equation}
    a_H^{\rm RN}
    =
    \frac{1}{4}
    \left(1+\sqrt{1-q^2}\right)^2\,,
    \qquad
    t_H^{\rm RN}
    =
    \frac{4\sqrt{1-q^2}}
    {\left(1+\sqrt{1-q^2}\right)^2}\,.
\end{equation}
In the Schwarzschild limit $q=0$, one has
$a_H^{\rm RN}=t_H^{\rm RN}=1$, whereas the extremal RN black hole
$q=1$ satisfies
\begin{equation}
    a_H^{\rm RN}=\frac{1}{4}\,,
    \qquad
    t_H^{\rm RN}=0.
\end{equation}

\subsubsection{Smarr relation}

The total mass of an asymptotically flat stationary spacetime can be related
to the Komar integral associated with the asymptotically normalized timelike
Killing vector $\xi^\mu=(1,0,0,0)$ which satisfies $\nabla_\mu\xi_\nu+\nabla_\nu\xi_\mu=0$.

The corresponding Komar integral is
\begin{equation}
    K_\xi
    =
    \frac{1}{8\pi}
    \int_{\Sigma}
    \nabla_\mu\xi_\nu\,dS^{\mu\nu}\,,
\end{equation}
up to the orientation convention adopted for the integration surface.

For the present EEH black holes, the resulting mass relation can be written
as
\begin{equation}
M={}
\frac{1}{2}T_HA_H
+\int_{r_H}^{\infty}
\left(
\frac{P^2}{r^2}+r^2V'^2
\right)dr
+
\epsilon
\int_{r_H}^{\infty}
\left(
2r^2V'^4
-\frac{6P^4}{r^6}
-\frac{10P^2V'^2}{r^2}
\right)dr \,.
\label{eq:Smarr_EEH}
\end{equation}
The first term is the horizon contribution, whereas the remaining terms
represent the electromagnetic contribution outside the event horizon,
including the nonlinear Euler--Heisenberg corrections.

In the EM limit $(\epsilon=0)$, using
$V'=-Q/r^2$, Eq.~\eqref{eq:Smarr_EEH} reduces to
\begin{align}
M
&=
\frac{1}{2}T_HA_H
+
\int_{r_H}^{\infty}
\left(
\frac{P^2}{r^2}+r^2V'^2
\right)dr \,,
\nonumber\\
&=
\frac{1}{2}T_HA_H
+\Phi_mP+\Phi_eQ\,,
\end{align}
which is the standard Smarr relation for the dyonic RN black
hole~\cite{Rasheed:1997ns}. Here, $\Phi_e=Q/r_H$ and $\Phi_m=P/r_H$
are the electric and magnetic horizon potentials, respectively.

For the purely electric configuration $(P=0)$, the mass relation becomes
\begin{equation}
M
=
\frac{1}{2}T_HA_H
+
\int_{r_H}^{\infty}
\left(
r^2V'^2+2\epsilon r^2V'^4
\right)dr \,.
\label{eq:Smarr_electric}
\end{equation}
Because $V'(r)$ is itself determined by the nonlinear cubic equation, this
integral generally doesn't have a closed form \cite{Gulin:2017ycu}.

For the purely magnetic configuration $(Q=0)$, one obtains
\begin{align}
M
&=
\frac{1}{2}T_HA_H
+
\int_{r_H}^{\infty}
\left(
\frac{P^2}{r^2}
-\frac{6\epsilon P^4}{r^6}
\right)dr \,,
\nonumber\\
&=
\frac{1}{2}T_HA_H
+\Phi_mP
-\frac{6\epsilon}{5P}\Phi_m^5 \,.
\end{align}

\subsubsection{Weak energy condition}

For an anisotropic energy--momentum tensor,
\begin{equation}
T^\mu{}_\nu
=
\operatorname{diag}(-\rho,p_r,p_t,p_t),
\end{equation}
where $\rho$ is the energy density, $p_r$ is the radial pressure and $p_t$ is the tangential pressure. The weak energy condition (WEC) requires
\begin{equation}
\rho\geq0 \,,\qquad
\rho+p_r\geq0 \,,\qquad
\rho+p_t\geq0 \,.
\label{eq:WEC_general}
\end{equation}
These inequalities ensure that the local energy density measured by any
timelike observer is nonnegative.

For the EEH black holes, the equality of the $tt$ and $rr$ components of
the Einstein equations implies
\begin{equation}
T^t{}_t=T^r{}_r \,,
\end{equation}
and therefore
\begin{equation}
p_r=-\rho \,,
\qquad
\rho+p_r=0 \,.
\end{equation}
Thus, the radial WEC inequality is identically saturated.

The energy density $\rho$ is
\begin{equation}
\rho
=
\frac{1}{8\pi}
\left(
\frac{P^2}{r^4}+V'^2
\right)
+
\frac{\epsilon}{4\pi}
\left(
3V'^4
-\frac{P^4}{r^8}
+\frac{5P^2V'^2}{r^4}
\right) \,,
\label{eq:WEC_rho}
\end{equation}
whereas the tangential pressure $p_t$ is
\begin{equation}
p_t
=
\frac{1}{8\pi}
\left(
\frac{P^2}{r^4}+V'^2
\right)
+
\frac{\epsilon}{4\pi}
\left(
V'^4
-\frac{3P^4}{r^8}
-\frac{5P^2V'^2}{r^4}
\right) \,.
\label{eq:WEC_pt}
\end{equation}
It follows that
\begin{equation}
\rho+p_t
=
\frac{1}{4\pi}
\left[
V'^2+\frac{P^2}{r^4}
+4\epsilon
\left(
V'^4-\frac{P^4}{r^8}
\right)
\right] \,.
\label{eq:WEC_tangential}
\end{equation}

Consequently, the WEC for the present solutions is completely determined
by the two nontrivial inequalities
\begin{equation}
\rho\geq0 \,,
\qquad
\rho+p_t\geq0 \,,
\end{equation}
while $\rho+p_r=0$ is automatically satisfied. In particular, a negative
energy density is sufficient to demonstrate WEC violation, but a positive
energy density alone is not sufficient to establish that the full WEC is
satisfied. The WEC properties of the purely electric, purely magnetic, and
dyonic black holes will be examined separately in the following sections.

\section{Three types of EEH black holes}
\label{sec:Re}

\subsection{Purely electric case $(P=0)$}

Setting $P=0$ in the discriminant, we find that
$\Delta>0$ for $\epsilon>0$. Hence, the two remaining Cardano roots given by Eqs.~\eqref{eq:root2} and \eqref{eq:root3}
are complex, and the purely electric solutions are completely determined by Eqs.~\eqref{eq:dyn_mprime} and \eqref{eq:root1}.

We begin by rewriting the physical electric-field branch $E_1(r)$.
Using
\begin{equation}
    E_1^{+}E_1^{-}=-\frac{1}{12\epsilon}\,,
\end{equation}
we obtain
\begin{align}
    E_1(r)
    &=
    E_1^{+}
    -\frac{1}{12\epsilon E_1^{+}} \,,
    \nonumber\\
    &=
    \sqrt[3]{
        \frac{Q}{8\epsilon r^2}
        +
        \sqrt{
            \left(\frac{Q}{8\epsilon r^2}\right)^2
            +
            \left(\frac{1}{12\epsilon}\right)^3
        }
    }
    -
    \frac{1}{12\epsilon}
    \left[
    \sqrt[3]{
        \frac{Q}{8\epsilon r^2}
        +
        \sqrt{
            \left(\frac{Q}{8\epsilon r^2}\right)^2
            +
            \left(\frac{1}{12\epsilon}\right)^3
        }
    }
    \right]^{-1}\,.
\end{align}

Introducing the dimensionless radial coordinate
\begin{equation}
    z^2=\frac{r^2}{\sqrt{27\epsilon Q^2}} \,,
\end{equation}
the electric field can be written as
\begin{equation}
    E_1(z)
    =
    \frac{1}{2\sqrt{3\epsilon}}
    \left[
    \sqrt[3]{\frac{1+\sqrt{z^4+1}}{z^2}}
    -
    \left(
    \sqrt[3]{\frac{1+\sqrt{z^4+1}}{z^2}}
    \right)^{-1}
    \right]\,.
    \label{eq:electro}
\end{equation}
Since $E_1=-dV_1/dr$, the potential reads
\begin{equation}
    \frac{dV_1}{dz}
    =
    -(27\epsilon Q^2)^{1/4}E_1(z)\,.
\end{equation}

To integrate Eq.~\eqref{eq:electro}, we introduce a second
dimensionless variable,
\begin{equation}
    u=
    \frac{\sqrt{z^4+1}-1}
         {\sqrt{z^4+1}+1}\,,
\end{equation}
The range $u\in[0,1)$ corresponds to $r\in[0,\infty)$. In terms of
$u$, the electric field takes the simple form
\begin{equation}
    E_1(u)
    =
    \frac{1}{2\sqrt{3\epsilon}}
    \left(u^{-1/6}-u^{1/6}\right)\,.
\end{equation}
The corresponding differential relations are
\begin{equation}
    dz=
    \frac{1}{2\sqrt{2}}
    (1+u)(1-u)^{-3/2}u^{-3/4}\,du\,,
    \qquad
    dr=(27\epsilon Q^2)^{1/4}dz\,.
\end{equation}
It follows that
\begin{align}
    V_1(u)
    &=
    -\frac{(27\epsilon Q^2)^{1/4}}
           {4\sqrt{6\epsilon}}
    \int
    (1+u)(1-u)^{-3/2}
    \left(u^{-11/12}-u^{-7/12}\right)\,du \,,
    \nonumber\\
    &=
    -\frac{(27\epsilon Q^2)^{1/4}}
           {4\sqrt{6\epsilon}}
    \int
    (1-u)^{-3/2}
    \left(
        u^{1/12}
        +u^{-11/12}
        -u^{-7/12}
        -u^{5/12}
    \right)\,du \,.
\end{align}

Each term can be integrated analytically using
\begin{equation}
    \int u^s(1-u)^p\,du
    =
    \frac{u^{s+1}}{s+1}
    {}_2F_1(-p,s+1;s+2;u)\,.
\end{equation}
Imposing the gauge condition $V_1(\infty)=0$, we obtain
\begin{equation}
\label{eq:elec_V}
    V_1(u)
    =
    \frac{\pi^{3/2}Q^{1/2}}
         {3^{1/4}\epsilon^{1/4}
          \Gamma(7/12)\Gamma(11/12)}
    -
    \frac{(27\epsilon Q^2)^{1/4}}
         {4\sqrt{6\epsilon}}
    \sum_{k=1}^{4}
    c_k u^{a_k}
    {}_2F_1
    \left(
        a_k,\frac{3}{2};b_k;u
    \right)\,,
\end{equation}
where the coefficients appearing in the sum are listed in Table~\ref{tab:coeff}.

\begin{table}
    \centering
    \begin{tabular}{c c c c}
        \hline\hline
        $k$ & $a_k$ & $b_k$ & $c_k$ \\
        \hline
        $1$ & $1/12$  & $13/12$ & $12$ \\
        $2$ & $5/12$  & $17/12$ & $-12/5$ \\
        $3$ & $13/12$ & $25/12$ & $12/13$ \\
        $4$ & $17/12$ & $29/12$ & $-12/17$ \\
        \hline\hline
    \end{tabular}
    \caption{
        Coefficients entering the hypergeometric representation
        of the electric potential in Eq.~\eqref{eq:elec_V}.
    }
    \label{tab:coeff}
\end{table}

\begin{figure}
\centering
\mbox{
    (a)\,
    \includegraphics[angle=0,scale=0.32]{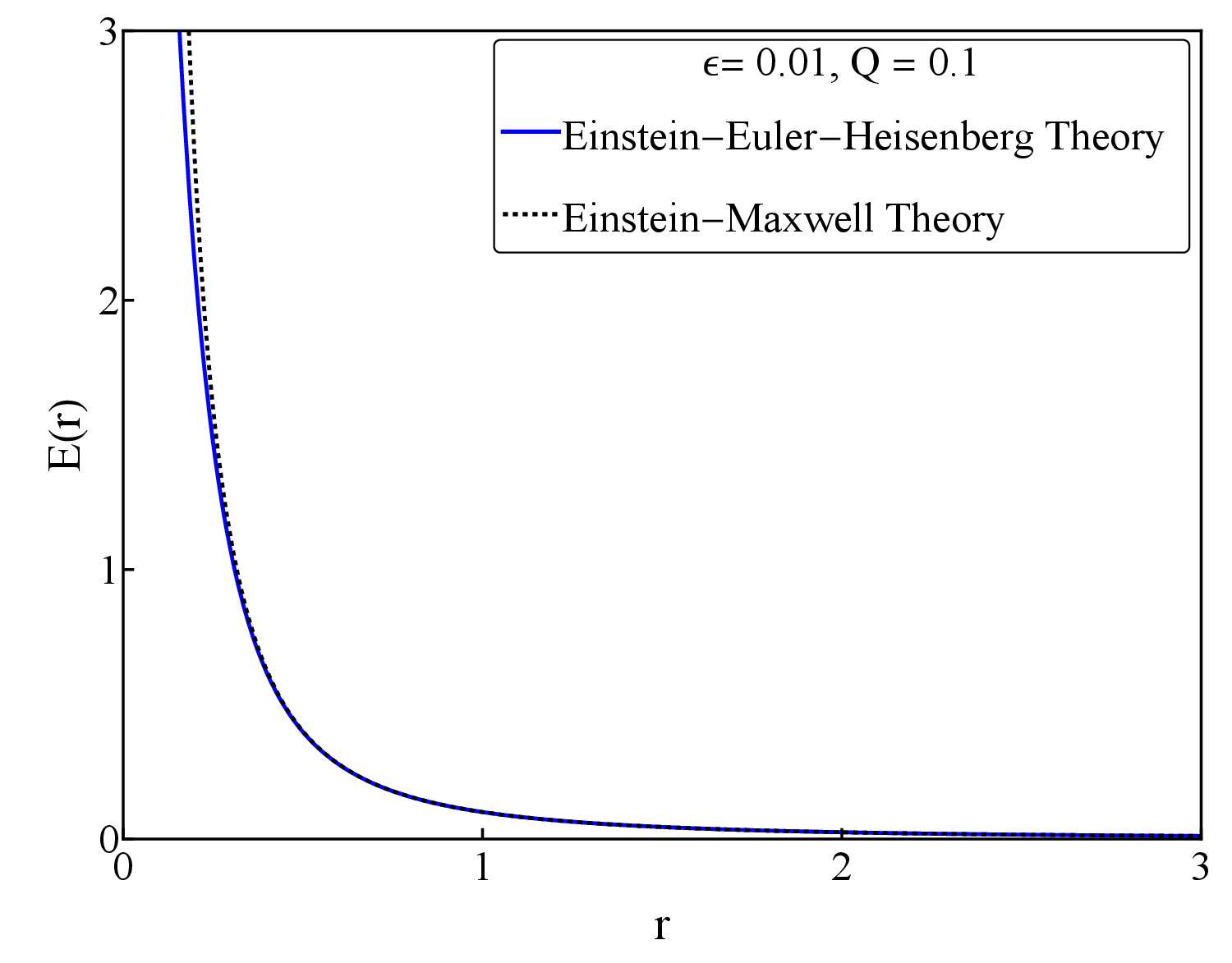}
    \qquad
    (b)\,
    \includegraphics[angle=0,scale=0.32]{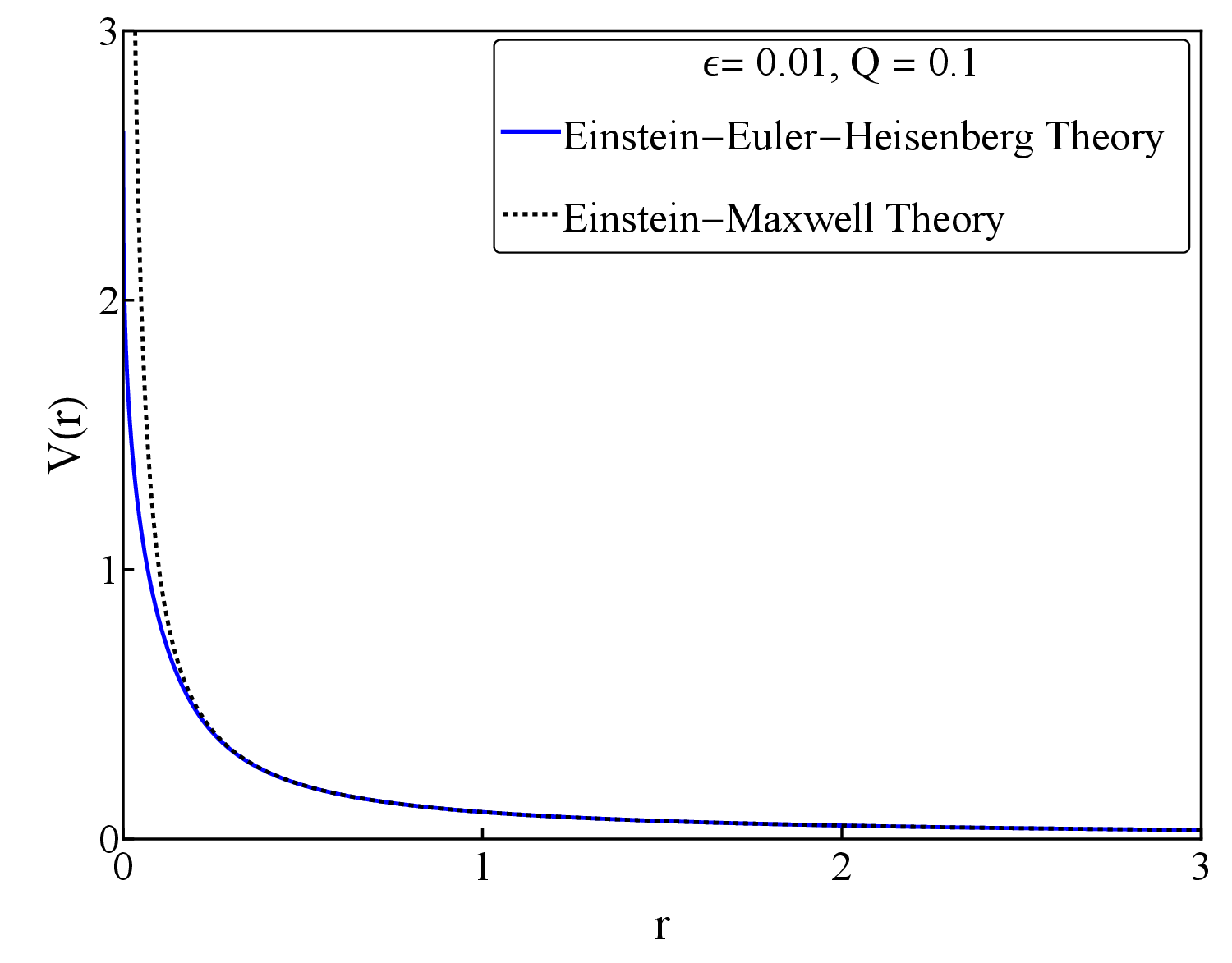}
}
\caption{
    Comparison of the radial profiles of
    (a) the electric field $E(r)$ and
    (b) the electric potential $V_1(r)$
    in the EM and EEH theories,
    for $\epsilon=0.01$ and $Q=0.1$.
}
\label{fig:e_E_V}
\end{figure}

The Gaussian hypergeometric function admits the series representation
\begin{equation}
\label{eq:hypergeo}
    {}_2F_1(a,b;c;u)
    =
    \sum_{k=0}^{\infty}
    \frac{(a)_k(b)_k}{(c)_k}
    \frac{u^k}{k!}\,,
\end{equation}
where $(a)_k$ denotes the Pochhammer symbol. Expanding
Eq.~\eqref{eq:elec_V} around $u=1$, corresponding to $r\rightarrow\infty$, gives
\begin{equation}
    V_1(r)
    =
    \frac{Q}{r}
    -\frac{4\epsilon Q^3}{5r^5}
    +\frac{16\epsilon^2Q^5}{3r^9}
    +O(r^{-13})\,.
\end{equation}
The approximate electric field is obtained by truncating the expansion to first order in $\epsilon$, as given by Eq.~\eqref{eq:elec_field_approx},
\begin{equation}
    E_1(r)=-\frac{dV_1(r)}{dr}\approx\frac{Q}{r^2}-\frac{4\epsilon Q^3}{r^6}\,.
\end{equation}
Fig.~\ref{fig:e_E_V} compares the electric field and electric potential in the EM and EEH theories. Both $E_1(r)$ and $V_1(r)$ decrease monotonically with increasing
$r$. At large distances, the EH potential approaches the Coulomb potential, and the deviation from the RN result becomes negligible. Near the origin, however, the NED correction qualitatively changes the behavior of the solution. Although $E_1(r)$ remains divergent as $r\rightarrow0$, the divergence is weaker than the
Maxwell behavior $Q/r^2$, and the potential approaches the finite value
\begin{equation}
    V_1(0)
    =
    \frac{\pi^{3/2}Q^{1/2}}
         {3^{1/4}\epsilon^{1/4}
          \Gamma(7/12)\Gamma(11/12)}.
\end{equation}

We next construct the spacetime metric by integrating
Eq.~\eqref{eq:dyn_mprime}. In the purely electric case, it reads
\begin{equation}
\label{eq:elec_mprime}
    m'(r)
    =
    \frac{1}{2} r^2V_1'^2
    \left(1+6\epsilon V_1'^2\right)\,.
\end{equation}
Since the two terms entering $E_1$ are mutually reciprocal, the cross term generated upon squaring is independent of $u$. Using the expression for the electric field, Eq.~\eqref{eq:elec_mprime}
becomes
\begin{equation}
    m'(r)
    =
    \frac{\sqrt{27\epsilon Q^2u}}
         {12\epsilon(1-u)}
    \left(u^{-1/6}-u^{1/6}\right)^2
    \left[
        1+
        \frac{1}{3}
        \left(u^{-1/6}-u^{1/6}\right)^2
    \right]\,.
\end{equation}
After transforming the radial integration to the variable $u$, we
find
\begin{equation}
    m(r)
    =
    \frac{(27\epsilon Q^2)^{3/4}}
         {48\sqrt{2}\epsilon}
    \int
    (1-u)^{-5/2}u^{-1/4}(1+u)
    \left(u^{1/3}+u^{-1/3}\right)
    \left(u^{1/3}+u^{-1/3}-2\right)\,du \,.
\end{equation}
Evaluating the integral by the same hypergeometric procedure yields
the exact mass function
\begin{equation}
    m(r)
    =
    M
    +
    U_1(u)
    \Bigg[
        U_2(u)
        -
        8\Bigg(
        {}_2F_1
        \left(
            \frac{1}{2},\frac{7}{12};
            \frac{3}{2};1-u
        \right)
        +
        {}_2F_1
        \left(
            \frac{1}{2},\frac{11}{12};
            \frac{3}{2};1-u
        \right)
        \Bigg)
    \Bigg]\,,
\end{equation}
where
\begin{equation}
    U_1(u)
    =
    \frac{Q^{3/2}\sqrt{2(1-u)}}
         {24\,3^{3/4}\epsilon^{1/4}}\,, \quad
    U_2(u)
    =
    \frac{
        3\left(
        11+8u^{1/3}+11u^{2/3}
        \right)u^{1/12}
    }{
        \left(
        1+u^{1/3}+u^{2/3}
        \right)^2
    }\,.
\end{equation}

To investigate the geometry, we examine the metric function $N(r)$ in the limits $r\rightarrow0$ and $r\rightarrow\infty$. Near the
origin, its expansion is
\begin{equation}
    N(r)
    =
    1-\frac{2M}{r}
    +
    \frac{
        4\pi^{3/2}Q^{3/2}
    }{
        3^{5/4}\epsilon^{1/4}
        r\Gamma(7/12)\Gamma(11/12)
    }
    -
    \frac{9Q^{4/3}}
         {2^{5/3}\epsilon^{1/3}r^{2/3}}
    +
    \frac{3Q^{2/3}r^{2/3}}
         {10\,2^{1/3}\epsilon^{2/3}}
    +O(r^2).
\end{equation}
Thus, despite the finiteness of the electric potential, the metric
function remains divergent at the origin.

This conclusion is confirmed by the Kretschmann scalar. Its
near-origin behavior is
\begin{equation}
    K
    =
    \frac{48}{r^6}
    \left[
        M+
        \frac{
            8\pi^{3/2}Q^{3/2}
        }{
            3^{1/4}\epsilon^{1/4}
            \Gamma(-1/12)\Gamma(7/12)
        }
    \right]^2
    +O(r^{-17/3})\,,
\end{equation}
which diverges as $r\rightarrow0$. Therefore, the spacetime retains a curvature singularity at the center.

For comparison with the RN geometry, one may also expand the Kretschmann scalar perturbatively in $\epsilon$:
\begin{equation}
    K
    =
    \frac{
        8\left(
            7Q^4-12MQ^2r+6M^2r^2
        \right)
    }{r^8}
    +
    \frac{
        64\epsilon
        \left(
            -19Q^6+14MQ^4r
        \right)
    }{5r^{12}}
    +O(\epsilon^2)\,.
\end{equation}
The leading term is precisely the Kretschmann scalar of the RN spacetime.

At spatial infinity, the metric function behaves as
\begin{equation}
    N(r)
    =
    1-\frac{2M}{r}
    +\frac{Q^2}{r^2}
    -\frac{2\epsilon Q^4}{5r^6}
    +\frac{16\epsilon^2Q^6}{9r^{10}}
    +O(r^{-14}).
\end{equation}
The first three terms reproduce the RN metric, showing that the purely electric EEH black hole is
asymptotically RN-like. The truncation through order $r^{-6}$ has been widely employed as an approximate purely electric EEH metric in studies of black-hole phenomenology
\cite{Ruffini:2013hia, Kruglov:2017ymn, Magos:2020ykt, Guerrero:2020uhn, Li:2021ygi,
Breton:2021mju, Magos:2023nnb, Gogoi:2023wih,
Zhang:2025msi, Kala:2025kkm, Guo:2025ksj,
Zhang:2026bqu, Li:2026ebc, Guo:2026qib, Becar:2026doz,Awal:2026vfu,Ditta:2026yvp,
Tan:2026itp,Feng:2026yhj}. Some studies have additionally introduced phenomenological
extensions of this asymptotic form.

Furthermore, we examine the WEC, which provides
an important criterion for assessing the physical viability of static spherically symmetric spacetimes \cite{Wang:2026jvo,Wang:2026sqr}. In the purely electric case, one finds
\begin{equation}
    \rho
    =
    \frac{1}{8\pi}
    V_1'^2
    \left(1+6\epsilon V_1'^2\right)\,, \quad
    \rho+p_r
    =0\,, \quad
    \rho+p_t
    =
    \frac{1}{4\pi}
    V_1'^2
    \left(1+4\epsilon V_1'^2\right)\,.
\end{equation}
Therefore, all WEC inequalities are satisfied for $\epsilon>0$.

\begin{figure}[t]
\mbox{
(a)
\includegraphics[angle =0,scale=0.32]{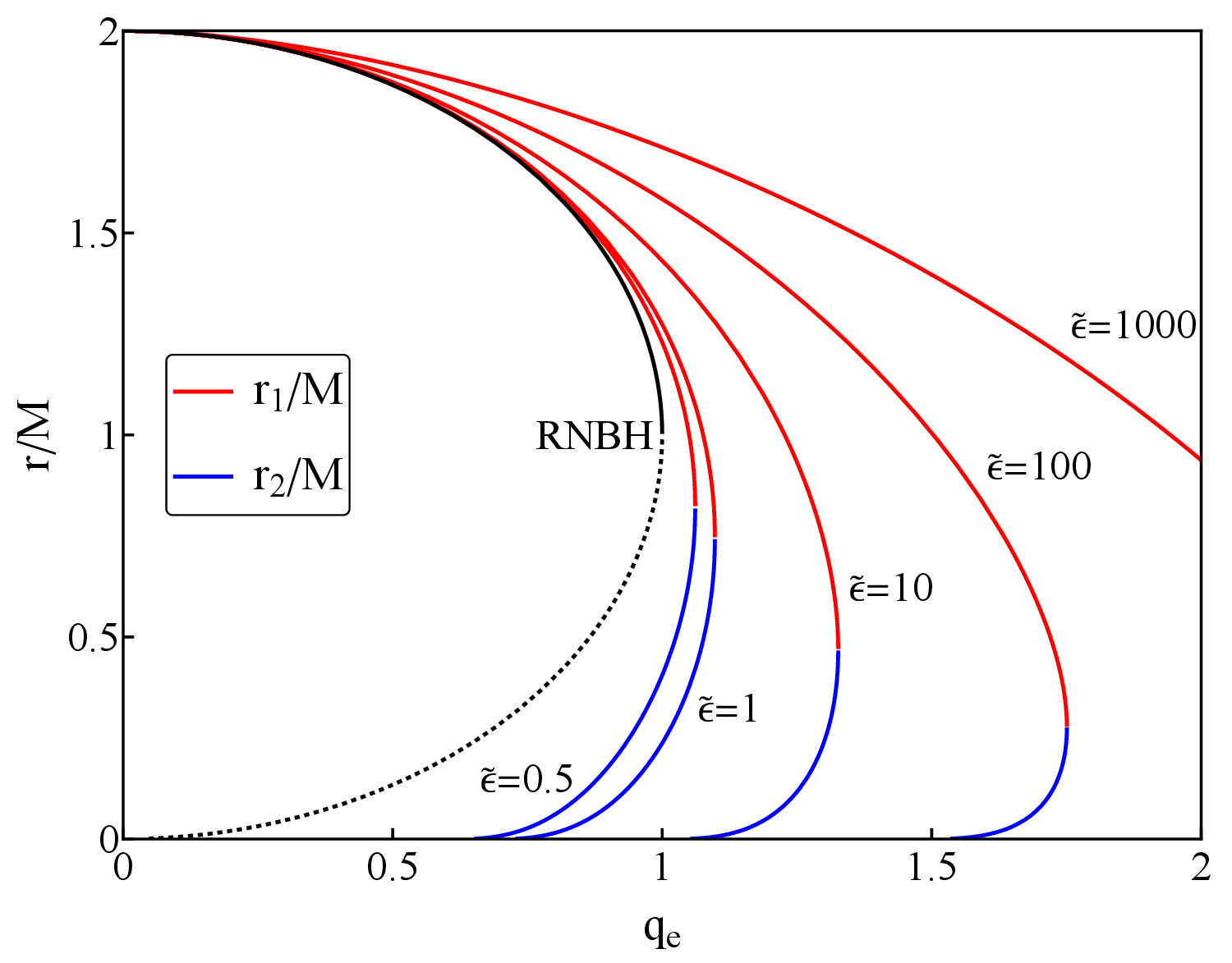}
(b)
\includegraphics[angle =0,scale=0.32]{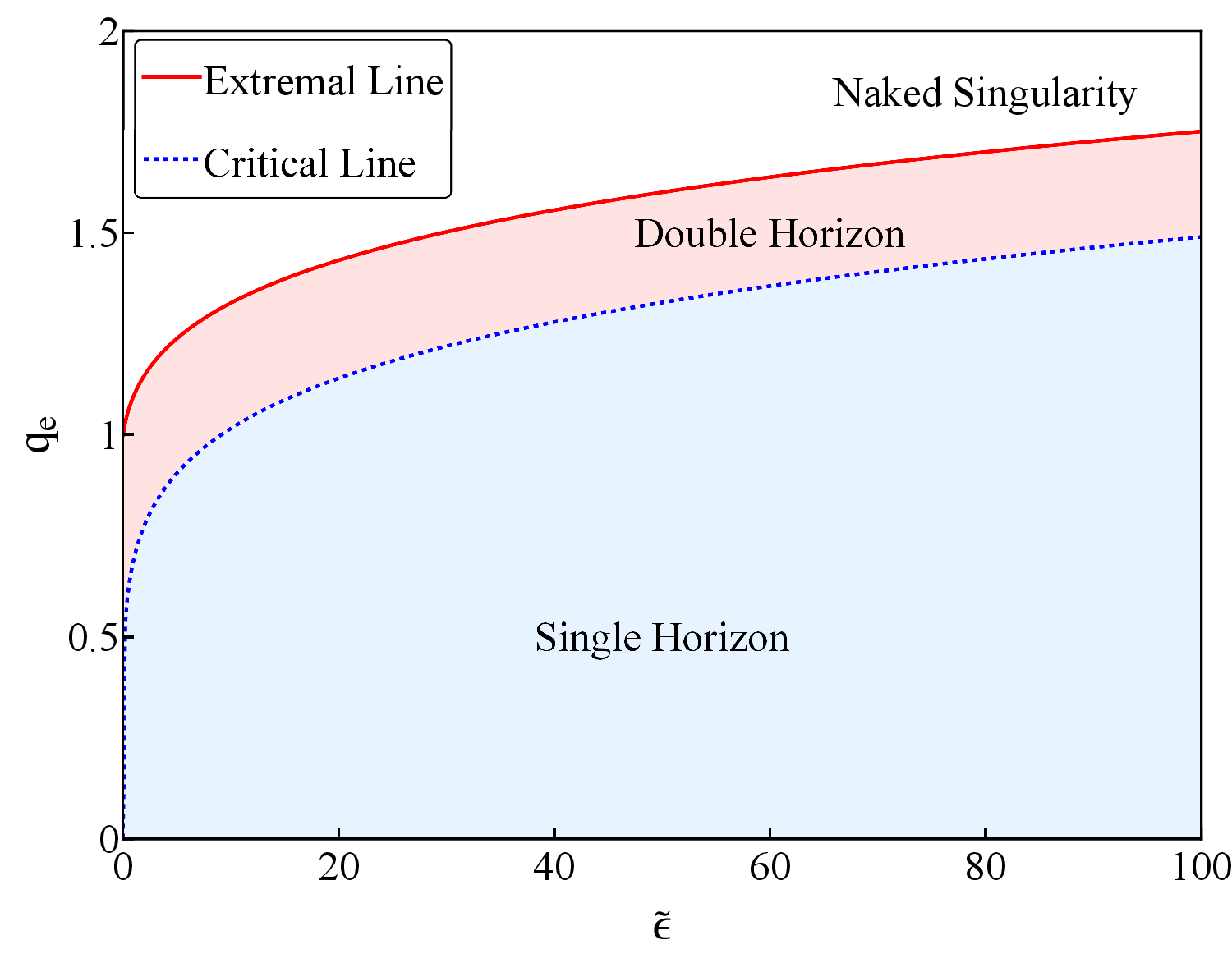}
}
\mbox{
(c)
\includegraphics[angle =0,scale=0.32]{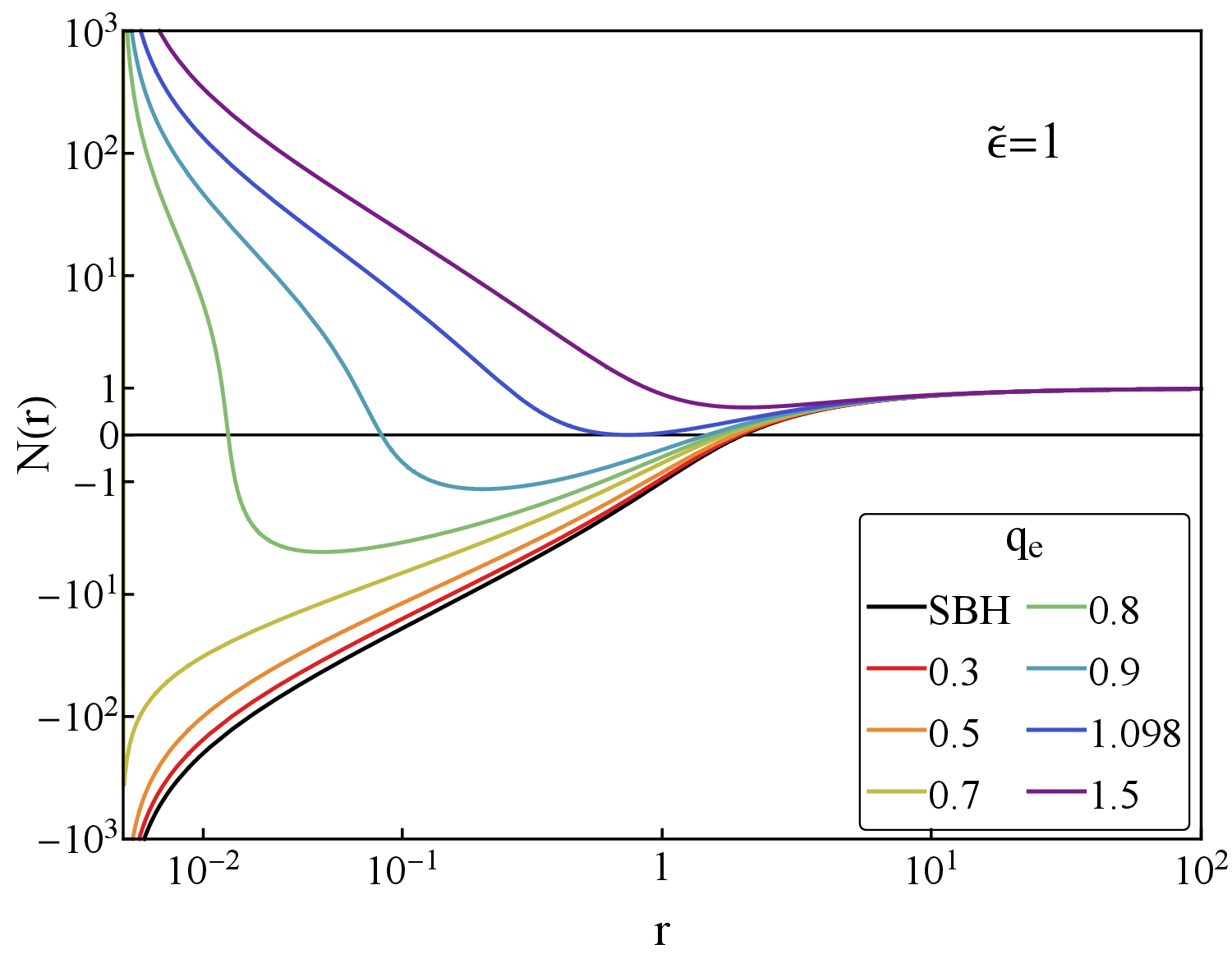}
(d)
\includegraphics[angle =0,scale=0.32]{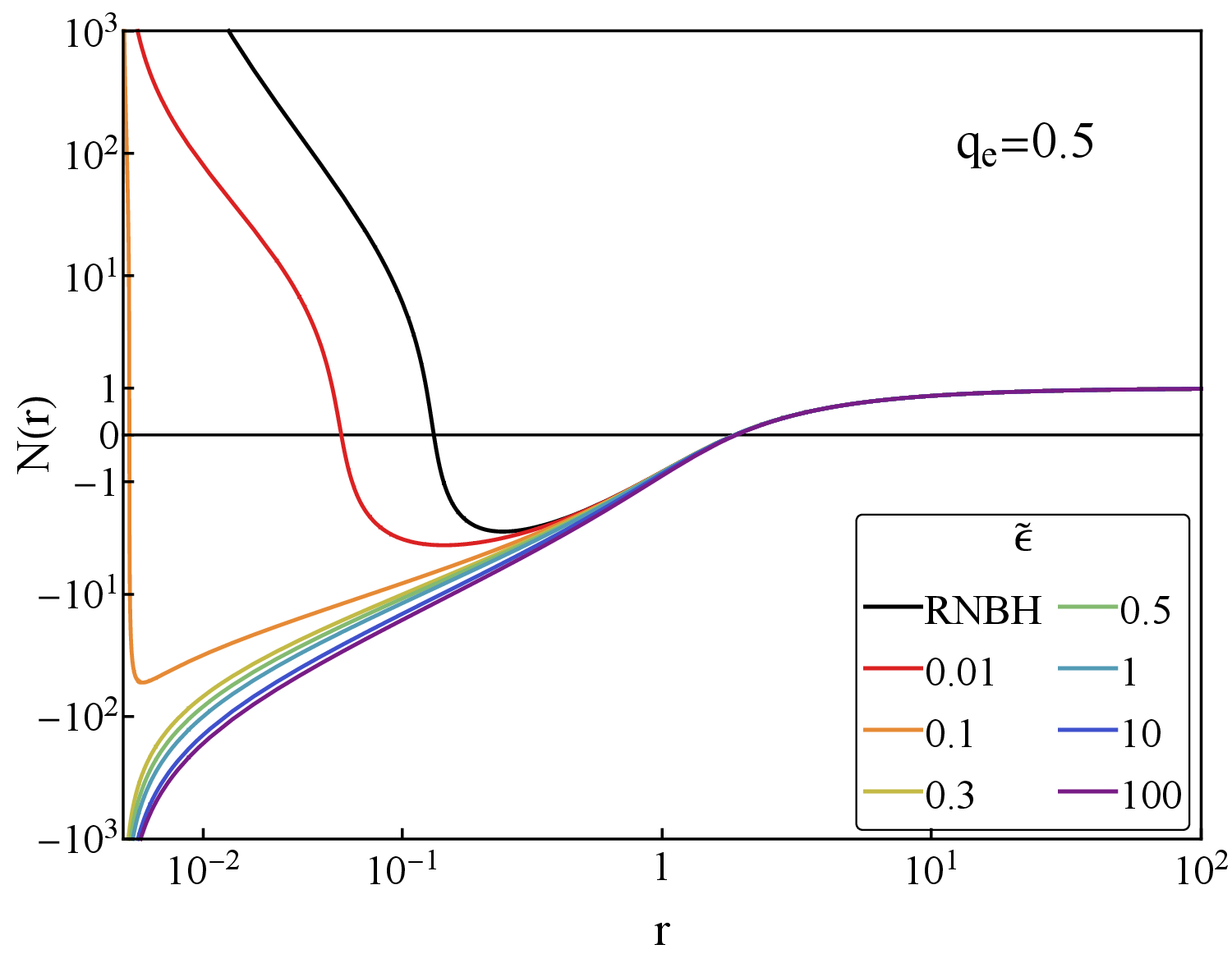}
}
\mbox{
(e)
\includegraphics[angle =0,scale=0.32]{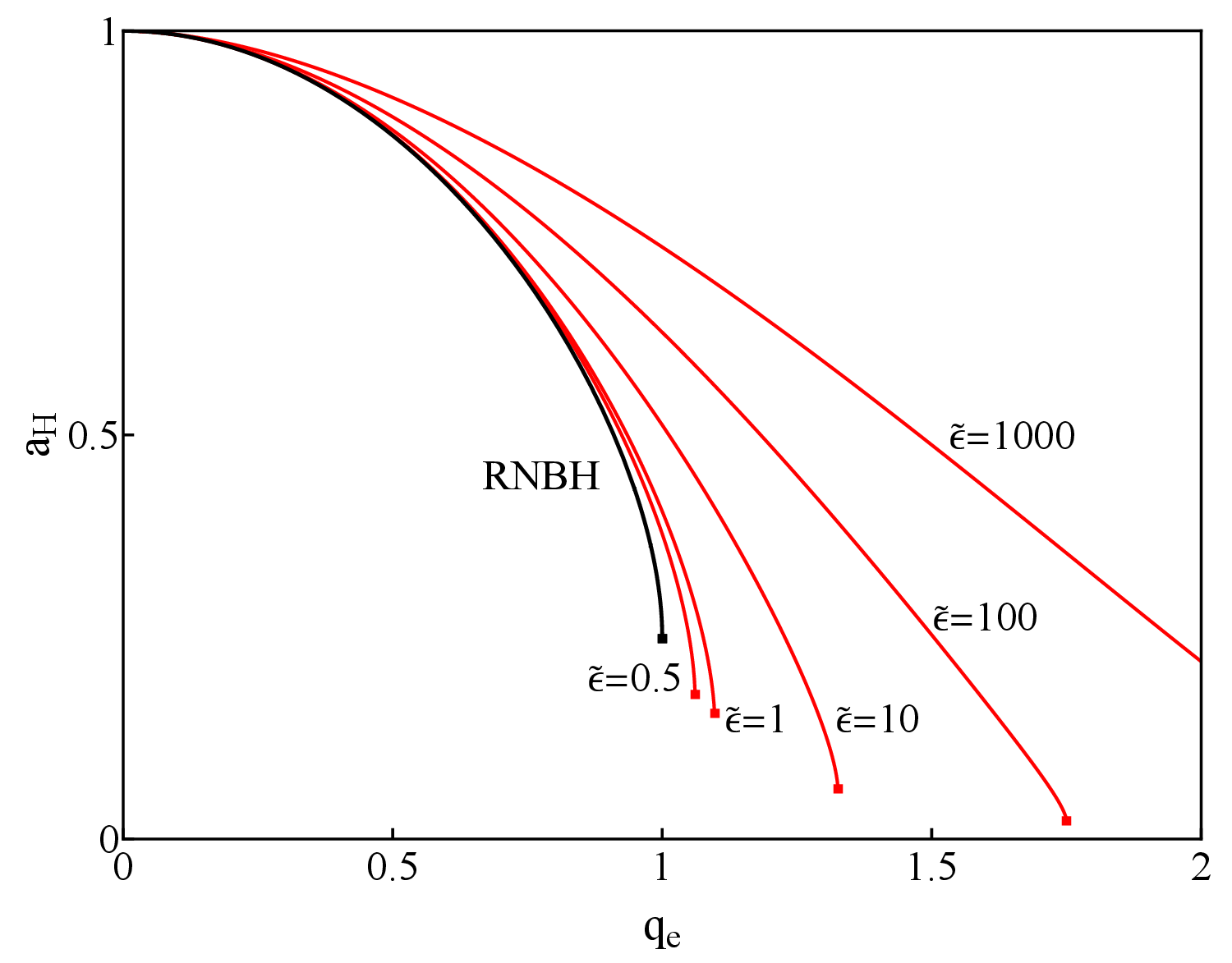}
(f)
\includegraphics[angle =0,scale=0.32]{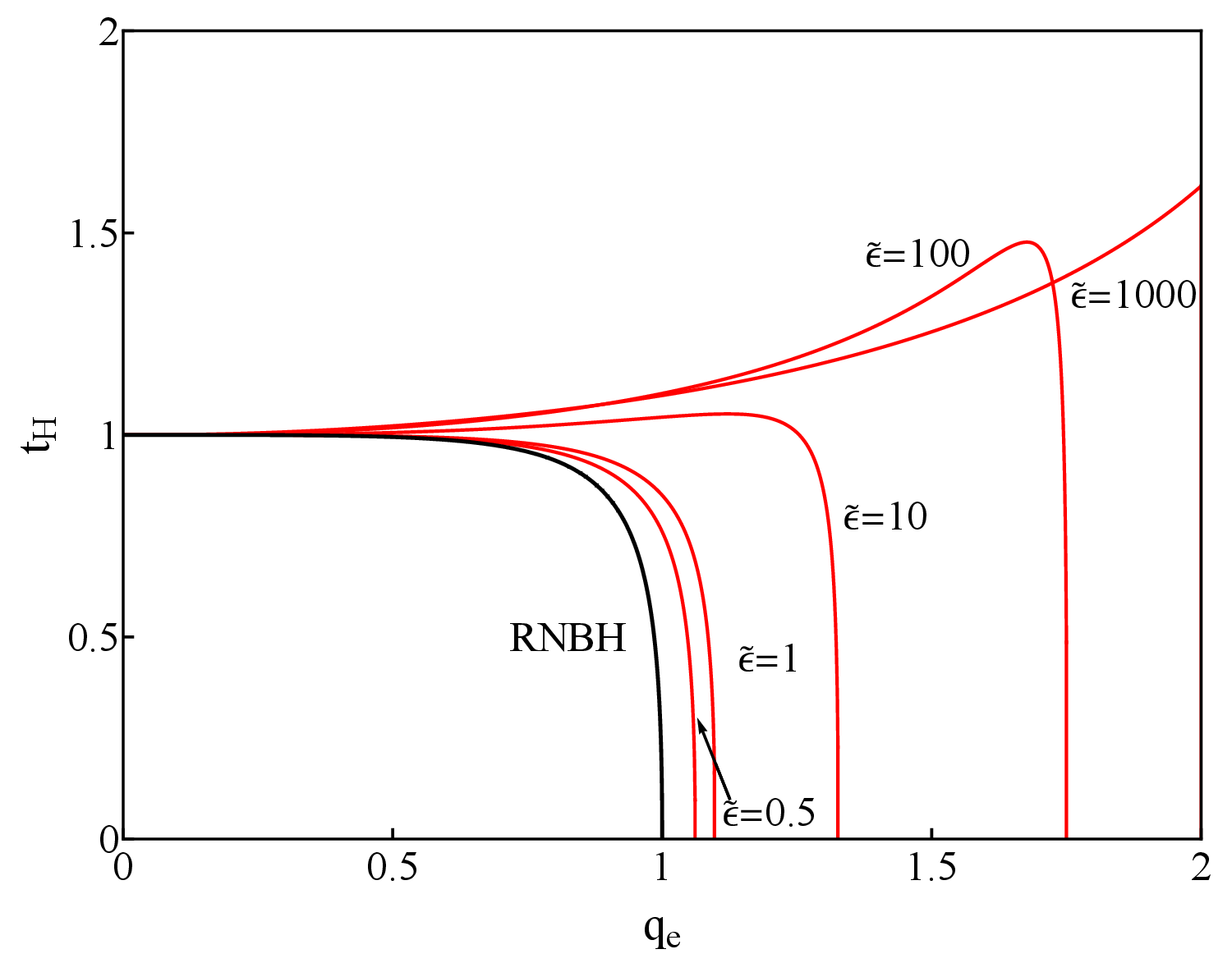}
}
\caption{(a) The locations of two scaled horizons $(r_1/M, r_2/M)$ as the function of $q_e$ for purely electric black holes with several values of $\tilde{\epsilon}$; (b) The domain of existence for single and double horizons, and naked singularity as functions of $(q_e,\tilde{\epsilon})$ for purely electric black holes; The profile of metric $N(r)$ in the radial coordinate $r$ for purely electric black holes (c) with fixed $\tilde{\epsilon}=1$, and several values of $q_e$; and (d) with fixed $q_e=0.5$, and several values of $\tilde{\epsilon}$; (e) The reduced area of horizon $a_H$ and (f) reduced Hawking temperature $t_H$ as the function of $q_e$ for purely electric black holes with several values of $\tilde{\epsilon}$.}
\label{fig:e_aH_tH_rH_Q_ep}
\end{figure}

Next, we investigate the horizon structure of the EEH black holes by solving the horizon equation $N(r)=0$ for different values of $\tilde{\epsilon}$. As shown in Fig.~\ref{fig:e_aH_tH_rH_Q_ep}(a), we consider $\tilde{\epsilon}=0.5, 1, 10, 100, 1000$. For $\tilde{\epsilon}=0$, the solution reduces to the RN black hole. In this case, the outer horizon $(r_1=r_+)$ decreases monotonically with increasing $q_e$, whereas the inner horizon $(r_2=r_-)$ increases monotonically. The two horizons eventually merge at the extremal limit $q_e=1$. For $\tilde{\epsilon}>0$, the qualitative behavior changes. The outer horizon $r_1/M$ always exists and gradually shrinks as $q_e$ increases. Unlike the RN case, however, the inner horizon $r_2/M$ is absent for sufficiently small $q_e$. It only emerges once $q_e$ exceeds a critical value, growing continuously from $r=0$ before eventually merging with the outer horizon at the extremal configuration. We further observe that increasing $\tilde{\epsilon}$ shifts both the onset of the inner horizon and the extremal point to larger values of $q_e$. This implies that stronger EH nonlinearities postpone the formation of the inner horizon and enlarge the parameter region in which the black hole possesses only a single event horizon.

Fig.~\ref{fig:e_aH_tH_rH_Q_ep}(b) illustrates the domain of existence in the $(\tilde{\epsilon},q_e)$ parameter space for $(0\leq\tilde{\epsilon}\leq100)$. For a fixed value of $\tilde{\epsilon}$, the black hole initially possesses only a single horizon, namely the event horizon $r_1/M$, as $q_e$ increases from zero. Once $q_e$ exceeds the critical line, an inner horizon $r_2/M$ emerges, then the spacetime contains two horizons. As $q_e$ is increased further, the two horizons approach each other and eventually coincide on the extremal line. Beyond this line, no horizons exist, and the solution describes a naked singularity. This figure also shows that both the critical line and the extremal line shift toward larger values of $q_e$ as $\tilde{\epsilon}$ increases, indicating that stronger $\tilde{\epsilon}$ delays the formation of the inner horizon and enlarge the parameter region occupied by single-horizon black holes.

Fig.~\ref{fig:e_aH_tH_rH_Q_ep}(c) displays the metric function $N(r)$ as a function of the radial coordinate $r$ for fixed $\tilde{\epsilon}=1$ and several values of the scaled electric charge $q_e$. For relatively small values of $q_e$, the metric closely resembles the Schwarzschild solution, with a single zero corresponding to the event horizon and $N(r)\rightarrow1$ in the asymptotic region. As $q_e$ increases, a second zero appears, signaling the formation of an inner horizon and the transition to a two-horizon configuration characteristic of the purely electric EEH black holes. At the extremal value $q_e\approx1.098$, the two horizons merge into a degenerate horizon, corresponding to the extremal black hole. For larger charges, $q_e>1.098$, the metric function has no real zeros, indicating that the horizons disappear and the central singularity becomes naked.

On the other hand, we fix $q_e=0.5$ and plot the metric function $N(r)$ as a function of the radial coordinate $r$ for several values of $\tilde{\epsilon}$ in Fig.~\ref{fig:e_aH_tH_rH_Q_ep}(d). For $\tilde{\epsilon}=0$, the solution reduces to the RN black hole, which possesses both an event horizon and an inner horizon. As $\tilde{\epsilon}$ increases, the inner horizon continuously moves inward toward $r=0$ and eventually disappears, whereas the outer horizon is only mildly affected. Hence, for sufficiently large values of $\tilde{\epsilon}$, the EEH black hole possesses only a single event horizon \cite{Yajima:2000kw}. This behavior demonstrates that $\tilde{\epsilon}$ suppresses the formation of the Cauchy horizon and provides a natural mechanism for removing the pathological inner horizon of the RN spacetime.

Fig.~\ref{fig:e_aH_tH_rH_Q_ep}(e) displays the reduced horizon area $a_H$ as a function of $q_e$ for $\tilde{\epsilon} = 0.5, 1, 10, 100, 1000$. For $\tilde{\epsilon}=0$, the black curve corresponds to the RN solution, where $a_H$ decreases monotonically from unity, representing the Schwarzschild limit, to $1/4$, corresponding to the extremal RN black hole. For $\tilde{\epsilon}>0$, $a_H$ also decreases monotonically as $q_e$ increases and terminates at the corresponding extremal configuration. We find that the EEH black holes continuously bifurcate from the Schwarzschild solution as the $q_e$ is turned on and evolve toward extremality at a critical value of $q_e$. Furthermore, the critical $q_e$ increases with the nonlinear coupling $\tilde{\epsilon}$, while the reduced horizon area of the extremal solution decreases and approaches zero in the large-$\tilde{\epsilon}$ limit. 

Fig.~\ref{fig:e_aH_tH_rH_Q_ep}(f) presents the reduced Hawking temperature $t_H$ as a function of the charge-to-mass ratio $q_e$. For $\tilde{\epsilon}=0$, the black curve corresponds to the RN black hole. In this case, $t_H$ remains nearly constant as $q_e$ increases from zero, before decreasing rapidly to zero as the extremal limit is approached. For $\tilde{\epsilon}>0$, the qualitative behavior remains similar. However, as $\tilde{\epsilon}$ increases, the $t_H$ first rises to a maximum before decreasing sharply to zero at the extremal configuration. This non-monotonic behavior becomes increasingly pronounced for larger values of $\tilde{\epsilon}$.

Interestingly, qualitatively similar thermodynamic behavior has also been reported for purely charged hairy black holes in Einstein-Maxwell-dilaton theory~\cite{Astefanesei:2019pfq}. In both theories, $a_H$ and $t_H$ bifurcate continuously from the Schwarzschild solution as the electric charge is increased and evolve toward an extremal configuration. Furthermore, $a_H$ approaches zero in the extremal limit, while $t_H$ exhibits either monotonic or non-monotonic evolution depending on the model parameters before ultimately vanishing at extremality.

\subsection{Purely magnetic case $(Q=0)$}
Here we set $Q=0$, hence Eq.~\eqref{eq:dyn_cubic} becomes
\begin{eqnarray}
V'^3 + \frac{1}{4\epsilon} \left( 1 + \frac{10 \epsilon P^2}{r^4} \right) V' = 0 \,. \label{eq:mag_cubic}
\end{eqnarray}
It gives rise to two conditions, which are
\begin{equation}
    V'(r) = 0 \,, \quad \text{or} \quad  V'(r) = \pm\sqrt{-\dfrac{1}{4\epsilon}\left(1+\dfrac{10\epsilon P^2}{r^4}\right) } \,.
\end{equation}
Next, we derive and then discuss properties of two distinct solutions separately.

\subsubsection{Case with $V'(r)=0$}

In the purely magnetic configuration, the electric charge vanishes $(Q=0)$, and therefore the gauge potential satisfies $V(r)=0$. Hence,  Eq.~\eqref{eq:dyn_mprime} reduces to
\begin{eqnarray}
m'(r)=\frac{P^2}{2r^2}\left(1-\frac{2\epsilon P^2}{r^4}\right)\,.
\end{eqnarray}
Integrating this equation gives
\begin{eqnarray}
m(r)=M-\frac{P^2}{2r}+\frac{\epsilon P^4}{5r^5}\,,
\end{eqnarray}
or equivalently,
\begin{eqnarray}
N(r)=1-\frac{2m(r)}{r}
=1-\frac{2M}{r}
+\frac{P^2}{r^2}
-\frac{2\epsilon P^4}{5r^6}\,.
\end{eqnarray}
The first three terms reproduce the magnetically charged RN solution, whereas the last term originates from the EH correction and modifies the horizon structure. This purely magnetic solution has previously been obtained in Refs.~\cite{Yajima:2000kw, Ruffini:2013hia, Maceda:2018zim, Karakasis:2022xzm, Bakopoulos:2024hah, Liang:2025hzr, Huang:2025jfa, Myung:2025dey, Kala:2025kkm, Croney:2026kly, Alencar:2026cer}. Thus, the present work does not aim to derive a new magnetic solution. Instead, we perform a systematic investigation of its horizon phase structure, geometrical properties, and thermodynamic behavior within the present formulation. In particular, we demonstrate the existence of a novel three-horizon phase and establish the corresponding phase diagrams, allowing a direct comparison with the purely electric and dyonic sectors. 

\begin{figure}[t]
\mbox{
(a)
\includegraphics[angle =0,scale=0.32]{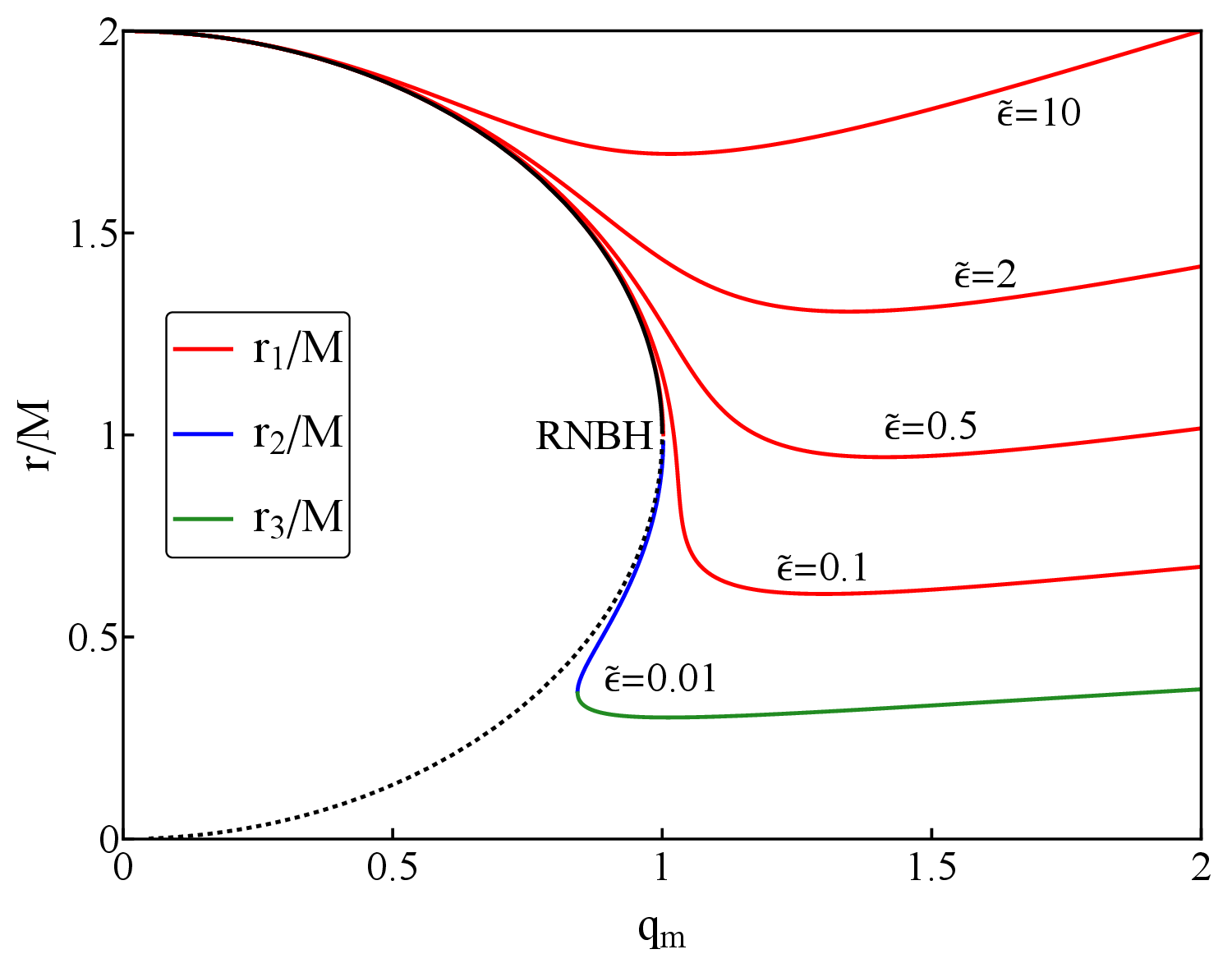}
(b)
\includegraphics[angle =0,scale=0.32]{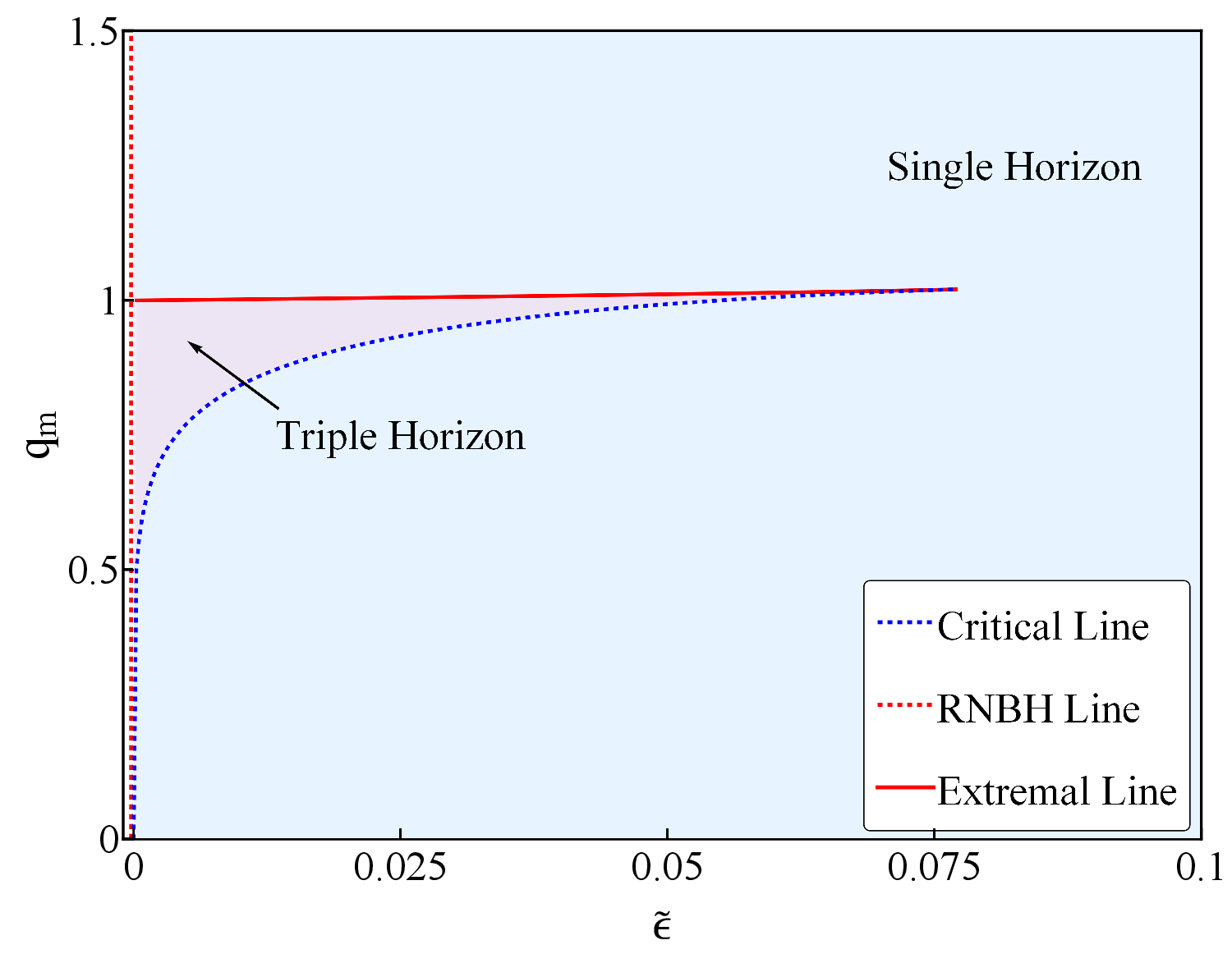}
}
\mbox{
(c)
\includegraphics[angle =0,scale=0.32]{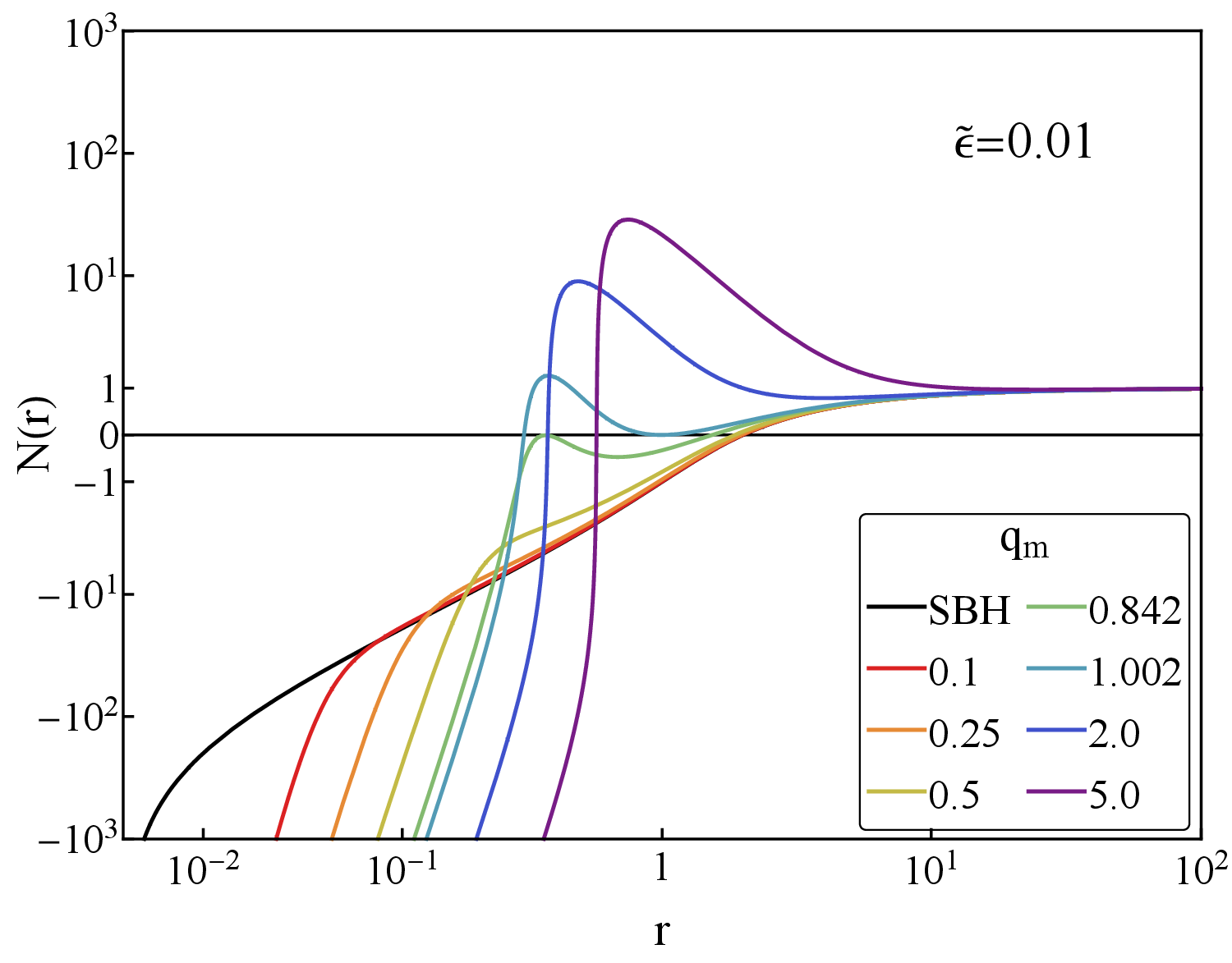}
(d)
\includegraphics[angle =0,scale=0.32]{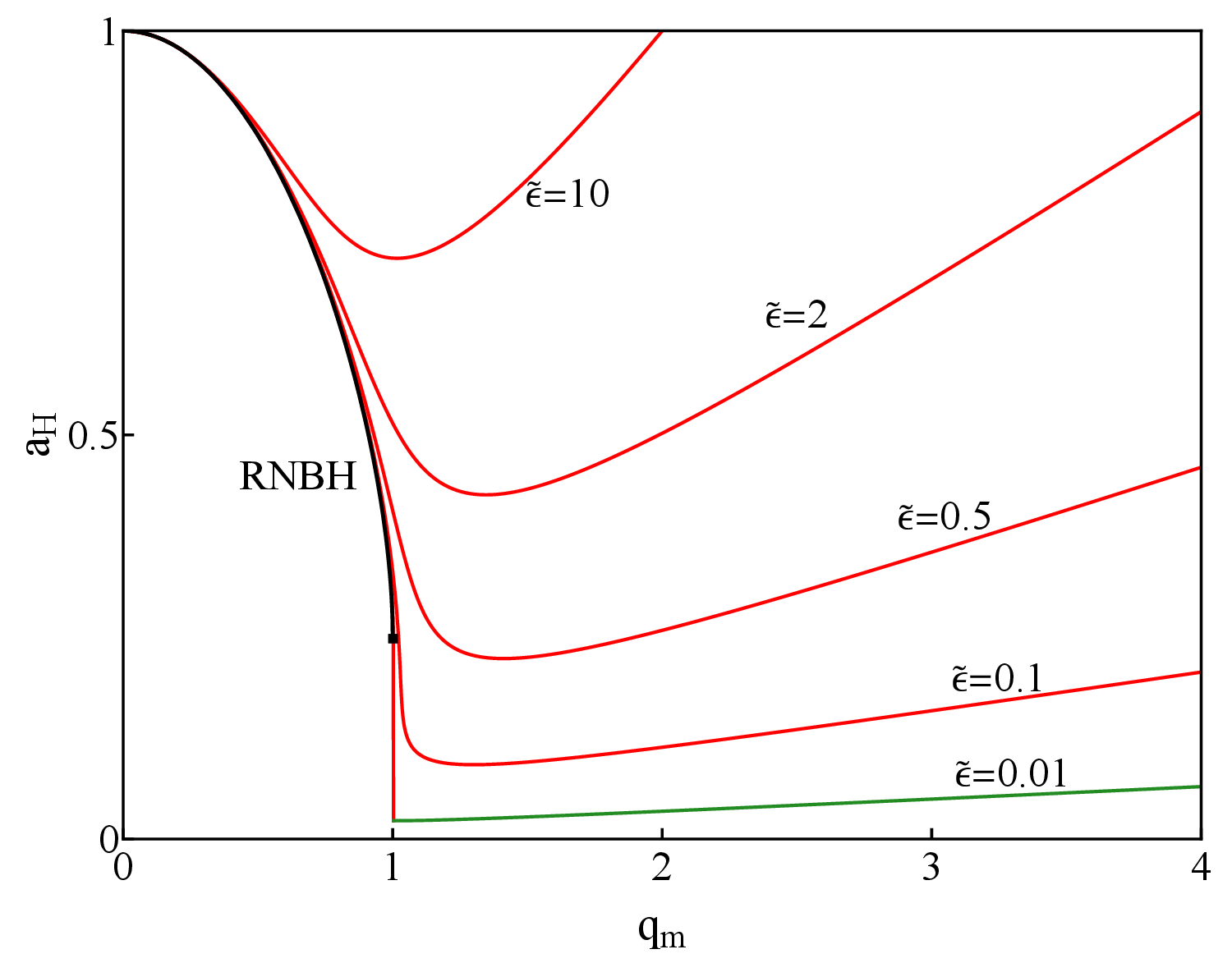}
}
\mbox{
(e)
\includegraphics[angle =0,scale=0.32]{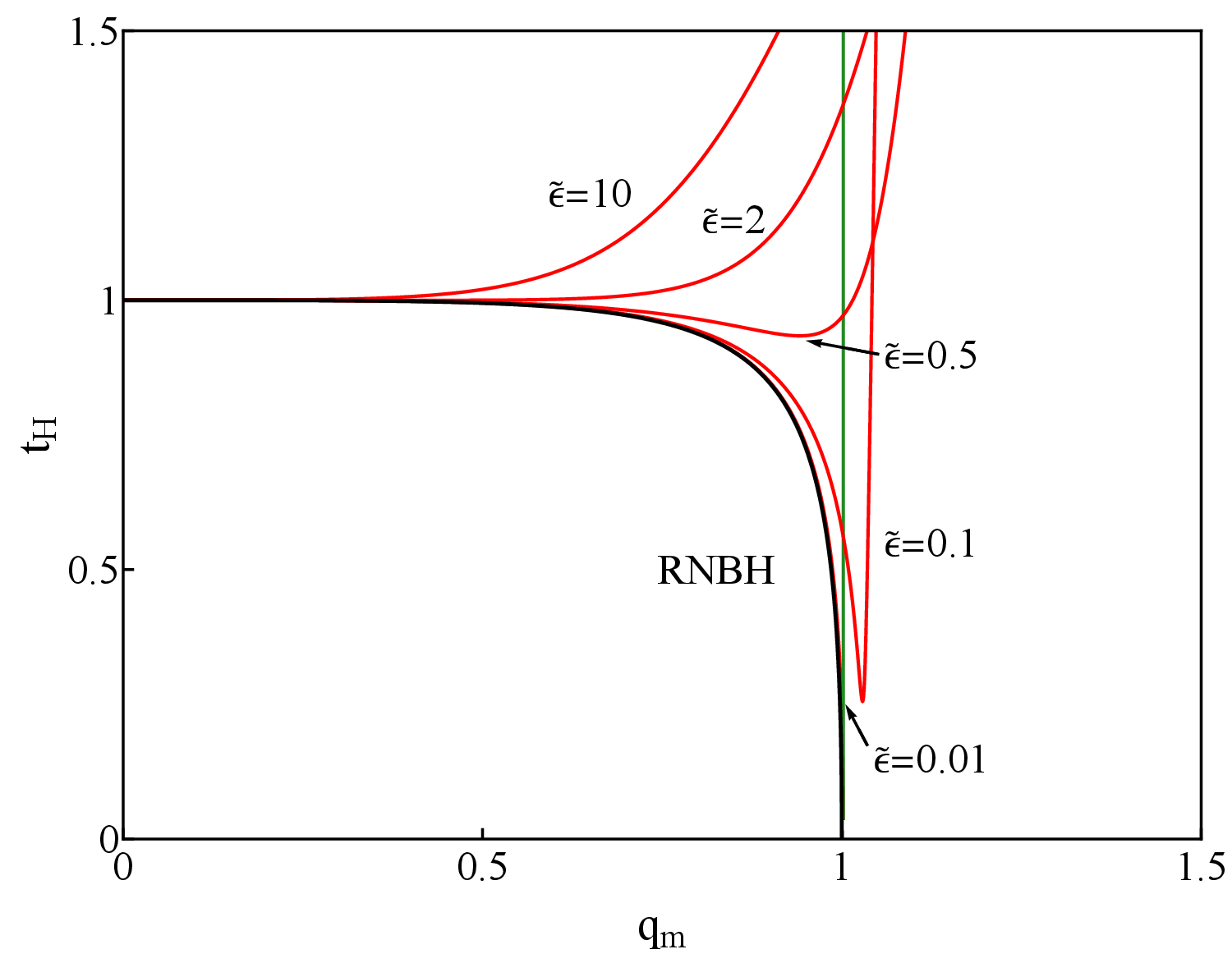}
}
\caption{(a) The locations of three scaled horizons $(r_1/M, r_2/M, r_3/M)$ as the function of $q_m$ for purely magnetic black holes with several values of $\tilde{\epsilon}$; (b) The domain of existence for single and triple horizons as functions of $(q_m,\tilde{\epsilon})$ for purely magnetic black holes; (c) The profile of metric $N(r)$ in the radial coordinate $r$ for purely magnetic black holes with fixed $\tilde{\epsilon}=0.01$ and several values of $q_m$; (d) The reduced area of horizon $a_H$ and (e) reduced Hawking temperature $t_H$ as the function of $q_m$ for purely magnetic black holes with several values of $\tilde{\epsilon}$.
}
\label{fig:m_aH_tH_rH_q_ep}
\end{figure}

The horizons are determined by the equation $N(r)=0$, which is a sixth-order polynomial in $r$. Although six roots exist in principle, only the positive real roots correspond to physical horizons. Depending on the values of $(q_m,\epsilon)$, up to three positive real roots may exist, corresponding to one event horizon and two distinct inner horizons. Their behavior is illustrated in Fig.~\ref{fig:m_aH_tH_rH_q_ep}(a), where the horizon radii are shown as functions of the magnetic charge-to-mass ratio $q_m$ for several values of the nonlinear coupling $\tilde{\epsilon}$.

As in the purely electric case, the purely magnetic EEH black hole initially possesses only a single event horizon $r_1/M$, whose radius decreases monotonically from the Schwarzschild value $r_1/M=2$ as the magnetic charge-to-mass ratio $q_m$ increases. Throughout its evolution, the event horizon $r_1/M$ remains very close to the outer horizon of the magnetically charged RN black hole for small $q_m$. For relatively small $\tilde{\epsilon}=0.01$, two additional inner horizons, denoted by $r_2/M$ (blue curve) and $r_3/M$ (green curve), emerge once $q_m$ exceeds a critical value. As $q_m$ increases further, the intermediate horizon $r_2/M$ moves outward, closely approaching the inner horizon of the RN solution, and eventually merges with the event horizon $r_1/M$ at $q_m=1$, corresponding to an extremal black hole. Meanwhile, the innermost horizon $r_3/M$ continues to increase monotonically and remains as the only event horizon for $q_m>1$. In contrast, for larger nonlinear couplings $(\tilde{\epsilon}=0.1, 0.5, 2, 10)$, the additional inner horizons do not develop, and the spacetime possesses only a single event horizon in the entire parameter range. These results indicate that sufficiently strong $\tilde{\epsilon}$ suppress the formation of multiple-horizon configurations.

Fig.~\ref{fig:m_aH_tH_rH_q_ep}(b) illustrates the domain of existence in the $(\tilde{\epsilon}, q_m)$ parameter space for $0\leq\tilde{\epsilon}\leq0.1$. In the EM limit $(\tilde{\epsilon}=0)$, the solution reduces to the magnetically charged RN black hole, which exists only for $0\leq q_m\leq1$, with the extremal configuration located at $q_m=1$. For $\tilde{\epsilon}\neq0$, the phase diagram is bounded by two characteristic curves: the critical line, where the additional inner horizons first emerge, and the extremal line, where the event horizon and the intermediate horizon merge. Hence, the three-horizon configuration exists only within the finite region enclosed by these two boundaries. As $\tilde{\epsilon}$ increases, the critical line approaches the extremal line, causing the three-horizon region to shrink continuously until it terminates at the critical value $\tilde{\epsilon}\approx0.076$. Beyond this point, EEH black holes possess only a single event horizon in their domain of existence. Interestingly, the extremal line remains fixed at $q_m=1$, indicating that the extremality condition is unaffected by the EH nonlinear interaction in the purely magnetic sector.

Fig.~\ref{fig:m_aH_tH_rH_q_ep}(c) displays the metric function $N(r)$ for the purely magnetic EEH black holes with fixed $\tilde{\epsilon}=0.01$ as a function of the radial coordinate $r$ for several values of $q_m$. For small values of $q_m$, the metric function increases monotonically from the vicinity of the origin and approaches unity asymptotically, indicating the existence of a single event horizon $r_1/M$. In contrast to the purely electric case, however, $N(r)$ gradually develops a local turning point as $q_m$ increases. When this turning point first intersects the $N(r)=0$ axis at $(q_m\approx0.842)$, two additional horizons, $r_2/M$ and $r_3/M$, emerge simultaneously, giving rise to a three-horizon configuration. As $q_m$ is increased further, the event horizon $r_1/M$ moves inward while the intermediate horizon $r_2/M$ moves outward, and the two horizons eventually merge at $q_m\approx1.002$. Beyond this critical value, the branches corresponding to $r_1/M$ and $r_2/M$ terminate, leaving $r_3/M$ as the only remaining positive root of $N(r)=0$. Consequently, $r_3/M$ becomes the event horizon for larger values of $q_m$. This behavior provides a clear geometric illustration of the transition from a one-horizon phase to a three-horizon phase and finally back to a single-horizon phase.

Fig.~\ref{fig:m_aH_tH_rH_q_ep}(d) presents the reduced horizon area $a_H$ as a function of $q_m$ for $\tilde{\epsilon}=0.01$, $0.1$, $0.5$, $2$, and $10$. For all nonlinear couplings considered, $a_H$ initially decreases from the Schwarzschild value $a_H=1$ as $q_m$ increases. For the weakest nonlinear coupling $(\tilde{\epsilon}=0.01)$, the event horizon radius $r_1/M$ decreases monotonically until it merges with the intermediate horizon $r_2/M$ at the extremal point, where this branch terminates when $q_m\leq 1$. The subsequent increase in $a_H$ (green curve) does not correspond to a re-expansion of the original event horizon. Instead, it is associated with the outer horizon $r_3$, which becomes the event horizon for $q_m>1$. By contrast, for larger $\tilde{\epsilon}$, the additional inner horizons do not form. In these cases, the event horizon remains unique within the parameter range considered, and $a_H$ decreases initially before increasing monotonically due to the re-expansion of the event horizon.

Fig.~\ref{fig:m_aH_tH_rH_q_ep}(e) presents the reduced Hawking temperature $t_H$ as a function of $q_m$. For relatively weak nonlinear couplings, $\tilde{\epsilon}=0.01$, $0.1$, and $0.5$, $t_H$ remains nearly constant as $q_m$ increases, decreases to a minimum value at intermediate $q_m$, and subsequently rises rapidly for larger $q_m$. For the weakest coupling $(\tilde{\epsilon}=0.01)$, the increase in $t_H$ originates from the transition between different horizon branches: for $q_m<1$, $t_H$ is evaluated on the event horizon $r_1$, whereas for $q_m>1$, it is determined by the outermost event horizon $r_3$ (green curve). Hence, the minimum of $t_H$ marks the branch transition rather than a turning point of the event horizon. By contrast, for $\tilde{\epsilon}=0.1$ and $0.5$, the non-monotonic behavior is associated with the re-expansion of the event horizon radius $r_1/M$. For stronger nonlinear couplings, $\tilde{\epsilon}=2$ and $10$, only a single horizon exists within the parameter range considered. In this case, $t_H$ increases monotonically with $q_m$, with the rate of increase becoming progressively larger as $\tilde{\epsilon}$ increases.

We now investigate several additional geometrical and physical properties of the purely magnetic EEH black holes. From Eq.~\eqref{eq:scalar_Ricci}, the Ricci scalar is given by
\begin{equation}
R=\frac{8\epsilon P^4}{r^8}\,,
\end{equation}
which vanishes in the EM limit $(\epsilon=0)$, as expected for the magnetically charged RN solution. Thus, the nonvanishing Ricci scalar provides a direct measure of the nonlinear EH correction to the spacetime geometry.

Similarly, the Kretschmann scalar obtained from Eq.~\eqref{eq:scalar_Kretsh} is
\begin{equation}
K=\frac{8\left(7P^4-12MP^2r+6M^2r^2\right)}{r^8}
+\frac{32\epsilon P^4\left(-190P^2r^4+140Mr^5+239\epsilon P^4\right)}{25r^{16}}\,,
\end{equation}
where the first term corresponds to the magnetically charged RN solution, while the second term represents the EH correction. Both curvature invariants diverge as $r\rightarrow0$, indicating that the spacetime still possesses a central curvature singularity despite the modification of its horizon structure by the nonlinear electromagnetic interaction.

Finally, the energy density $\rho$ is given by
\begin{equation}
\rho
=
\frac{P^2}{8\pi}
\left(
\frac{1}{r^4}
-\frac{2\epsilon P^2}{r^8}
\right)\,.
\end{equation}
Hence, $\rho$ becomes negative for $r<(2\epsilon P^2)^{1/4}$.
In addition, Eq.~\eqref{eq:WEC_tangential} reduces to
\begin{equation}
\rho+p_t
=
\frac{P^2}{4\pi r^4}
\left(
1-\frac{4\epsilon P^2}{r^4}
\right)\,,
\end{equation}
which becomes negative for $r<(4\epsilon P^2)^{1/4}$.
Since $(4\epsilon P^2)^{1/4}>(2\epsilon P^2)^{1/4}$, the tangential
condition $\rho+p_t\geq0$ is violated before the energy density itself becomes negative as one approaches the black-hole center. Therefore,
the WEC is violated sufficiently close to the center as a direct consequence of the nonlinear EH correction. In the EM limit $\epsilon\to0$, both $\rho$ and $\rho+p_t$ remain non-negative.

\subsubsection{Case with $V'(r)=\pm\sqrt{-\dfrac{1}{4\epsilon}\left(1+\dfrac{10\epsilon P^2}{r^4}\right)}$}

For $\epsilon>0$, the above expression implies that $V'$ is purely imaginary. Consequently, this branch does not correspond to a real purely magnetic in the electromagnetic configuration, since the associated electric potential $V(r)$ becomes complex. Nevertheless, the Einstein field equations remain well defined and admit a formal black hole solution. To the best of our knowledge, this branch has received little attention in the existing literature.

Substituting $V'$ into Eq.~\eqref{eq:dyn_mprime}, we obtain
\begin{equation}
m'(r)=\frac{7P^2}{4r^2}\left(1+\frac{3\epsilon P^2}{r^4}\right)+\frac{r^2}{16\epsilon} \,.
\end{equation}
Integrating this equation yields
\begin{equation}
m(r)=M-\frac{7P^2}{4r}-\frac{21\epsilon P^4}{20r^5}+\frac{r^3}{48\epsilon}\,,
\end{equation}
or equivalently,
\begin{equation}
N(r)=1-\frac{2m(r)}{r}
=1-\frac{2M}{r}
+\frac{7P^2}{2r^2}
+\frac{21\epsilon P^4}{10r^6}
-\frac{r^2}{24\epsilon}\,.
\end{equation}
The first three terms resemble those of the magnetically charged RN solution, with an effective rescaling of the magnetic-charge contribution, whereas the last two terms originate from the EH correction. In particular, the quadratic term in $r$ can be identified with an effective cosmological constant through
\begin{equation}
\frac{\Lambda}{3}=\frac{1}{24\epsilon}\,,
\end{equation}
indicating that this branch is asymptotically de-Sitter. Hence, the corresponding solution may be interpreted as a magnetically charged de-Sitter black hole within the EEH theory.

For completeness, we also examine the weak energy condition for this branch of solution. The corresponding energy density is
\begin{equation}
\rho(r)
=
\frac{7P^2}{16\pi r^4}
+\frac{1}{64\pi\epsilon}
+\frac{21\epsilon P^4}{16\pi r^8} \,.
\end{equation}
Moreover, substituting this branch into
Eq.~\eqref{eq:WEC_tangential} yields
\begin{equation}
\rho+p_t
=
\frac{7P^2}{8\pi r^4}
+\frac{21\epsilon P^4}{4\pi r^8}\,.
\end{equation}
Thus, for $\epsilon>0$, both $\rho$ and $\rho+p_t$ are manifestly positive. Together with $\rho+p_r=0$, this shows that all the inequalities defining the WEC are formally satisfied within
the spacetime.

Although the EH effective theory derived from QED requires a $\epsilon>0$, it is nevertheless instructive to consider the formal extension to $\epsilon<0$. In this case, the effective cosmological constant becomes negative, giving rise to an asymptotically anti de-Sitter (AdS) branch. The corresponding metric function is
\begin{equation}
N(r)=1-\frac{2m(r)}{r}
=1-\frac{2M}{r}
+\frac{7P^2}{2r^2}
-\frac{21|\epsilon| P^4}{10r^6}
+\frac{r^2}{24|\epsilon|}\,.
\end{equation}
Moreover, the electric field becomes real-valued only when $r^4\geq 10|\epsilon|P^2$,
\begin{equation}
V'(r)=\pm\sqrt{\frac{1}{4|\epsilon|}\left(1-\frac{10|\epsilon|P^2}{r^4}\right)}\,,
\qquad \epsilon<0\,,
\end{equation}
which approaches the constant value $V'\rightarrow \pm1/(2\sqrt{|\epsilon|})$ as $r\rightarrow\infty$. The direct integration yields the closed form of gauge field,
\begin{equation}
    V(r) = \mp \sqrt{\frac{r^4-10|\epsilon|P^2}{4|\epsilon|r^2}} {}_2F_1\left(-\frac{1}{2}, 1,\frac{3}{4},\frac{r^4}{r^4-10|\epsilon|P^2}\right)- V_H\,.
\end{equation}

The corresponding energy density is
\begin{equation}
\rho(r)=
\frac{7P^2}{16\pi r^4}
-\frac{1}{64\pi|\epsilon|}
-\frac{21|\epsilon|P^4}{16\pi r^8}\,.
\end{equation}
Unlike the case $\epsilon>0$, $\rho$ is no longer positive definite. The condition $\rho(r)\ge0$ is satisfied only within the finite radial interval
\begin{equation}
2(7-2\sqrt7)|\epsilon|P^2
\le r^4
\le
2(7+2\sqrt7)|\epsilon|P^2 \,.
\end{equation}
Thus, the formal AdS extension with $\epsilon<0$ satisfies the WEC only over a finite range of radii, while it is violated both in the vicinity of the black-hole center and in the asymptotic region.

Although this branch lies outside the physical parameter range of the EH effective theory, it provides an interesting mathematical extension of the EEH system. The comparison between the positive- and negative-coupling branches demonstrates that the sign of the EH coupling governs not only the asymptotic geometry but also the admissible electromagnetic configurations and the validity of the energy conditions.

\subsection{Dyonic case $(Q\neq0, P\neq0)$}

We first analyze the discriminant $\Delta$ of the cubic equation and find that $\Delta>0$. Hence, the remaining two branches, Eqs.~\eqref{eq:root2} and \eqref{eq:root3} are complex, and only the real branch given by Eq.~\eqref{eq:root1} needs to be considered together with Eq.~\eqref{eq:dyn_mprime} to construct the dyonic black hole solutions.

Due to the highly nonlinear coupling between the gravitational and electromagnetic fields, analytical dyonic solutions don't appear to admit closed-form expressions. Therefore, we solve the coupled field equations numerically. Since both the purely electric and purely magnetic sectors admit multiple horizon configurations, determining the correct event horizon a priori is nontrivial. To circumvent this difficulty, we integrate the ODEs backward from the asymptotic region toward the origin. Compared with the conventional shooting method that integrates outward from the event horizon, this strategy avoids specifying the horizon location in advance and allows the horizon structure to emerge naturally from the numerical solutions, without any prior assumption regarding the number or locations of the horizons. Specifically, we choose a sufficiently large outer boundary, $r_{\mathrm{max}}=10^{4}M$, and impose the asymptotically flat boundary conditions 
$N(\infty)=1$, $V_1(\infty)=0$. Without loss of generality, we set the black hole mass to $M=1$. During the numerical integration, the zeros of the metric function $N(r)$ are monitored to determine both the number and the locations of the horizons. This procedure enables a systematic classification of the one-, two-, and three-horizon phases within a single numerical framework.

The asymptotic boundary conditions are obtained from the large-$r$ expansion of the field equations,
\begin{align}
    N(r) &= 1-\frac{2M}{r}+ \frac{\left(1+\beta^2\right)Q^2}{r^2} - \frac{2\epsilon \left(1+5\beta +\beta^2\right)Q^4}{r^6} + \frac{2\epsilon^2\left(2+5\beta^2\right)^2Q^6}{9r^{10}} + O\left(r^{-14}\right)\,, \\
    V_1(r) &= \frac{Q}{r} - \frac{2\epsilon \left(2+5\beta^2\right)Q^3}{5 r^5} + \frac{4\epsilon^2\left(2+5\beta^2\right)\left(6 + 5\beta^2\right) Q^5}{9r^9} + O(r^{-13})  \,.
\end{align}
When $\epsilon=0$, the first three terms reproduce the asymptotic behavior of the dyonic RN solution, whereas the limit $\beta=0$ reduces to the purely electric asymptotic expansion. These asymptotic expansions provide the initial conditions for the backward numerical integration.

\begin{figure}
\mbox{
(a)
\includegraphics[angle =0,scale=0.32]{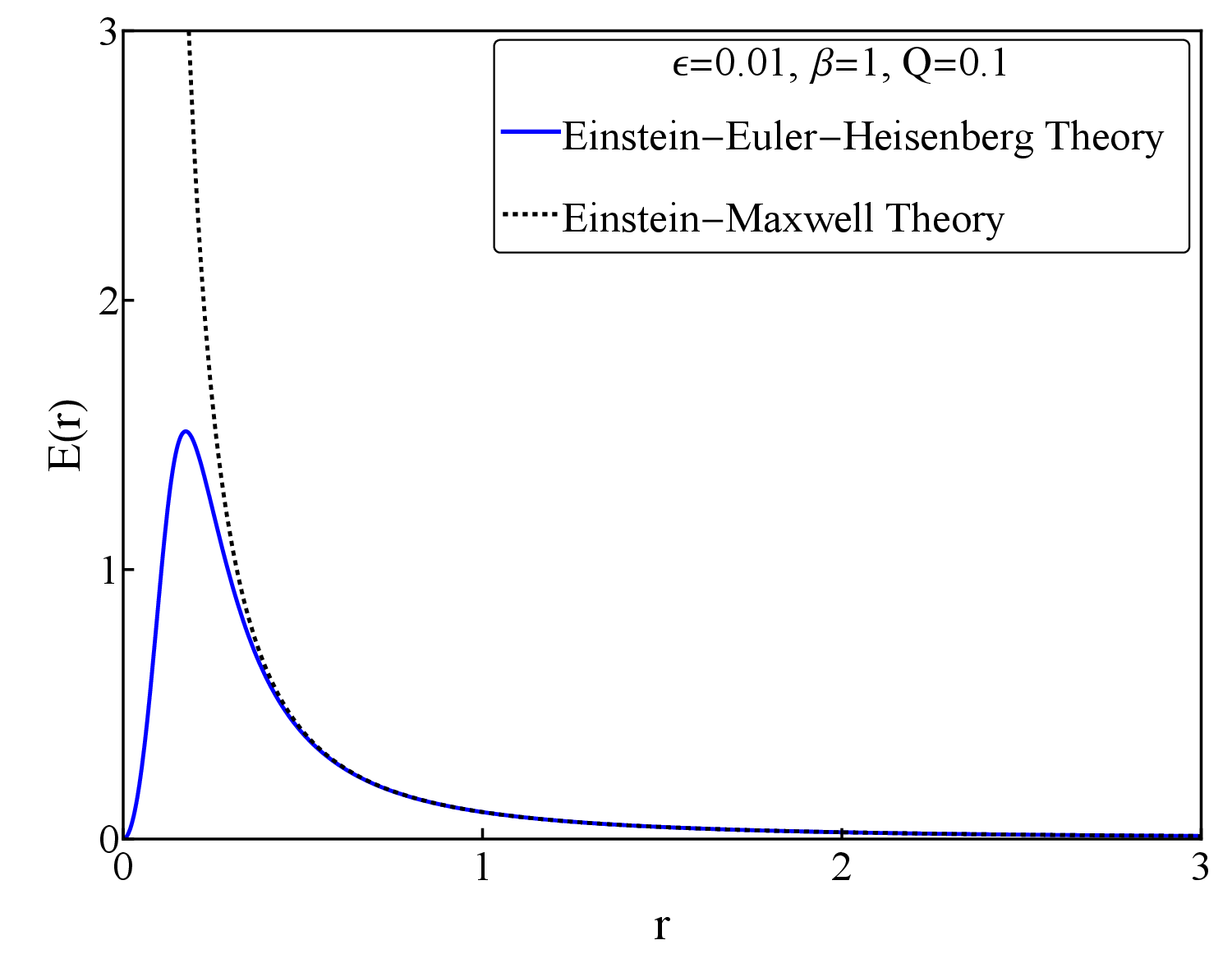}
(b)
\includegraphics[angle =0,scale=0.32]{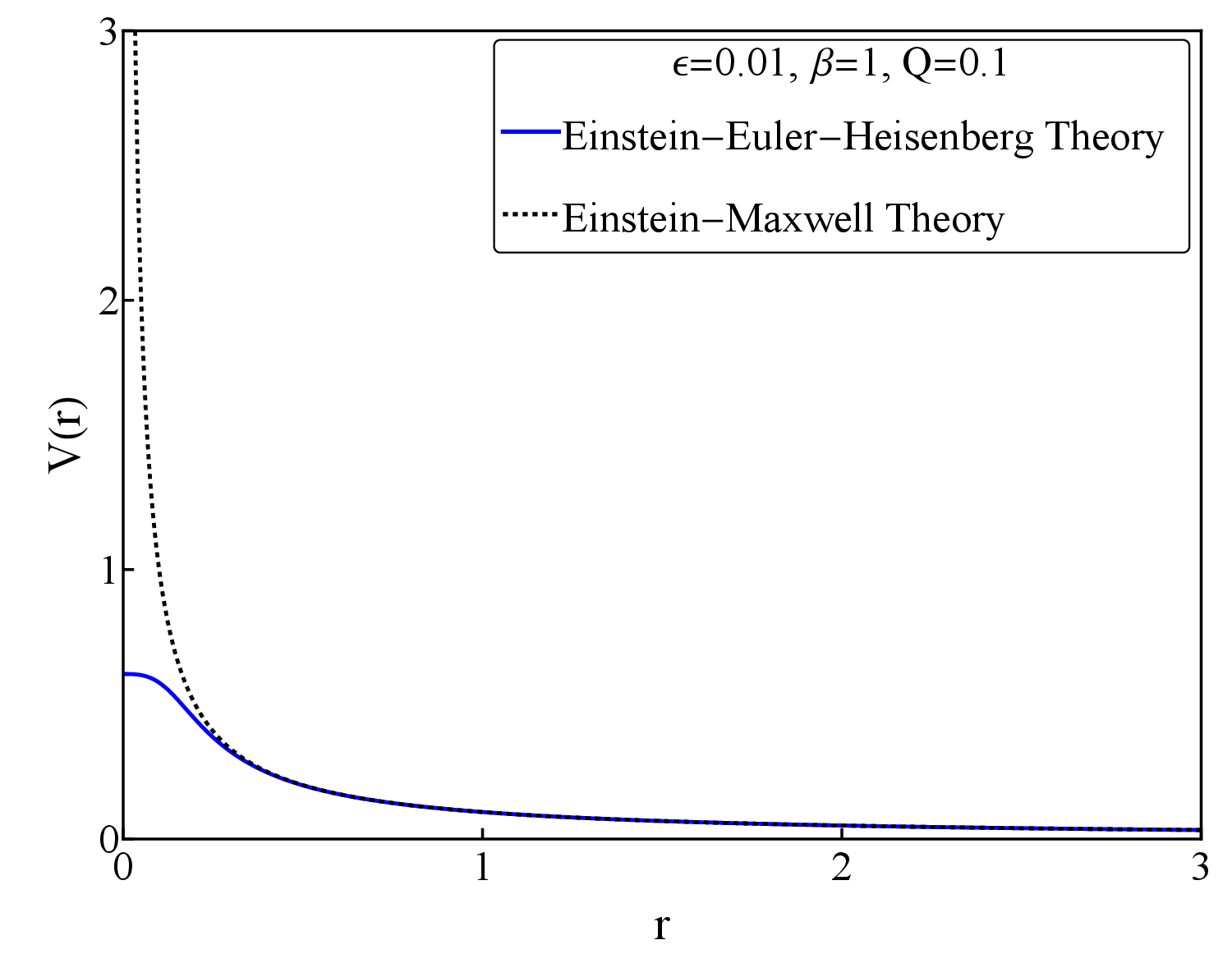}
}
\caption{The comparison between EM and EEH theories for profiles of (a) electric field $E(r)$ and (b) gauge field $V_1(r)$ with $\epsilon=0.01$, $\beta=1$ and $Q=0.1$ in the dyonic case.}
\label{fig:d_E_V}
\end{figure}

Since Eq.~\eqref{eq:root1} is decoupled from Eq.~\eqref{eq:dyn_mprime}, the electric field $E_1(r)$ can be determined independently before solving the remaining Einstein equations. The corresponding electric potential $V_1(r)$ is then obtained by numerical integration. Fig.~\ref{fig:d_E_V} presents the resulting both $E_1(r)$ and $V_1(r)$ for fixed $\epsilon=0.01$, $\beta=1$, and $Q=0.1$. In the EM theory, both $E_1(r)$ and $V_1(r)$ diverge at the origin and decrease monotonically to zero as $r\rightarrow\infty$. By contrast, in the EEH theory both quantities remain regular everywhere. In particular, $E_1(r)$ vanishes at the black-hole center, while both $E_1(r)$ and $V_1(r)$ approach their EM counterparts asymptotically as $r\rightarrow\infty$. Consequently, the EH nonlinear interaction removes the classical Coulomb singularity while preserving the EM limit at large distances. The regularity of $E_1(r)$ can be demonstrated analytically. Following Ref.~\cite{Kruglov:2017ymn}, Eq.~\eqref{eq:root1} can be rewritten as
\begin{equation}
E_1(r)
=
\sqrt{B}\,
\sinh\!\left(
\frac13\operatorname{arcsinh}D
\right)\,,
\end{equation}
where
\begin{equation}
B=
\frac{r^4+10\epsilon\beta^2Q^2}
{3\epsilon r^4}\,, \quad
D=
\frac{Q}{8\epsilon r^2} B ^{-3/2} \,.
\end{equation}
Since $D\rightarrow0$ as $r\rightarrow0$, we obtain 
\begin{align} 
\lim_{r\rightarrow0}E_1(r)
&=
\lim_{r\rightarrow0}
\frac{\sinh\!\left(\frac13\operatorname{arcsinh}D\right)}{{\operatorname{arcsinh}D}} \
\frac{\operatorname{arcsinh}D}{D}
\sqrt{B}\,D \,,
\nonumber\\
&=
\frac13
\lim_{r\rightarrow0}
\sqrt{B}\,D
=
\lim_{r\rightarrow0}
\frac{Qr^2}
{2r^4+20\epsilon\beta^2Q^2}
=0\,.
\end{align}
Hence, unlike the EM solution, the electric field not only remains finite but vanishes smoothly at the black-hole center. Thus, the electric contribution to the electromagnetic energy density is completely regularized by the EH nonlinear interaction. Nevertheless, this regularization does not remove the spacetime singularity in the dyonic case, since the magnetic field retains the Maxwell behavior $B=P/r^2$, whose divergent contribution dominates the electromagnetic invariant $\mathcal{F}$ and causes the curvature invariants to diverge at the origin.

\begin{figure}[t]
\mbox{
(a)
\includegraphics[angle =0,scale=0.32]{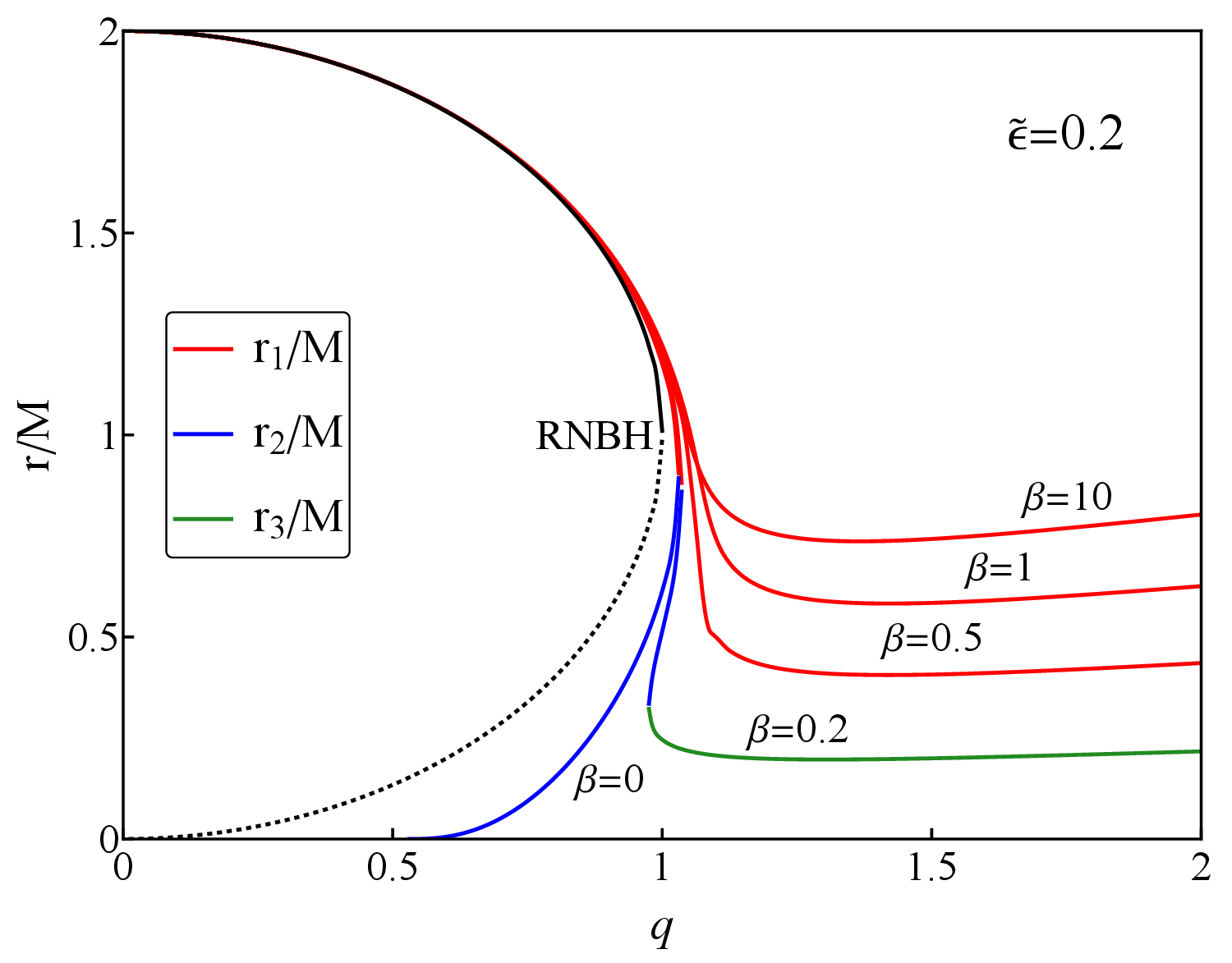}
(b)
\includegraphics[angle =0,scale=0.32]{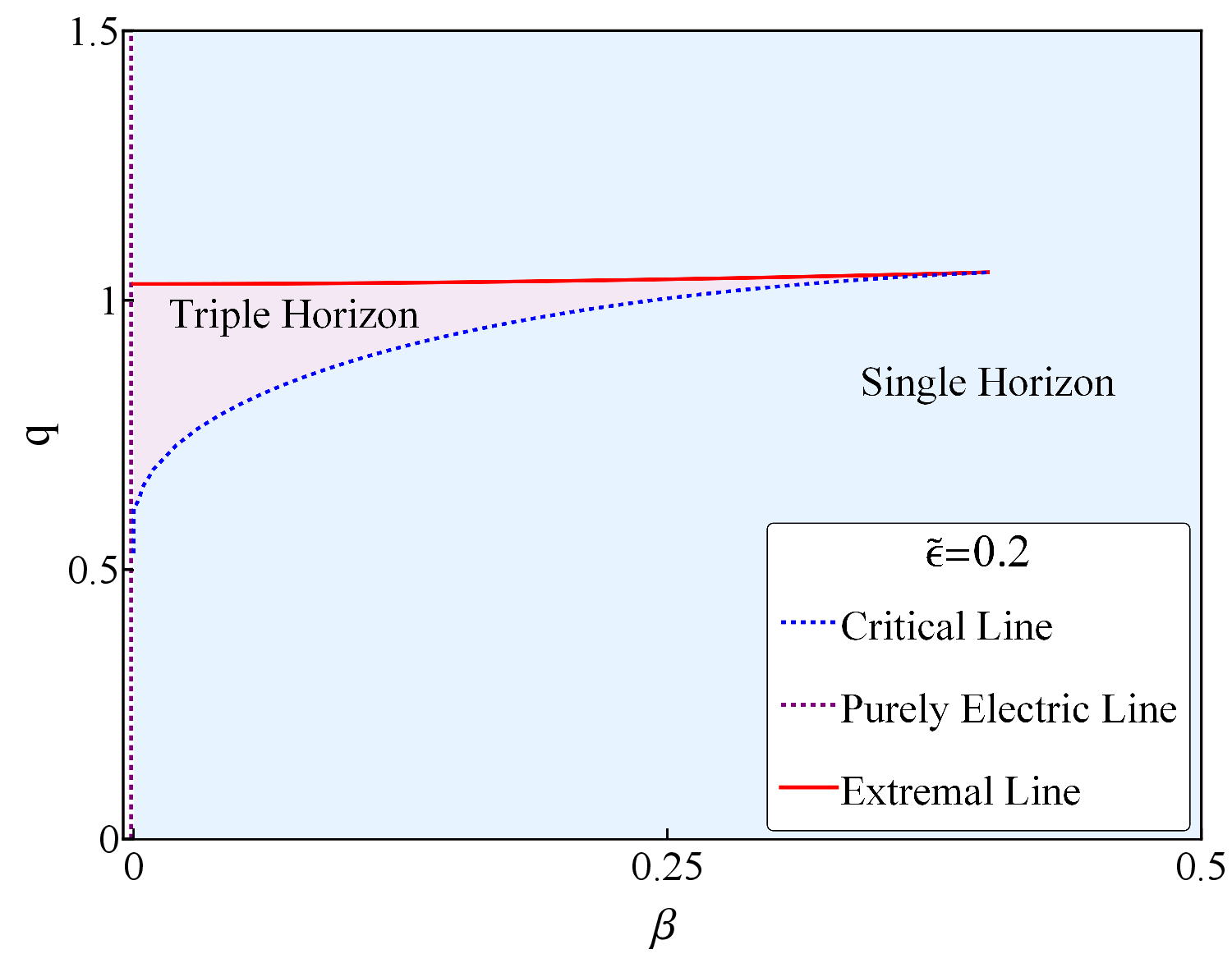}
}
\mbox{
(c)
\includegraphics[angle =0,scale=0.32]{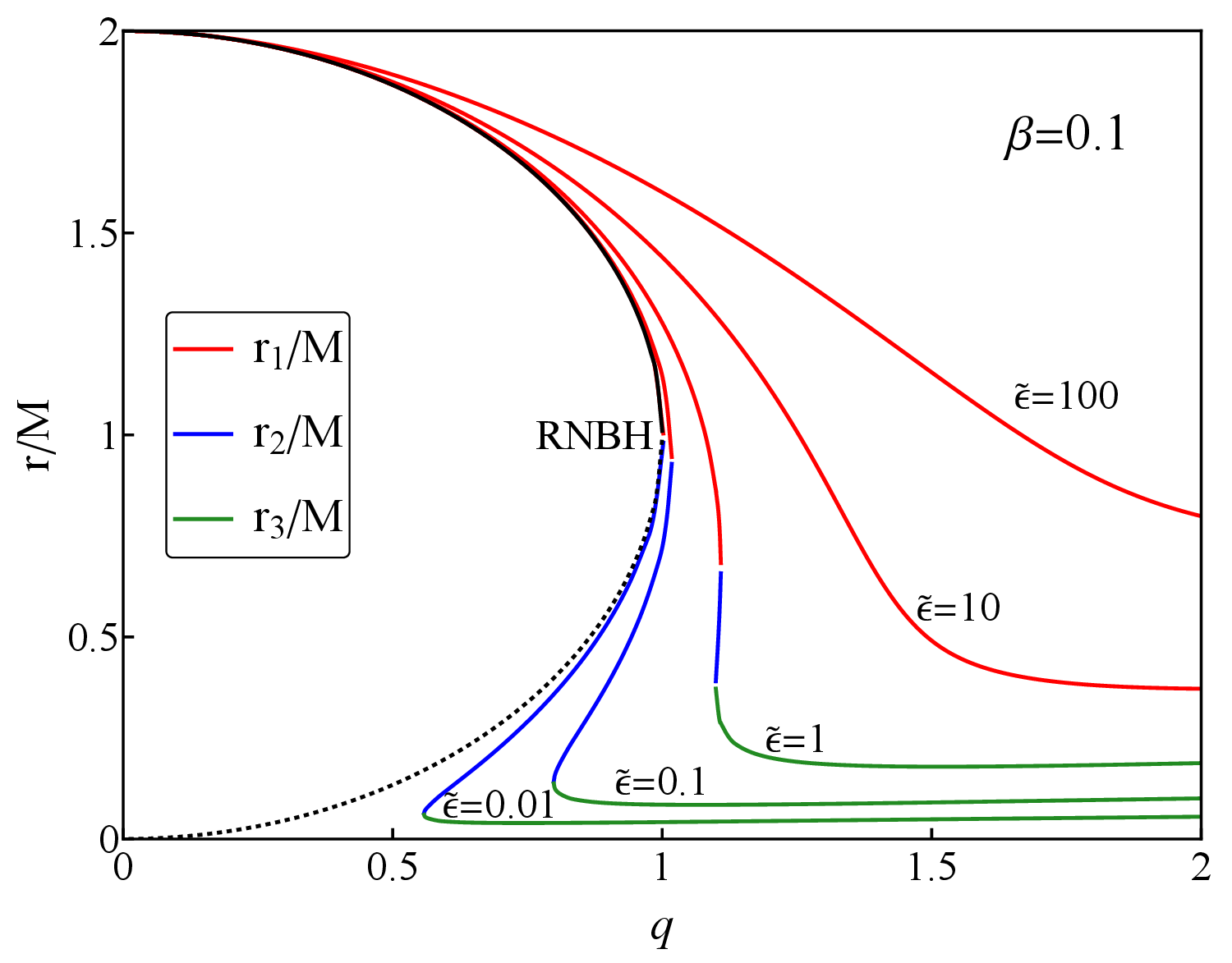}
(d)
\includegraphics[angle =0,scale=0.32]{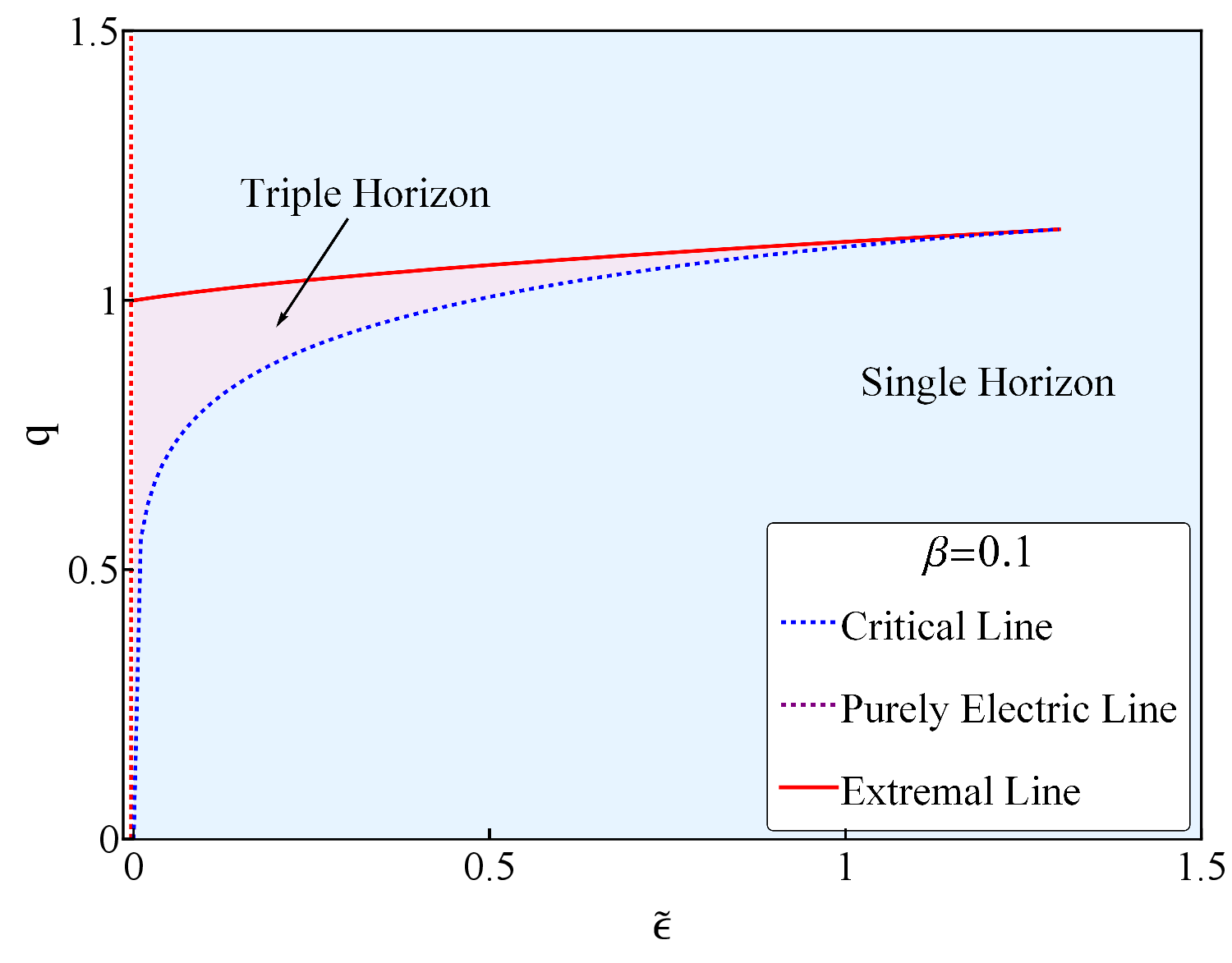}
}
\mbox{
(e)
\includegraphics[angle =0,scale=0.32]{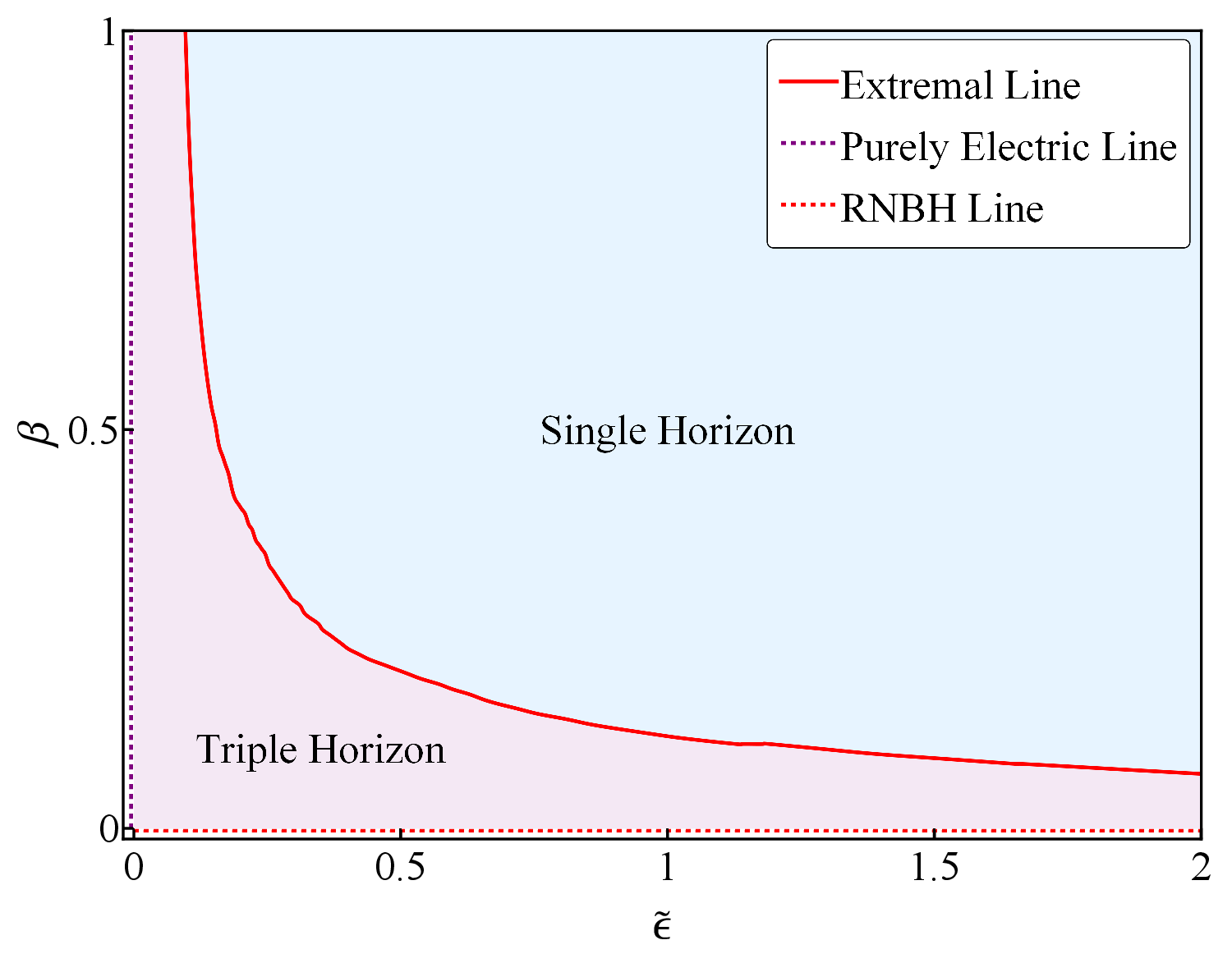}
(f)
\includegraphics[angle =0,scale=0.32]{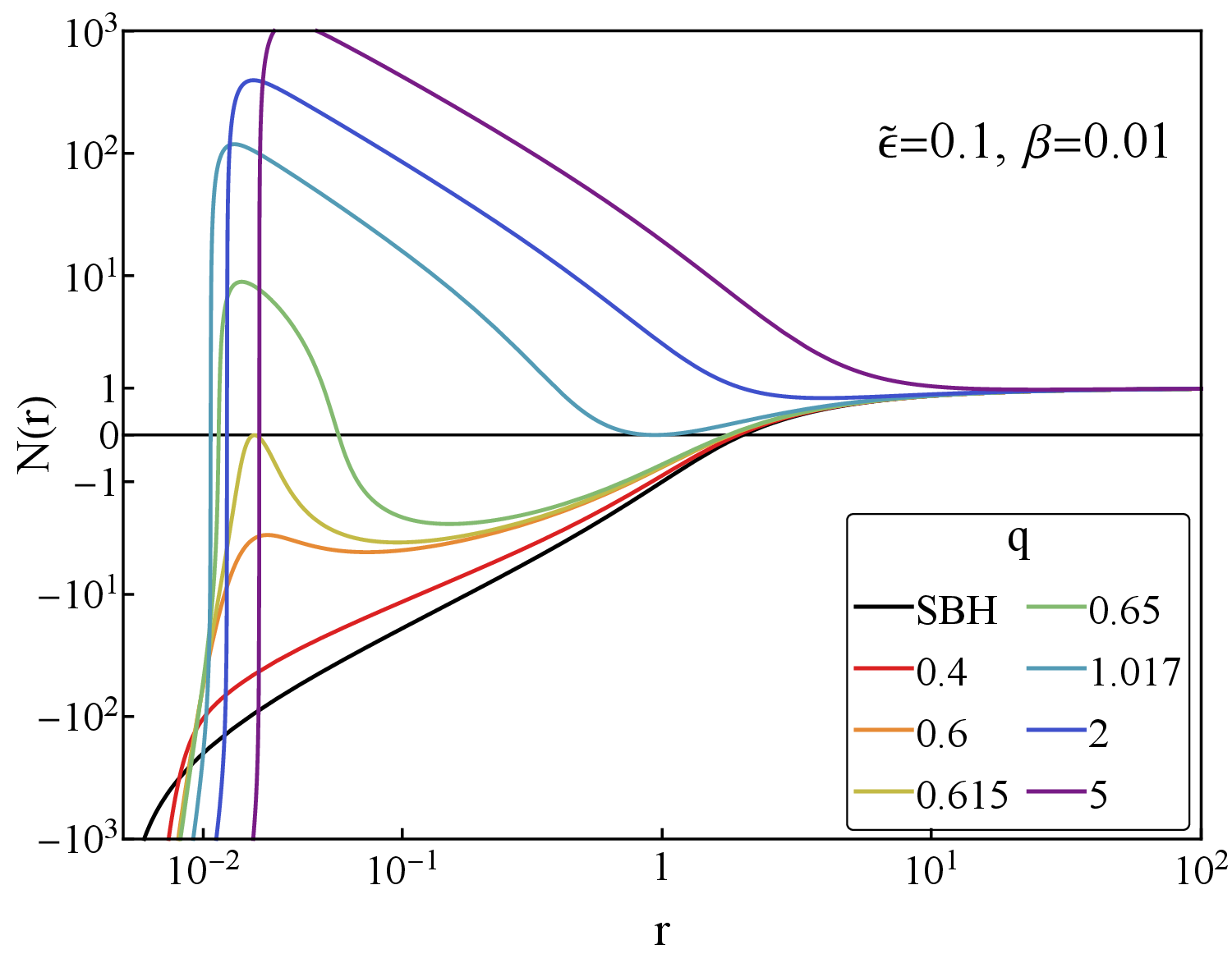}
}
\caption{(a) The locations of three scaled horizons $(r_1/M, r_2/M, r_3/M)$ as the function of $q$ for dyonic black holes with fixed $\tilde{\epsilon}=0.2$ and several values of $\beta$; (b) The domain of existence for single and triple horizons as functions of $(\beta,q)$ for dyonic black holes with fixed $\tilde{\epsilon}=0.2$; (c) The locations of three scaled horizons $(r_1/M, r_2/M, r_3/M)$ as the function of $q$ for dyonic black holes with fixed $\beta=0.1$ and several values of $\tilde{\epsilon}$; (d) The domain of existence for single and triple horizons as functions of $(q,\tilde{\epsilon})$ for dyonic black holes with fixed $\beta=0.1$; (e) The domain of existence for single and triple horizons as functions of $(\beta,\tilde{\epsilon})$ for dyonic black holes; (f) The profile of metric $N(r)$ in the radial coordinate $r$ for dyonic black holes with fixed $\tilde{\epsilon}=0.1$, $\beta=0.01$ and several values of $q$.
}
\label{fig:d_q_ep}
\end{figure}

Fig.~\ref{fig:d_q_ep}(a) displays horizon radii $(r_1/M,r_2/M,r_3/M)$ of dyonic EEH black holes as functions of the dimensionless charge parameter $q$ for $\tilde{\epsilon}=0.2$ and several values of the magnetic-to-electric charge ratio $\beta$. The case $\beta=0$ corresponds to the purely electric solution, where the spacetime possesses an outer horizon $r_1/M$ and an inner horizon $r_2/M$, which eventually merge at the extremal point ($q \approx 1.031$). For $\beta \neq 0$, the horizon structure becomes qualitatively similar to that of the purely magnetic solutions. As $q$ increases from zero, the dyonic black hole initially possesses only a single event horizon $r_1/M$, whose radius remains close to that of the corresponding RN solution. For a relatively small magnetic contribution $\beta=0.2$, two additional inner horizons, $r_2/M$ and $r_3/M$, emerge once $q$ exceeds a critical value $(q=0.975)$. As $q$ increases further, the intermediate horizon $r_2/M$ moves outward and eventually merges with the event horizon $r_1/M$ at $q \approx 1.036$, corresponding to an extremal black hole, while the innermost horizon $r_3/M$ becomes the outer horizon, which continues to grow monotonically. For larger magnetic charge ratios ($\beta=0.5, 1, 10$), the additional inner horizons do not develop. Hence, the spacetime possesses only a single event horizon throughout the parameter range considered, which initially decreases with increasing $q$, and then increases monotonically at larger values of $q$.

Fig.~\ref{fig:d_q_ep}(b) illustrates the domain of existence in the $(\beta,q)$ parameter space for $\tilde{\epsilon}=0.2$. Building on the horizon changes shown in Fig.~\ref{fig:d_q_ep}(a), this phase diagram summarizes the regions with one and three horizons. For sufficiently small $\beta$, the spacetime undergoes a transition from a single-horizon configuration to a three-horizon configuration once the charge parameter exceeds a critical value. As $\beta$ increases, the three-horizon region shrinks continuously, indicating that progressively larger values of $q$ are required for the additional inner horizons to form. The three-horizon phase terminates at approximately $(\beta,q)\approx(0.4,1.052)$, beyond which the dyonic EEH black hole possesses only a single event horizon throughout its domain of existence. These results demonstrate that increasing the magnetic contribution suppresses the formation of multiple-horizon configurations and continuously drives the dyonic solutions toward the horizon structure of the purely magnetic limit.

Fig.~\ref{fig:d_q_ep}(c) illustrates the horizon radii $(r_1/M,r_2/M,r_3/M)$ of dyonic EEH black holes with fixed $\beta=0.1$ as functions of the charge parameter $q$ for several values of the nonlinear coupling $\tilde{\epsilon}$. The overall behavior closely resembles that shown in Fig.~\ref{fig:d_q_ep}(a). For relatively weak nonlinear couplings $(\tilde{\epsilon}=0.01,\,0.1,\,1)$, the spacetime evolves from a single-horizon configuration to a three-horizon configuration once $q$ exceeds a critical value, with the event horizon $r_1/M$ and intermediate horizon $r_2/M$ eventually merging at the extremal point $(q=1)$. As $\tilde{\epsilon}$ is increased, the parameter region supporting three horizons shrinks. For sufficiently strong nonlinear couplings $(\tilde{\epsilon}=10$ and $100)$, the additional inner horizons no longer form, leaving a single event horizon $r_1/M$ whose radius decreases initially and then increases monotonically throughout the parameter range considered.

Fig.~\ref{fig:d_q_ep}(d) illustrates the domain of existence in the $(\tilde{\epsilon},q)$ parameter space for dyonic EEH black holes over the range $0\leq\tilde{\epsilon}\leq1.5$. Similar to the purely magnetic case shown in Fig.~\ref{fig:m_aH_tH_rH_q_ep}(b), the phase diagram consists of regions with one and three horizons, separated by the critical and extremal lines. The three-horizon configuration exists only within the finite region enclosed by these two boundaries. As $\tilde{\epsilon}$ increases, both the critical charge and the extremal charge shift toward larger values of $q$, indicating that the EH nonlinear interaction modifies the extremality condition of the dyonic solutions. At the same time, the separation between the two boundaries gradually decreases, causing the three-horizon region to shrink continuously until it terminates at a critical endpoint located at approximately $(\tilde{\epsilon},q)\approx(1.3,1.13)$. Beyond this point, the dyonic EEH black holes possess only a single event horizon within their domain of existence. These results demonstrate that sufficiently strong EH nonlinearities suppress the three-horizon phase and drive the dyonic solutions toward a single-horizon causal structure.

Fig.~\ref{fig:d_q_ep}(e) presents the unified phase diagram in the $(\tilde{\epsilon},\beta)$ parameter space. The horizontal axis $(\beta=0)$ corresponds to the purely electric EEH black holes, while the vertical axis $(\tilde{\epsilon}=0)$ represents the dyonic RN solutions. The red curve denotes the extremal boundary separating the single-horizon and triple-horizon phases. As shown in the figure, the triple-horizon configuration is confined to a finite wedge-shaped region bounded by the RNBH line, the purely electric line, and the extremal curve. Outside this region, the dyonic EEH black holes possess only a single event horizon. The phase boundary further shows that increasing either $\tilde{\epsilon}$ or $\beta$ eventually suppresses the triple-horizon phase, driving the solutions toward a single-horizon configuration. Remarkably, neither the purely electric EEH black holes nor the dyonic RN solutions exhibit a triple-horizon structure. Hence, the phase diagram demonstrates that the three-horizon configuration arises only through the combined effects of the EH nonlinear interaction and a nonvanishing magnetic charge, highlighting its genuinely dyonic origin.

Fig.~\ref{fig:d_q_ep}(f) displays the metric function $N(r)$ for a dyonic EEH black hole with fixed $\tilde{\epsilon}=0.1$ and $\beta=0.01$ as a function of the radial coordinate $r$ for several values of the charge parameter q. Similar to the purely magnetic case shown in Fig.~\ref{fig:m_aH_tH_rH_q_ep}(c), the metric function increases monotonically from the vicinity of the origin and approaches unity asymptotically for small values of $q$, indicating the existence of a single event horizon $r_1/M$. As $q$ increases, however, $N(r)$ gradually develops a local turning point. When this turning point first intersects the $N(r)=0$ axis at $q\approx0.615$, two additional horizons, $r_2/M$ and $r_3/M$, emerge simultaneously, giving rise to a three-horizon configuration. With a further increase in $q$, the event horizon $r_1/M$ moves inward while the intermediate horizon $r_2/M$ moves outward, and the two horizons eventually merge at $q\approx1.017$. Beyond this critical value, the branches corresponding to $r_1/M$ and $r_2/M$ terminate, leaving $r_3/M$ as the only remaining positive root of $N(r)=0$. Consequently, $r_3/M$ becomes the event horizon for larger values of $q$. This evolution of the metric function provides a geometric interpretation of the horizon phase transition identified in the phase diagrams, demonstrating that the dyonic solutions undergo the same sequence of one-, three-, and finally single-event-horizon phases as the purely magnetic black holes.

\begin{figure}[t]
\mbox{
(a)
\includegraphics[angle =0,scale=0.32]{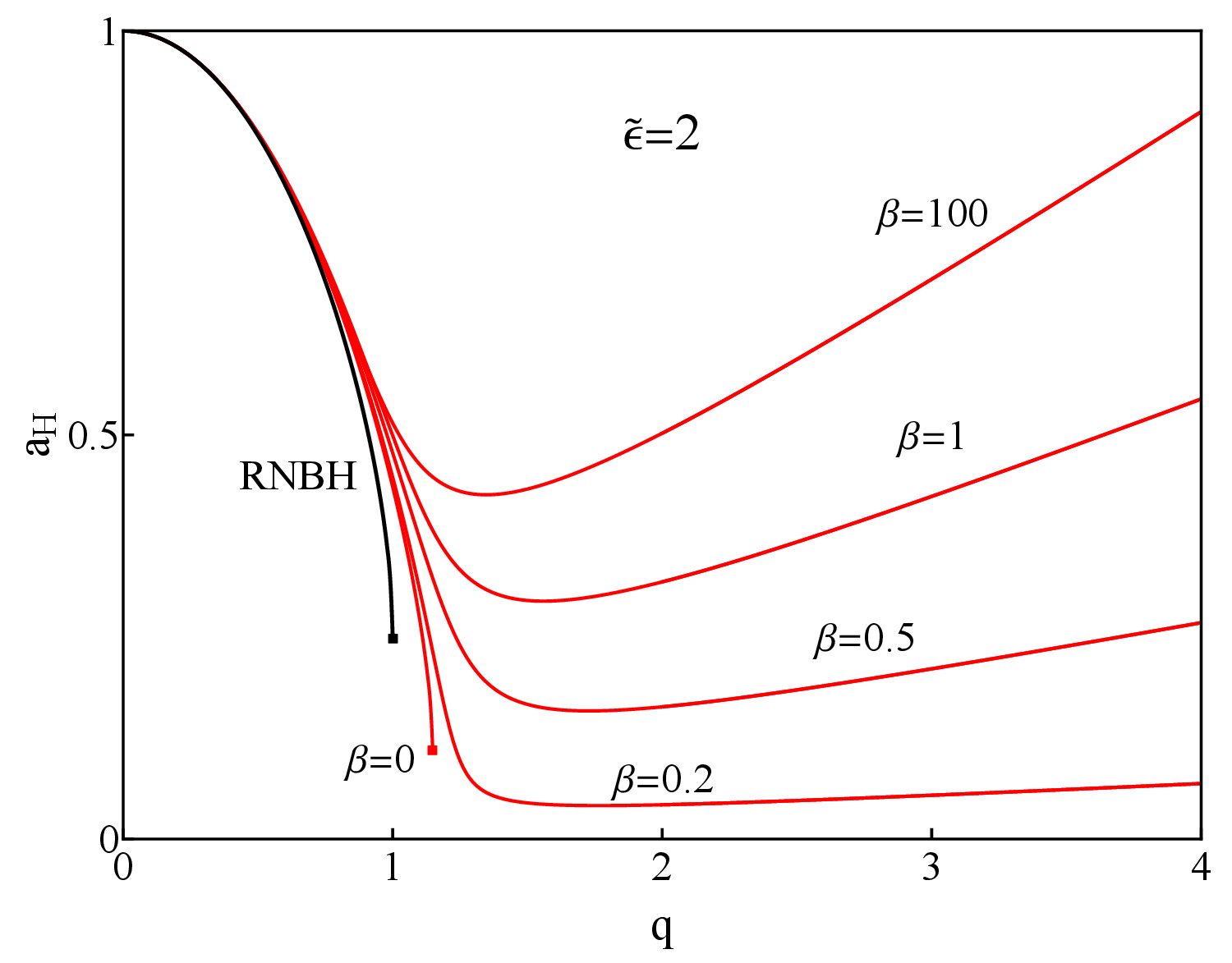}
(b)
\includegraphics[angle =0,scale=0.32]{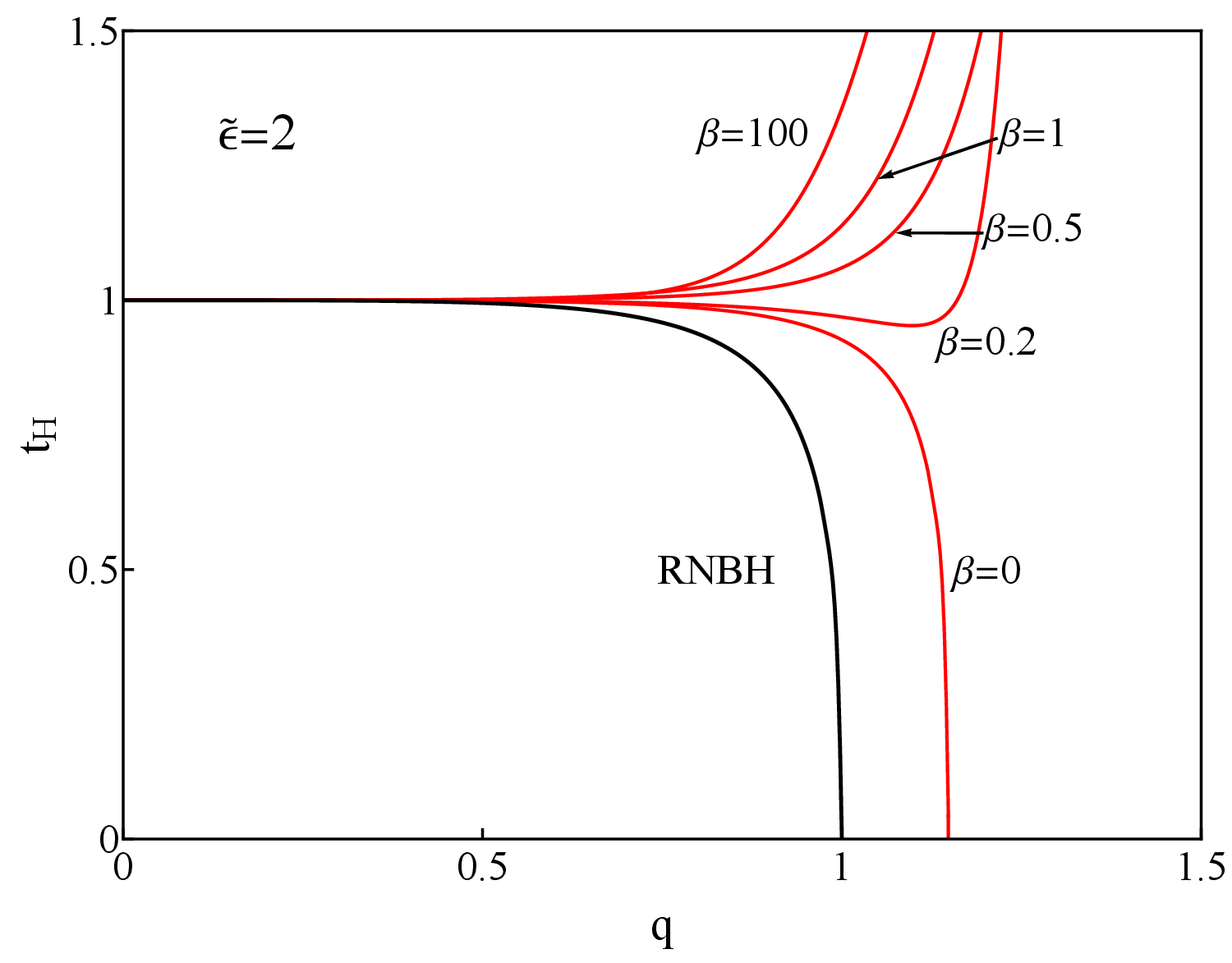}
}
\mbox{
(c)
\includegraphics[angle =0,scale=0.32]{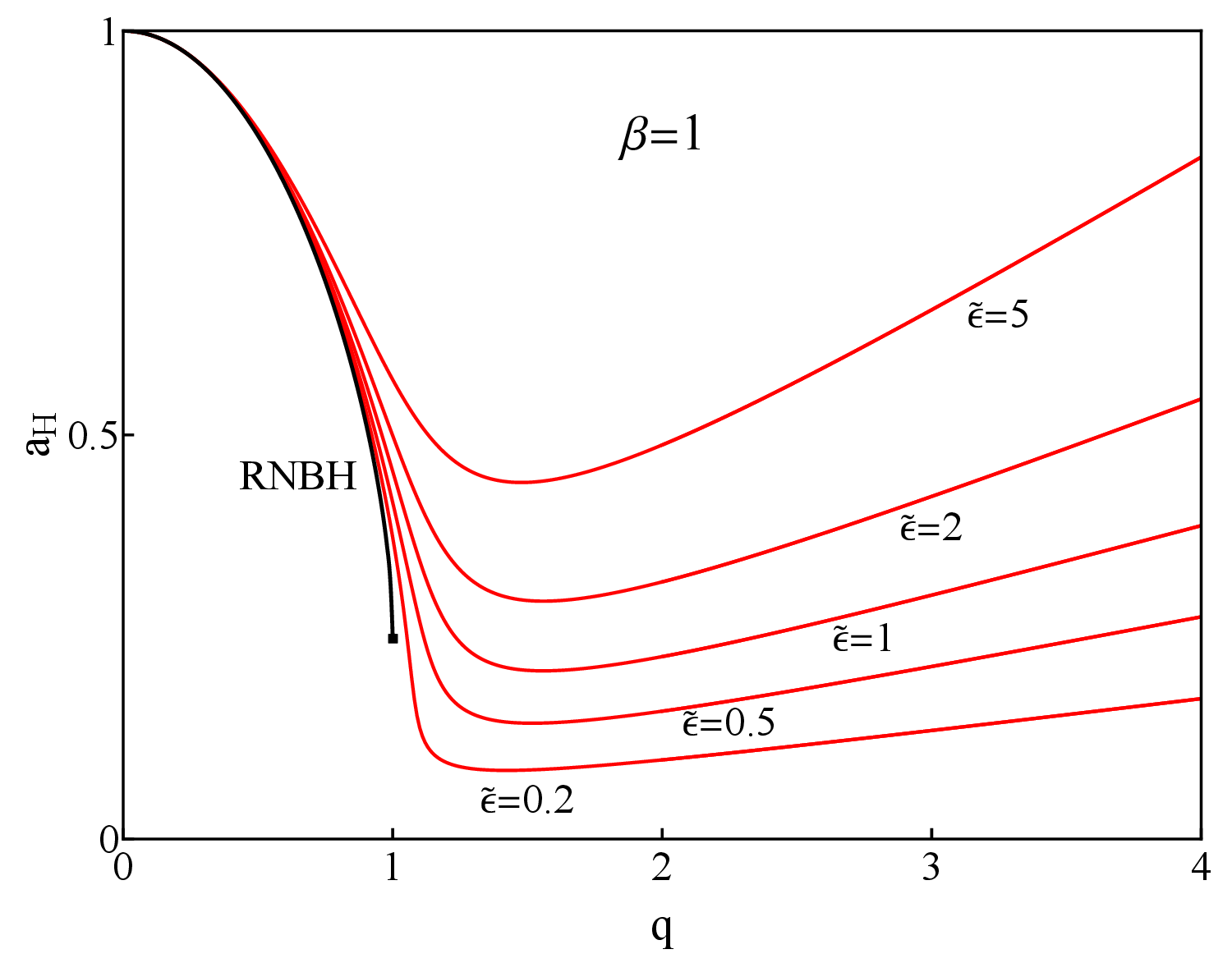}
(d)
\includegraphics[angle =0,scale=0.32]{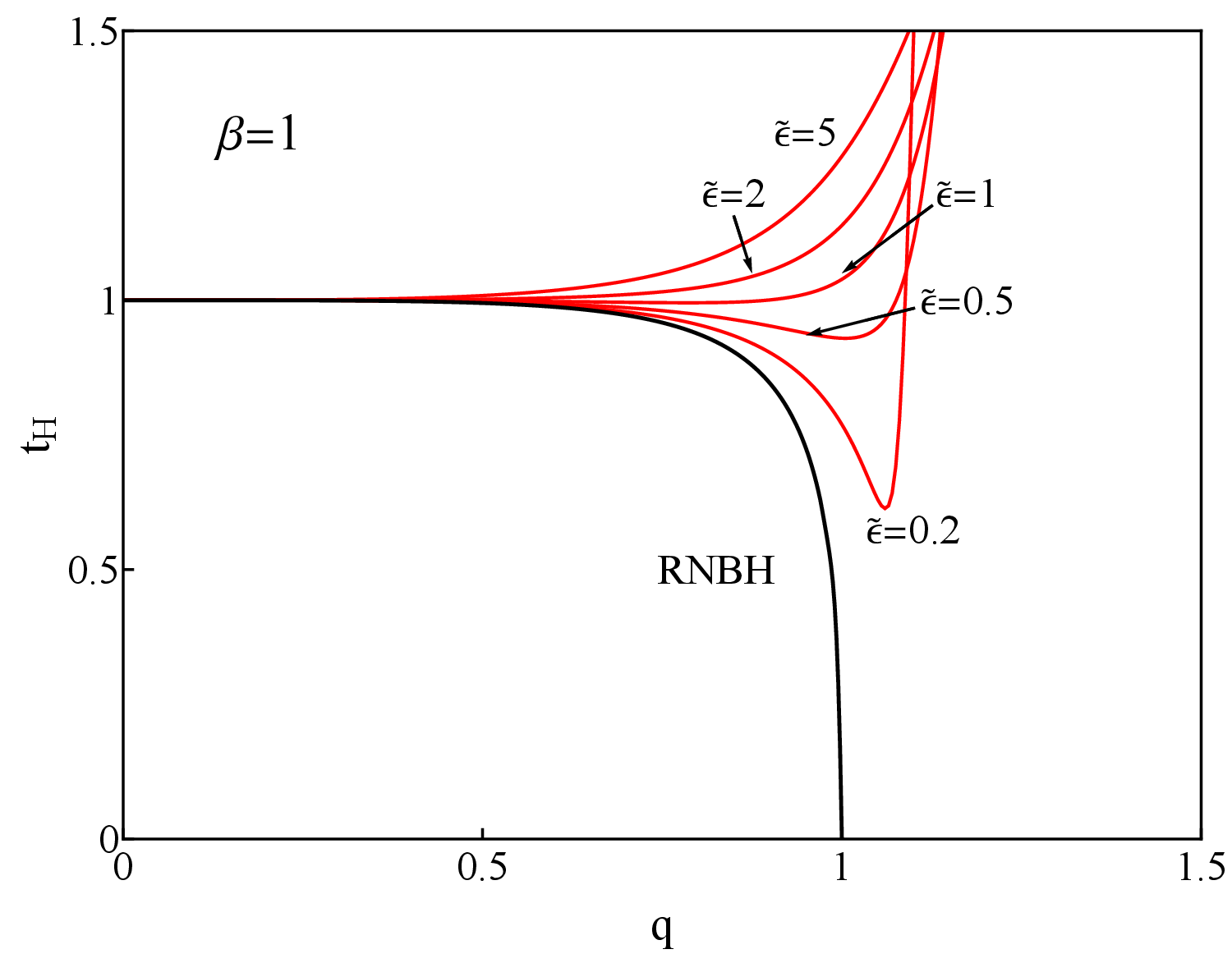}
}
\caption{(a) The reduced area of horizon $a_H$ and (b) reduced Hawking temperature $t_H$ as the function of $q$ for dyonic black holes with fixed $\tilde{\epsilon}=2$ and several values of $\beta$; (c) The reduced area of horizon $a_H$ and (d) reduced Hawking temperature $t_H$ as the function of $q$ for dyonic black holes with fixed $\beta=1$ and several values of $\tilde{\epsilon}$.}
\label{fig:d_rH_q_ep_be}
\end{figure}

Fig.~\ref{fig:d_rH_q_ep_be}(a) presents the reduced horizon area $a_H$ as a function of the charge parameter $q$ for dyonic EEH black holes with fixed $\tilde{\epsilon}=2$ and several values of the magnetic-to-electric charge ratio $(\beta=0, 0.2, 0.5, 1, 100)$. The case $\beta=0$ corresponds to the purely electric solution, where the black hole bifurcates smoothly from the Schwarzschild solution with $a_H=1$. As $q$ increases, $a_H$ decreases monotonically until the event horizon merges with the inner horizon at the extremal limit $(q\approx 1.148)$. For $\beta>0$, the dyonic solutions also bifurcate smoothly from the Schwarzschild solution. In contrast to the purely electric case, $a_H$ decreases initially, reaches a minimum, and then increases monotonically as $q$ is further increased. Since these dyonic EEH black holes possess only a single event horizon, the non-monotonic behavior originates from the re-expansion of the event horizon itself rather than from a transition between different horizon branches.

Fig.~\ref{fig:d_rH_q_ep_be}(b) shows the reduced Hawking temperature $t_H$ as a function of the charge parameter $q$ for the same set of solutions. For the purely electric case $(\beta=0)$, $t_H$ remains nearly constant for small $q$ before decreasing rapidly to zero at the extremal limit $(q\approx 1.148)$, reproducing the familiar RN behavior. Introducing a magnetic charge qualitatively changes the thermodynamic properties. For $\beta>0$, the temperature no longer approaches zero; instead, it increases rapidly at sufficiently large $q$. An intermediate behavior is observed for $\beta=0.2$, where $t_H$ first decreases to a minimum before rising again. This non-monotonic behavior reflects the changes of the single event horizon and is closely associated with its re-expansion. Overall, increasing the magnetic contribution drives the thermodynamic behavior away from the RN limit, suppresses the formation of extremal black holes, and leads to a distinct magnetic-dominated thermodynamic regime.

Fig.~\ref{fig:d_rH_q_ep_be}(c) presents the reduced horizon area $a_H$ as a function of the charge parameter $q$ for dyonic EEH black holes with fixed magnetic-to-electric charge ratio $\beta=1$ and several values of the nonlinear coupling $(\tilde{\epsilon}=0,0.2,0.5,1,2,5)$. The case $\tilde{\epsilon}=0$ corresponds to the dyonic RN solution. For $\tilde{\epsilon}>0$, the black holes bifurcate smoothly from the Schwarzschild solution with $a_H=1$. As the charge parameter increases, the horizon area decreases initially, reaches a minimum, and then increases monotonically owing to the re-expansion of the event horizon. Together with Fig.~\ref{fig:d_rH_q_ep_be}(a), these results demonstrate that both the magnetic charge and the EH nonlinear coupling suppress the formation of extremal configurations and substantially modify the horizon structure of dyonic black holes.

Fig.~\ref{fig:d_rH_q_ep_be}(d) shows the reduced Hawking temperature $t_H$ as a function of the charge parameter $q$ for dyonic EEH black holes with fixed magnetic-to-electric charge ratio $\beta=1$ and several values of the nonlinear coupling $(\tilde{\epsilon}=0,0.2,0.5,1,2,5)$. The case $\tilde{\epsilon}=0$ corresponds to the dyonic RN solution. For nonzero nonlinear coupling, the thermodynamic behavior changes qualitatively. When the coupling is relatively weak $(\tilde{\epsilon}=0.2,0.5)$, the temperature remains nearly constant at small $q$, decreases to a minimum, and subsequently rises as $q$ increases further. For stronger nonlinear couplings $(\tilde{\epsilon}\ge1)$, the minimum disappears, and $t_H$ increases monotonically after an initial plateau. Together with Fig.~\ref{fig:d_rH_q_ep_be}(b), these results show that both the magnetic charge and the EH nonlinear interaction progressively drive the thermodynamic behavior away from the RN limit, suppress the formation of extremal configurations, and establish a magnetic-dominated thermodynamic regime.

\begin{figure}[t]
\mbox{
(a)
\includegraphics[angle =0,scale=0.32]{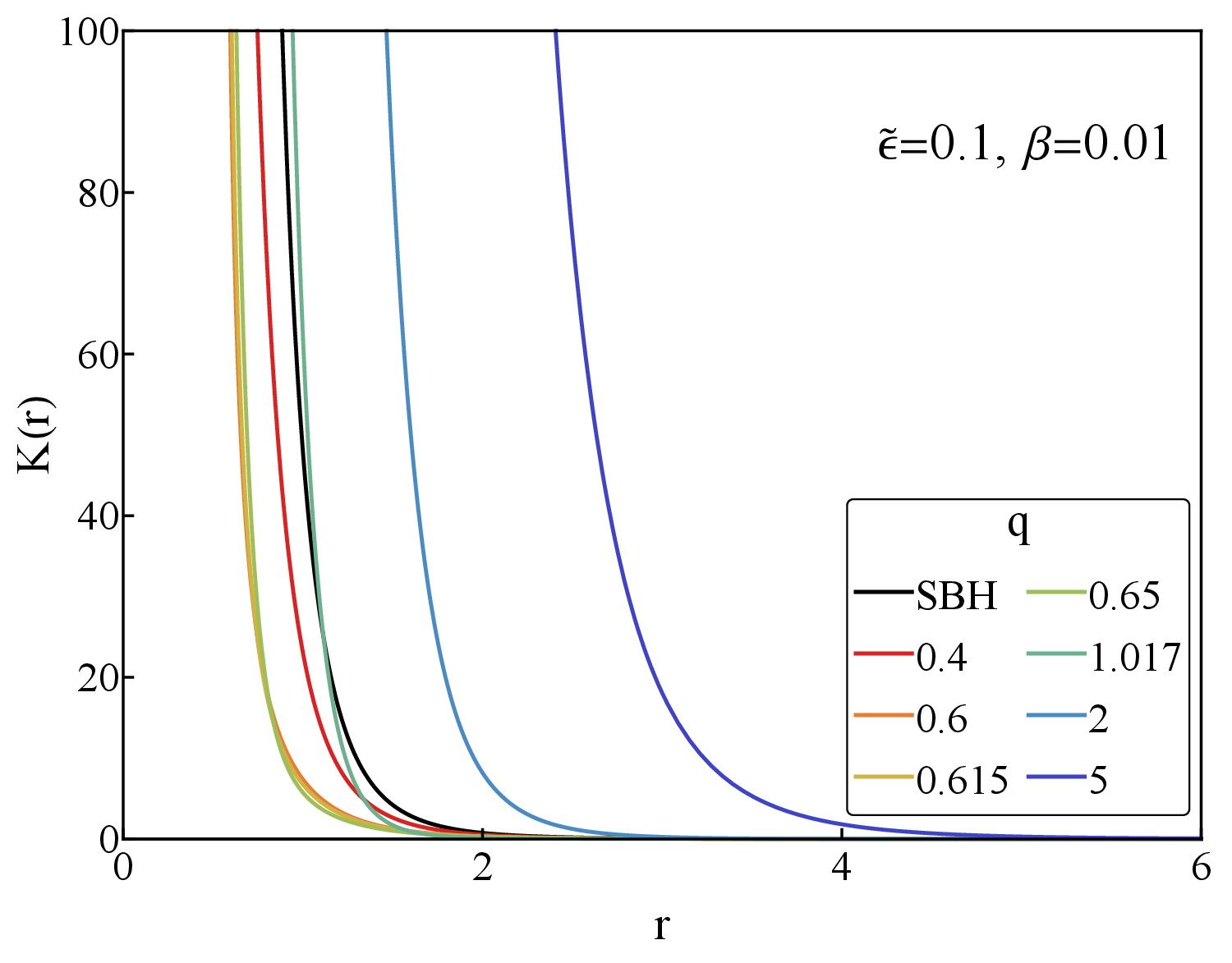}
(b)
\includegraphics[angle =0,scale=0.32]{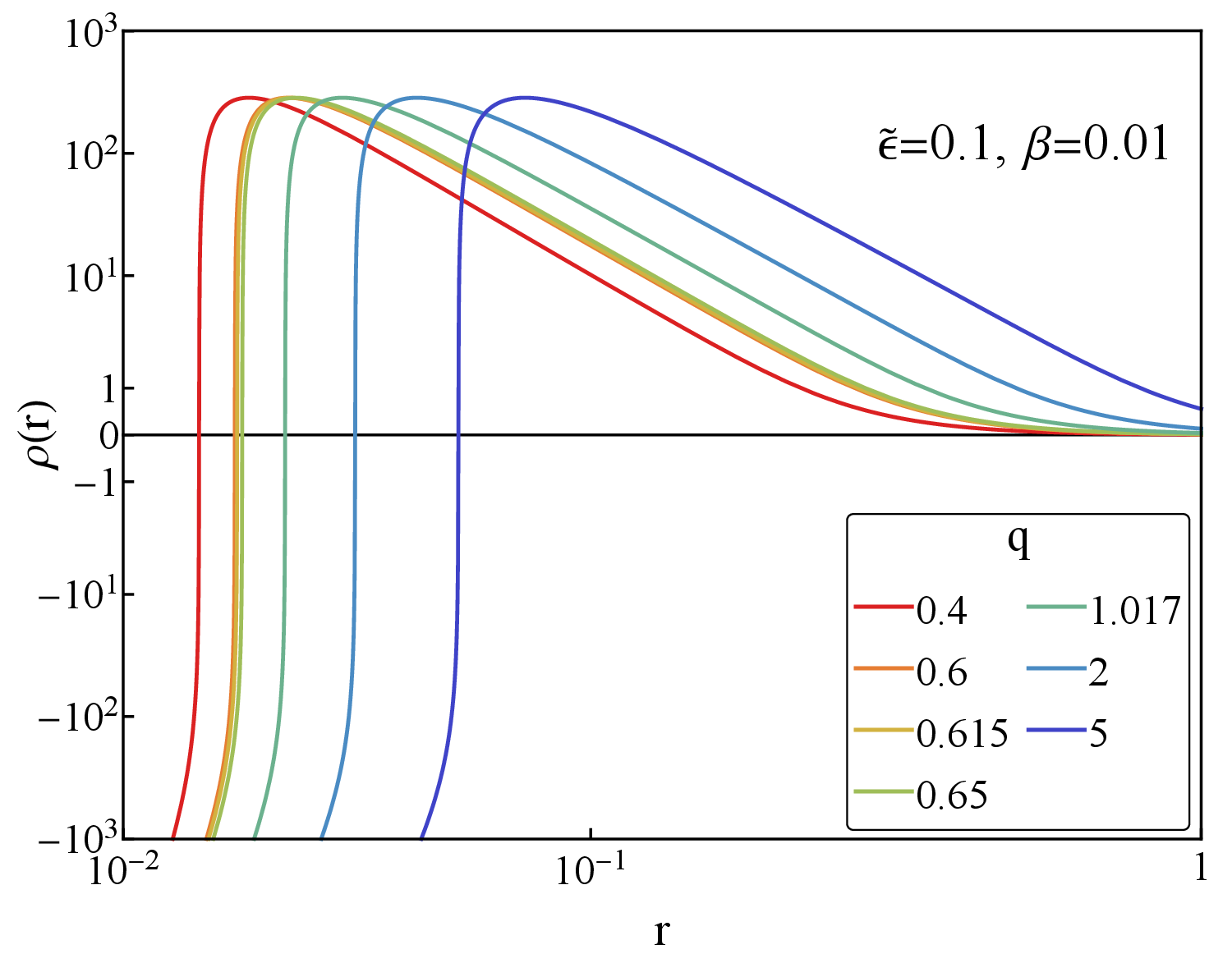}
}
\caption{(a) The Kretschmann scalar $K(r)$; and (b) Energy density $\rho(r)$ in the radial coordinate $r$ for dyonic black holes with fixed $\tilde{\epsilon}=0.1$, $\beta=0.01$ and several values of $q$ in the EEH theory.}
\label{fig:d_K_rho_vs_r}
\end{figure}

Fig.~\ref{fig:d_K_rho_vs_r}(a) presents the Kretschmann scalar $K$ for dyonic EEH black holes with fixed $\tilde{\epsilon}=0.1$, $\beta=0.01$, and several values of the charge parameter $q$. For all values of $q$, the Kretschmann scalar diverges as $r\rightarrow0$ and decreases monotonically with increasing radial coordinate, confirming the existence of a central curvature singularity. Hence, although the EH nonlinear interaction substantially modifies the horizon structure, it does not regularize the spacetime or eliminate the singularity at the origin.

Fig.~\ref{fig:d_K_rho_vs_r}(b) displays the corresponding energy density $\rho$. Near the singularity, the energy density becomes negative, indicating a violation of the weak energy condition in the strong-field region. As the radial coordinate increases, $\rho$ changes sign, reaches a positive maximum, and then gradually decreases toward zero, recovering the asymptotically flat behavior at large distances. This behavior shows that the nonlinear EH correction is significant only in the vicinity of the black-hole interior, while its influence becomes negligible far from the horizon.

\begin{figure}[t]
\mbox{
(a)
\includegraphics[angle =0,scale=0.3]{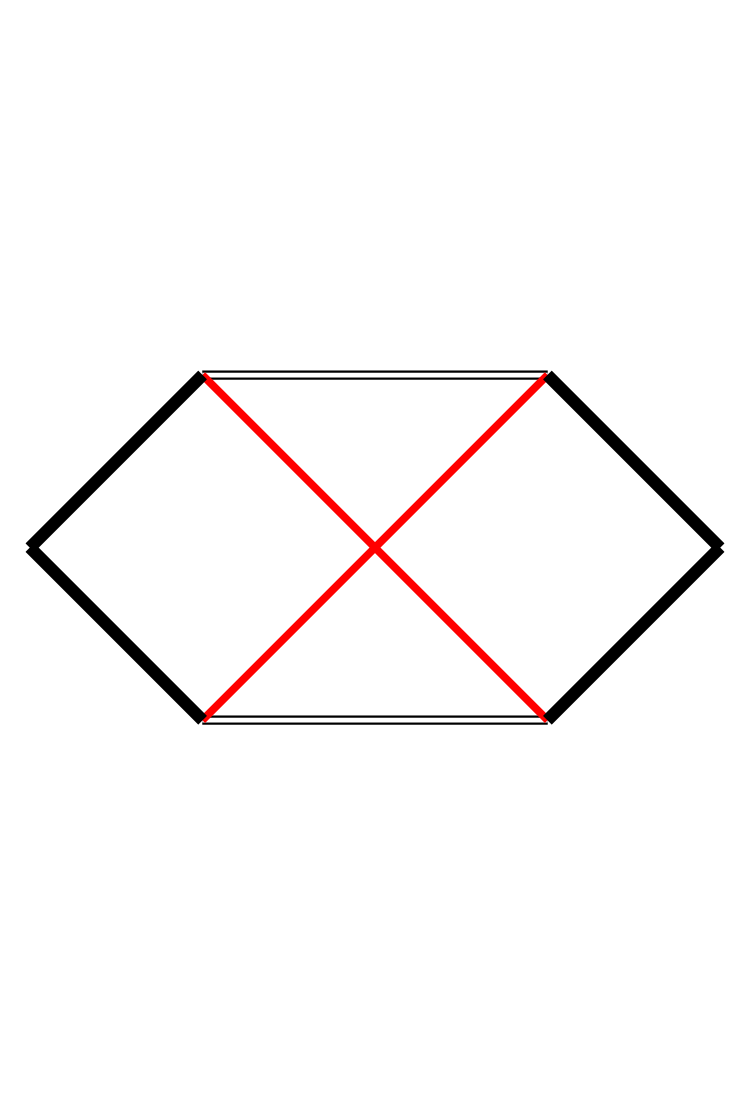}
(b)
\includegraphics[angle =0,scale=0.3]{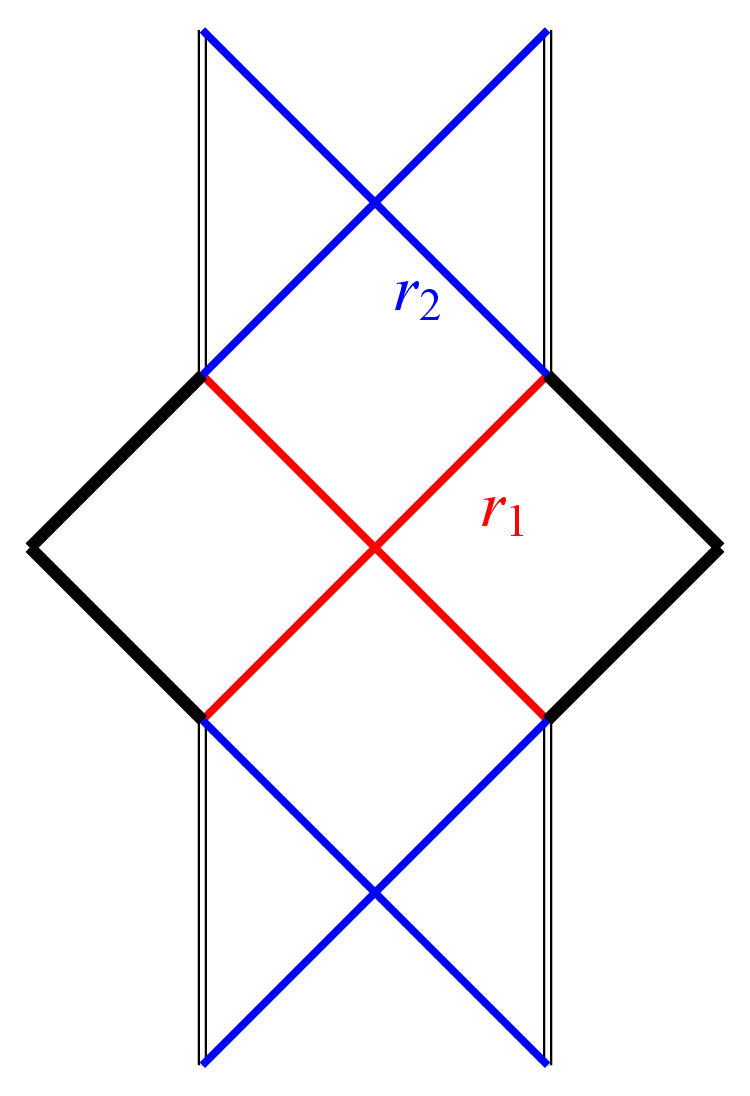}
(c)
\includegraphics[angle =0,scale=0.3]{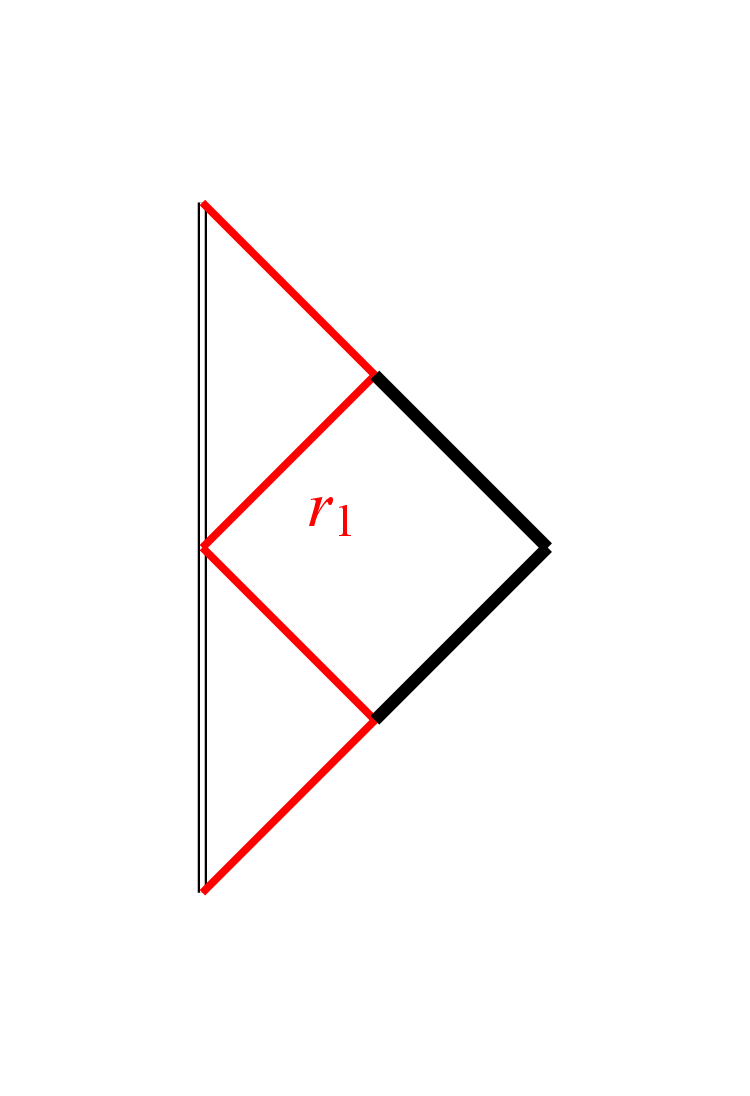}
(d)
\includegraphics[angle =0,scale=0.3]{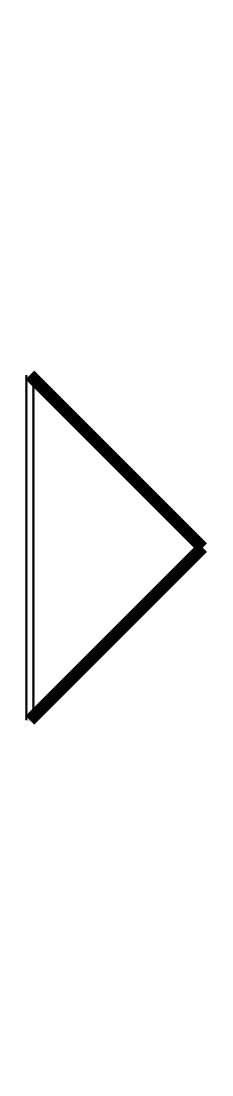}
}
\mbox{
(e)
\includegraphics[angle =0,scale=0.3]{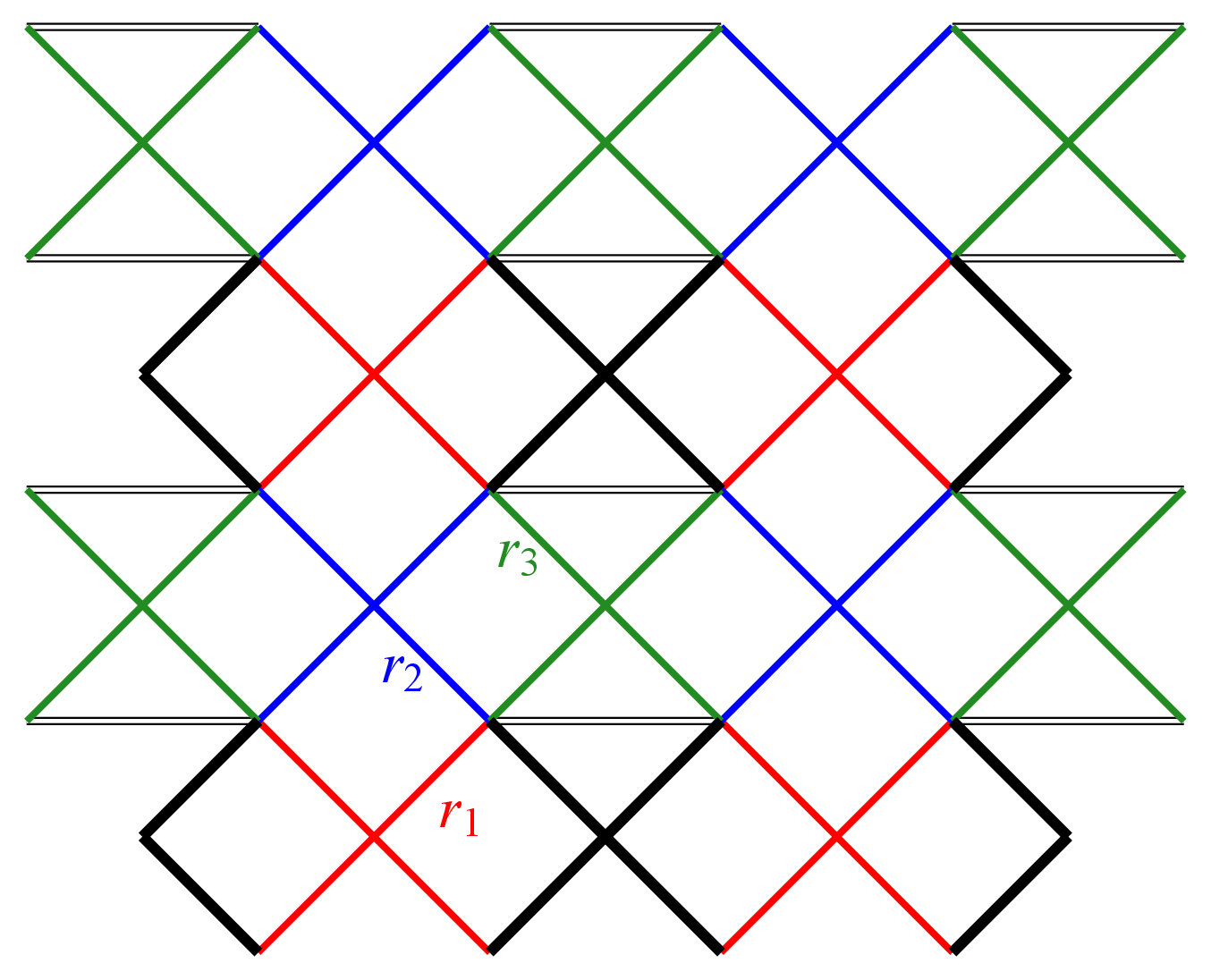}
(f)
\includegraphics[angle =0,scale=0.3]{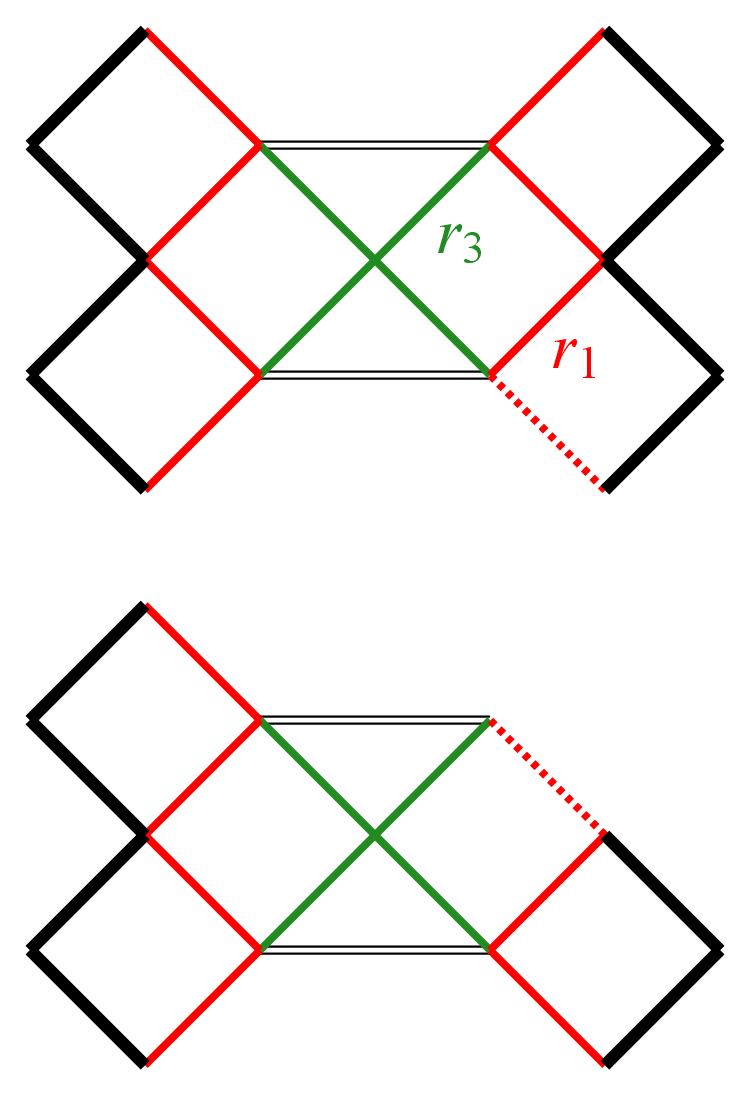}
}
\caption{Penrose diagrams illustrating the causal structures of EEH black holes: (a) Schwarzschild-type spacetime with a single event horizon; (b) RN-type spacetime with two horizons; (c) extremal RN-type spacetime with a degenerate horizon; (d) naked singularity; (e) three-horizon EEH black holes; and (f) two-horizon EEH black holes obtained from the degeneration of the three-horizon configuration.}
\label{fig:penrose_diagram}
\end{figure}

\subsection{Causal structures of black holes}

Lastly, we briefly discuss the causal structures of EEH black holes, which are summarized schematically in Fig.~\ref{fig:penrose_diagram}. Several limiting cases coincide with familiar electrovacuum black holes. When the spacetime possesses a single event horizon with a spacelike singularity, the causal structure is identical to that of the Schwarzschild black hole, as shown in Fig.~\ref{fig:penrose_diagram}(a). On the other hand, the purely electric EEH solutions inherit the causal structures of the RN spacetime, depending on whether the solution possesses two horizons, a degenerate horizon, or no horizon, as illustrated in Figs.~\ref{fig:penrose_diagram}(b)--(d), respectively.

The most distinctive causal structures arise for the purely magnetic and dyonic EEH black holes, whose singularities are spacelike. In this case, the spacetime may contain three distinct horizons: $r_1>r_2>r_3$, which $r_1$ denotes the event horizon, $r_2$ the intermediate horizon, and $r_3$ the innermost horizon surrounding the spacelike singularity. The corresponding maximally extended Penrose diagram is illustrated schematically in Fig.~\ref{fig:penrose_diagram}(e). After crossing the event horizon $r_1$, causal trajectories necessarily pass through the intermediate horizon $r_2$. Subsequently, they may either return to another asymptotically flat region by crossing $r_1$ again, reminiscent of the RN wormhole structure, or continue inward across $r_3$ toward the spacelike singularity, analogous to the Schwarzschild interior. Consequently, the elementary Penrose block shown in Fig.~\ref{fig:penrose_diagram}(e) is expected to repeat indefinitely in the maximally extended spacetime.

If two of the three horizons coincide, the spacetime reduces to the two-horizon configuration shown in Fig.~\ref{fig:penrose_diagram}(f). In this case, the exterior region resembles that of an extremal RN black hole, whereas the interior terminates at a spacelike singularity after crossing the remaining inner horizon. The resulting causal structure therefore interpolates between those of the extremal RN and Schwarzschild spacetimes. An interesting feature concerns the global extension of the two-horizon geometry. Although neighboring Penrose blocks are joined continuously across the event horizon, the remaining horizons need not connect to the same causal patch. Consequently, the maximally extended spacetime cannot, in general, be represented completely within a single two-dimensional Penrose diagram. Instead, its global structure is expected to consist of multiple Penrose blocks connected in a nontrivial multi-sheeted, or knot-like, arrangement.

Finally, when all three horizons coincide, the spacetime reduces to the Schwarzschild-type causal structure shown in Fig.~\ref{fig:penrose_diagram}(a). Therefore, the causal structures of EEH black holes extend the familiar Schwarzschild and RN geometries by admitting novel multi-horizon configurations that have no counterparts in EM theory.

\section{Conclusion and Outlook}
\label{sec:Conc}

In this work, we have investigated static, spherically symmetric purely electric, purely magnetic, and dyonic black holes in Einstein gravity coupled to Euler--Heisenberg (EH) nonlinear electrodynamics. Rather than employing the Hamiltonian formulation based on the auxiliary electromagnetic invariant $\mathcal{P}$, we formulated the Einstein--Euler--Heisenberg (EEH) field equations directly in terms of the original electromagnetic invariant $\mathcal{F}$. This formulation avoids the introduction of auxiliary variables and treats all charged configurations within the original EEH Lagrangian. As a consequence, we obtained an exact analytical solution for the purely electric configuration, recovered the purely magnetic solution directly from the original field equations, and constructed the dyonic black-hole solutions numerically.

We performed a systematic analysis of the horizon structure, geometrical properties, and thermodynamic behavior of these solutions. The purely electric black holes exhibit one- and two-horizon phases analogous to those of the Reissner--Nordstr\"om (RN) spacetime. By contrast, the purely magnetic solutions possess a novel three-horizon configuration consisting of one event horizon and two inner horizons. The dyonic solutions interpolate continuously between these limiting cases, displaying either one- or three-horizon phases depending on the magnetic-to-electric charge ratio and the EH nonlinear coupling. We further showed that increasing either the magnetic contribution or the nonlinear coupling progressively suppresses the three-horizon phase, eventually leaving only a single event horizon in the strong-coupling regime. The geometrical and thermodynamic properties were further characterized through the curvature invariants, reduced horizon area, Hawking temperature, and Smarr relation. The resulting phase diagrams reveal that both the magnetic charge and the EH nonlinear interaction substantially modify the horizon geometry and thermodynamic behavior, driving the solutions away from the RN limit. In particular, the magnetic field plays a central role in determining both the causal structure and thermodynamic properties of the spacetime. Overall, these results demonstrate that the EH nonlinear interaction gives rise to qualitatively new causal structures beyond Einstein--Maxwell (EM) theory while significantly reshaping the geometry and thermodynamics of charged black holes.

Several interesting directions remain for future study. In particular, it would be worthwhile to investigate the linear stability of these black holes through spherical perturbation and quasinormal-mode analyses, and to examine observational signatures such as black-hole shadows and ringdown spectra. Such studies may help to determine whether EH nonlinear electrodynamic effects can be distinguished from the EM prediction in the strong-field regime. Equally important is the study of their fully nonlinear evolution, for example by means of double-null simulations  \cite{Nakonieczna:2018tih,Chew:2023upu}, to determine the dynamical fate of the multiple-horizon configurations and to clarify the impact of EH nonlinear electrodynamics on mass inflation and the strong cosmic censorship conjecture.

\section*{Acknowledgment}
XYC is supported by National Science Foundation of China (No.~W2533026). DY was supported by the National Research Foundation of Korea (NRF) grant funded by the Korean government (No.~RS-2026-25476711). CC is supported by National Natural Science Foundation of China (No.~12503003, No.~1243300), National Key R\&D Program of China (No.~2021YFC2203100).

\bibliography{mybiblio}



\end{document}